\documentclass[11pt]{article}
\usepackage{graphicx} 
\usepackage{physics}

\usepackage{amsmath}
  \usepackage{subcaption}

  \usepackage[parsep]{collref}
  
 \usepackage{booktabs,float,slashed}
\usepackage[setpagesize=false,pagebackref=false, 
linktocpage, bookmarksopen=true, colorlinks=true, 
linkcolor=black,citecolor=black,urlcolor=black]{hyperref}

\DeclareFontFamily{OT1}{rsfs10}{}
\DeclareFontShape{OT1}{rsfs10}{m}{n}{ <-> rsfs10 }{}
\DeclareMathAlphabet{\mathscript}{OT1}{rsfs10}{m}{n}

\numberwithin{equation}{section}
\usepackage[T1]{fontenc}
\usepackage[dvipsnames]{xcolor}
\usepackage{tikz} 
\usetikzlibrary{arrows,decorations.pathreplacing,decorations.markings,snakes}
\usetikzlibrary{cd}

\newcommand{\ZZ}{{\mathbf{Z}}}

\newcommand{\bN}{{\mathbf N}}

\newcommand{\al}[1]{\begin{align}{#1}\end{align}}

\def\a{\alpha}
\def\b{\beta}

\def\c{\chi}

\def\f{\phi}

\def\z{\psi}
\def\k{\kappa}
\def\l{\lambda}

\def\r{\rho}
\def\s{\sigma}

\def\z{\zeta}

\def\G{\Gamma}
\def\J{\Psi}

\def\O{\Omega}

\def\U{\Upsilon}

\def\cE_{{\mathcal E}}

\def\bb{{\bar b}}

\def\gsim{ \lower .75ex \hbox{$\sim$} \llap{\raise .27ex \hbox{$>$}} }
\def\lsim{ \lower .75ex \hbox{$\sim$} \llap{\raise .27ex \hbox{$<$}} }
\def\be{\begin{equation}}
\def\ee{\end{equation}}
\def\bea{\begin{eqnarray}}
\def\eea{\end{eqnarray}}

\def \td {\tilde}

\def \ha {{1 \ov 2}}

\def \del{\partial}
\def \a {\alpha}

\def\ov{\over}
\def \ci {\cite}

\def \foot {\footnote}
\def \bi{\bibitem}
\def\la{\label}
\def\foot{\footnote}
\newcommand{\rf}[1]{(\ref{#1})}

\def \OO {{\cal  O}}\def \no {\nonumber}

\def \z {\zeta}

\def \C  {{\cal C}}

\def \ov {\over}

\def \iffa {\iffalse}

\def \iffa  {\iffalse}

\def \s {\sigma}  
\def \G {\Gamma} \def \four {{1\ov 4}}
\def \gs {g_s}

\def \te {\textstyle}
 
\def \ve {\varepsilon}

  \def \rT {{\rm T}}

\def \no {\nonumber}
 
  \def \rT  {{\cal T}} 

\def \ss {\tau}

\def \ba {\begin{align}}
\def \ea {\end{align}}

\def \nf {{\rm n}_{_{\rm F}}}

\def \T  {{\rm T}}

\def \ln {{\rm log\,}}

\def \edo {\newpage
\small
\bibliographystyle{JHEP-v2.9}
\bibliography{bib}
\end{document}}
\def\la{\label}\def\foot{\footnote} 
 
\def \ss {{\rm s}}    

\def \Z  {{\cal Z}}

\def \OO  {{\cal O}}

 \def \T {{\rm T}}

  \def \O {{\cal O}}

\def \C {{\cal Q}}
  \def \ls {\ell_s}  \def \gs {g_s}

  \def \s {\sigma} \def \fo {\tfrac{1}{4}}

  \def \z {{\cal Z}} \def  \ha {\tfrac{1}{2}}

    \def \bb {{\rm b}}
  
   \def \k {{\kappa}}
   
   \def \det {{\rm det\,}}

     \def \r   {\rho}
\def \Z  {\mathbb{Z}}       
     \def \tR {\tilde R}
\def \tR {\tilde R}

\def \z  {\zeta}

\def \l  {\lambda}

\def\re{\operatorname{Re}}
\def\li{\operatorname{Li}}
\def \B {\mathcal B}
\def \C {\mathcal C}

 \def\be{\begin{equation}}
 \def\ee{\end{equation}}
\def\ba{\begin{align}}
 \def\ea{\end{align}}

\def\bn{\begin{enumerate}}
\def\en{\end{enumerate}}
\def\bi{\begin{itemize}}
\def\ei{\end{itemize}}
\def\it{}

\def\rT{\mathrm T}
\def \Z{\mathcal Z }
\def\lra {\leftrightarrow}
\newcommand{\Oe}[1]{ \mathcal O(e^{-{#1}\pi R/\tilde R})}
\def\c {l}
\def\T{\mathrm T}

\def \Ee{\mathcal A}

\def \f {\mathcal E}
\def \tR {\tilde R}
\def \Bsumt {\mathcal G} 
\def \Csumt {\mathcal K} 

\def \U {\mathrm U}
\newcommand{\ntD}[1]{\Delta^{{#1}}} 
\newcommand{\tqD}[1]{\Delta^{(q){#1}}}

\def \E {{\cal E}}

\def \ed {
\small
\bibliography{biblio2.bib}
\bibliographystyle{JHEP-v2.9}
\end{document}
}
\def \edo {\end{document}}

\makeatletter
\renewcommand*{\@fnsymbol}[1]{\textit{\@alph{#1}}}
\makeatother
\usepackage{amsmath,amssymb,lmodern}    
\usepackage{caption}
\usepackage{dsfont}
\numberwithin{equation}{section}
\usepackage{slashed}

\pdfoutput=1
\usepackage{amsthm, amssymb,latexsym,amsfonts}
\usepackage[T1]{fontenc}
\usepackage{comment}
\usepackage{graphicx}
\usepackage{caption}
\usepackage{amsmath}
\usepackage{appendix}
\usepackage{tikz} 
\usetikzlibrary{cd}
\begin{document}
\begin{titlepage}

\hfill{\small Imperial-TP-2026-AAT-02}

\vspace{1cm}

\begin{center}
  {\LARGE Energy of toroidal M2 brane 
  in flat 11d background \par}
  
  \vspace{1.5cm}
  
  {\large 
    Arkady A. Tseytlin\footnote{Also at the Institute for Theoretical and Mathematical Physics (ITMP) of MSU and Lebedev Institute.  tseytlin@ic.ac.uk.} 
    \ \ \ \ \ and \ \ \ \ 
    Zihan Wang\footnote{zihan.wang18@imperial.ac.uk}
  \par}
  
  \vspace{0.6em}
  
  {\it \small  Abdus Salam Centre for Theoretical Physics \\
  
  Imperial College London, SW7 2AZ, UK \par}
\end{center}

\vspace{1cm}

\begin{abstract}
\noindent
The supermembrane action is non-linear and a priori non-renormalizable. Still, some quantities in this theory  may be  free of log UV divergences  and thus  defined unambiguously.  To explore this possibility  we compute, in static gauge,
the energy of an M2 brane wrapped on a 2-torus  in
flat 11d  space to two loops in the inverse-tension expansion.
The result is free of logarithmic  divergences and thus well defined in the
standard $\zeta$-function regularization. The same should hold also  at higher
loop orders.  For fermions periodic around both circles the
quantum corrections cancel, consistent with the BPS nature of the wrapped M2
brane state. For fermions antiperiodic around one or both circles the 
energy  is a 
non-trivial function of the radii $R_{10}$ and $R_{11}$, given by infinite
sums of modified Bessel functions.
We also  compute the 3-loop correction to the energy of the bosonic membrane on
$\mathbb R^2\times S^1$. In contrast to the string case, it contains, besides
powers of the 1-loop $\zeta(3)$  coefficient, a new term proportional to
$\zeta(9)$.
Summing  contributions of   all-loop   bubble  graphs   leads 
to a simple cubic equation for the membrane energy.  The
analogous resummation in the Nambu or GS 
string theory case reproduces the familiar exact
square-root expression
$
{\cal E}=\sqrt{(2\pi R\,T_1)^2+m_0^2}
$
found in
the orthogonal or light-cone gauge.
 We find that the    bubble-graph part of 
  the membrane  energy   does  not 
vanish for any value of the radius $R_{11}$.  
If  contributions of other non-trivial  diagrams in the supermembrane case 
 do not 
qualitatively  change this conclusion,  this  would 
disfavour the conjectured relation between 11d  theory on an antiperiodic circle and the strong-coupling limit of type 0A string theory  (which assumes  that  the wrapped-brane state
 should become massless at a critical value of  $R_{11}$ or string  coupling).
Finally, we generalize the 2-loop analysis to an M2 brane in a flat
11d  ``twisted''  background that interpolates
between the periodic and antiperiodic fermion boundary conditions.

\end{abstract}
\end{titlepage}

\tableofcontents

\def \bN  {\bar N} \def \Np { {\bar N_P}   } \def \Na  {   {\bar N_A}  }
\def \ss {{\rm s}}
\def \lp {\ell_{\mathrm P}}
\def \Ee {{\cal A}}
\def \tD {\Sigma}
\def\hal {\frac{1}{2}}
\def \lra {\leftrightarrow}
\def \tR {\tilde R}
 \def \UU {\Phi}

 \def \bE {\bar \f} 
\def \cJ {{\cal J}} \def \nF {\nf}
 \def \cE {{\cal E}}
 \def \Rt {\tR} \def \cT {{\rm T}}
\def \bT {\bar T} \def \diag {{\rm diag}} \def \bb {{\rm bb}}
\def \J  {{\cal X}}
\def \ve {\varepsilon}
\newcommand{\angl}[1]{\big\langle #1 \big\rangle_{\rm angl}}

\setcounter{footnote}{0}
\section{Introduction }

Semiclassical quantization of Green-Schwarz    string   in curved RR  backgrounds  provides  an important source of information about  strong-coupling expansion in dual gauge theories, in some cases 
  allowing non-trivial  checks of AdS/CFT  duality.  At the same time,   GS    action  is formally non-renormalizable 
  and thus may  require  a  particular prescription of how to define   higher loop corrections  or how to fix    counterterms. In  the most symmetric cases (and at  the leading order in  string coupling expansion) one expects   that requiring that  integrability  should apply to
  the  full quantum   theory   may    provide such a  UV definition of  the theory. 
  In the simplest case of  the bosonic  Nambu string  expanded  about  long-string vacuum in static gauge 
 it is necessary  to add particular $(\int  K^4, \ \ K= \del\del X + ..$)  counterterms  in order for the 2-loop world-sheet 
  S-matrix   to be given by  the elastic  pure-phase expression   \cite{Dubovsky:2012sh,Conkey:2016qju}
  required for the equivalence  of the spectrum  with the  l.c. gauge one.  The same applies  to the case of the GS string \ci{Seibold:2023zkz,Seibold:2024oyr,Beccaria:2025xry}. Demanding   quantum integrability thus selects a particular UV completion of a formally non-renormalizable world-sheet theory.

  In the case of the bosonic membrane or  BST M2 brane \ci{Bergshoeff:1987cm}  
   there are no  log divergences at the 1-loop order  but 
  they do appear at the 2-loop order \ci{Beccaria:2025xry}  and,  in contrast to the string case,  
  here  there is no  known principle that  would  
   fix  the  required UV counterterms (like the 2-loop $\int  (\del K)^2 K^2$ one).
  This   raises the question   about  predictive power of a semiclassical  M2 brane  quantization as a tool to test the AdS/CFT   correspondence  on the 
  M-theory side. At the same time,  while the M2  brane action   is non-renormalizable in a general  3d world-volume 
  background,   its  2-loop   free energy may still be free of log  UV  divergences 
    when expanded  near a particular M2   solution.  This is the case, in particular, 
  in the case of  M2 brane on $AdS_3\subset  AdS_7 \times S^4$  \ci{Beccaria:2025ahf}.
  
  \subsection{Setup} 
  
  Here  we will 
  consider an example where  the same happens in a much simpler   setting: 
  loop    corrections to   free  energy of a toroidal  M2 brane  in flat  11d  target   space background.
  It is found to be free of 1-loop and 2-loop 
  log UV  divergences (consistent  with the vanishing of  potential $\int  (\del K)^2 K^2$  counterterm) 
  and   thus well  defined  in the standard $\z$-function regularization. 
  The same  is found to be true also at the  3-loop  order in the bosonic membrane case, and, in fact, 
   should be true to all loop orders 
 as  one cannot construct a   covariant   counterterm that is non-vanishing 
  on such wrapped   membrane background.

  We shall assume that the 11d space  has two  circular directions ($X^{10}$  and $X^{11}$)   and that the 
  M2 brane   is wrapped on  these  circles  with  their  radii  denoted as\foot{We shall  use the
   static gauge   with the  time 
   coordinate  on the world volume  identified with the  target space time (we shall  formally assume  that the  time is 
  Euclidean  in both  target  space and world volume,  with  a continuation  to 
  Minkowski  case being straightforward). As a result, the target space energy   will be the same as the energy of the world-volume theory. }
  \be  R_{10} \equiv R \ , \ \ \ \ \ \ \ \ \   R_{11} \equiv \td R \ . \la{1} \ee
  In the case when  the M2 brane fermionic   coordinates are  periodic in both circles this    configuration is supersymmetric  and 
  thus 
  all quantum corrections   to the 
   free energy should cancel. This was checked  in the 1-loop  approximation in \ci{Duff:1987cs}  using  the l.c. 
  gauge  formulation. Here we will  demonstrate this cancellation  
   also  at the  2-loop  order using the static gauge  and adapted $\kappa$-symmetry gauge   approach.
  
 In the case   when the fermions  are taken to be  antiperiodic in one or two 
  circles   so that supersymmetry is broken,  
  the  1-loop and 2-loop 
  contributions to free energy will  no longer  vanish. The 1-loop  correction was  computed   using the  l.c. gauge formulation 
   in \ci{Russo:2001vh,Russo:2001tf}  (cf. also \ci{Bytsenko:1989bv}).  Here we  
will      find the 2-loop   correction to the M2 brane energy. 
  This case  may be related to  a  strong-coupling limit 
  of type 0 string  which was conjectured to correspond to 
  M-theory  on an antiperiodic  circle \ci{Bergman:1999km}. 
  
    In contrast to the string case, the   M2 brane   theory  remains   non-linear
  and   not exactly solvable   in the  l.c. gauge \ci{Bergshoeff:1987qx,deWit:1988wri}; 
   thus there is no  known  analog 
  of the  familiar    square root  expression  for the energy  in the string case (see \rf{5} below).  
  This precludes one from drawing   definite conclusions   about  critical values of  radii
    or onset  of instability (when  the energy $\E$   vanishes) 
   just from   the  knowledge of the first few orders in  inverse tension expansion of $\E$.



We shall  define the  free energy density $\f$  by
 a semiclassical expansion  of   M2  brane  path integral 
\be\la{6}
Z=\int D X D \theta \ e^{-S[X, \theta]}=e^{-F}\ , \qquad \qquad F=V \f\ , \ \ \ \ \   \ \f= \f_0+\f_1+\f_2+...\ .
\ee
Here  $V$ is the volume of 
 non-compact world-volume  directions (i.e.  length of  Euclidean time interval   in the case of  membrane wrapped on a 2-torus). 
$S$  is the static-gauge  M2  brane action proportional  to the tension $T_2$. \   $\E_0, \E_1, \E_2, ...$ stand  for  the classical, 1-loop, 2-loop, etc., terms in   inverse tension expansion. 
The  actual  expansion parameter is a dimensionless   tension 
$\rT_2$ that may be   defined as
\be
\rT_2\equiv (2\pi)^2 \tR^3T_2={\tR^3\ov \ell_{\mathrm P}^3}\ ,\qquad \qquad T_2= \frac 1{(2\pi)^2 \ell_{\mathrm P}^3}\ , \qquad\ \ \ \  \tR \equiv R_{11} \ .  \la{7}
\ee
Explicitly, fixing the  static  gauge   and the associated $\k$-symmetry
 gauge   conditions  as
 \al{
& X^1= \sigma^1\ ,\qquad \qquad X^{10}\equiv X^{10} + 2\pi R =\s^2\ ,\qquad\qquad X^{11}\equiv X^{11} + 2 \pi\tR=\s^3\ , \label{14}
\\
&  \mathcal P \theta =\theta,\qquad\qquad \quad  \mathcal P=\tfrac{1}{2} (1+\Gamma^{123})\label{15}\ , 
 }
 one finds   the following expansion \cite{Seibold:2024oyr,Beccaria:2025xry} of the M2 brane  BST  action\footnote{Here we keep the relevant quartic terms in fluctuation fields (that are rescaled by $\sqrt{T_2}$)  and  drop terms proportional to the equations of motion which will not contribute to the free energy. $\Gamma^M$ are 11d  Dirac matrices.}
\al{&S[X, \theta]=\int d^{\,3} \sigma \;\Big(  \mathcal L_B[X]  + \mathcal L_F[X, \theta] \Big)\ ,      \qquad \qquad  a=1,2,3,  \qquad \  i=1, ..., N, \ \ \   N=8\ , \la{101}\\
&\te \mathcal L_B=T_2+\frac 12\partial_a X^i \partial^a X^i+\frac 1{T_2}\Big(\frac 1{8} \partial_a X^i \partial^a X^i\partial_b X^j \partial^b X^j
-\frac 1{4} \partial_a X^i \partial_b  X^i\partial^a X^j \partial^b X^j\Big)+...\ ,\label{11}\\
&\te \mathcal L_F= -\frac 12 \bar{\theta} \Gamma^a \partial_a \theta
  +\frac 1 {T_2}\Big(\frac{1}{4} \partial_a X^i \partial_b X^i \bar{\theta} \Gamma^a \partial^b \theta
+\frac{i}{8} \epsilon^{a b c} \partial_a X^i \partial_b X^j \bar{\theta} \Gamma_{i j} \partial_c \theta
-\frac{1}{16} \bar{\theta} \Gamma_a \partial_b \theta \bar{\theta} \Gamma^b \partial^a \theta\Big)+...\ .\label{12}
} 
Note that there are no cubic vertices in \rf{11},\rf{12} so that  
 the  2-loop contribution comes solely from the bubble  diagrams (see Fig. \rf{fig0})  involving  products of  derivatives of free propagators at  coinciding points. 
 \begin{figure}[H]
    \centering
        \centering
        \includegraphics[width=0.3\textwidth]{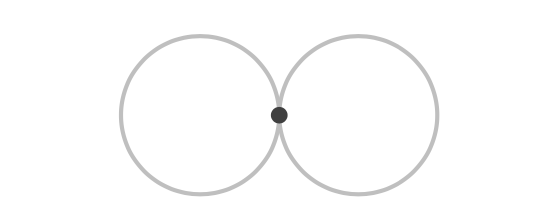}
    \caption{\small 
  2-loop vacuum double bubble diagram.
    }
    \label{fig0}
\end{figure}
 Since the 3d propagators at coinciding points 
  have no log UV divergences,  the resulting 2-loop  contribution is  guaranteed not to have UV logs either. The  remaining power   divergences  can   be  redefined away using the $\z$-function  regularization. 
 As a result, one  finds  a   finite expression for  the free energy $\E$  of  wrapped M2  brane. 

It is useful also to draw analogy with a similar computation in the string theory case. 
The expression for the GS superstring action  is related to  \rf{101}--\rf{12}   by  a double  dimensional reduction \ci{Duff:1987bx}.
The    10d string  parameters  $(\ls,\gs)$   corresponding to reduction on the 11th circle  are  related to $\tR$ and $\lp$ in \rf{7} as
\be
 T_1 =2\pi  \tR\,  T_2=\frac 1 {2\pi \ell_s^2}\ ,\qquad\qquad  \tR=\gs^{2/3} \lp= g_s\ell_s\ ,\qquad\qquad   \ell_{\mathrm P}=g_s^{1/3}\ell_s \ , \qquad \ \ 
  \rT_2 = {\tR^3\ov \lp^3}= { \gs^2} \ . \label{144}
\ee
  The energy  of  bosonic or  GS string  wrapped on a circle  of radius $R$   can  also be computed    in the standard  way \ci{Arvis:1983fp}  using the  orthogonal or l.c. gauge 
  (and the $\zeta$-function regularization)\foot{The type 0 string  expression may be obtained  by starting with GS string action and imposing  antiperiodic  boundary conditions  on the  fermions around  the  circle (see \ci{Skrzypek:2021eue}).}
  \al{
 &\te  \f^2 = (2\pi R\, T_1)^2 + m^2_0 \ , \ \ \ \ \ \  \ \ \   m^2_0=4 \pi T_1 \zeta(-1) \bN= - {1\ov 6 \a' } \bN \ , \qquad 
     \ \   \ \ \  T_1 = {1\ov 2\pi \ls^2 } \ , \ \ \ \  \ \ \  \ls= \sqrt{ \a'} \ , 
 \la{2} \\
& \te   \bN\big|_{\rm bose } =  N \ ,  \ \ \ \ \ \ \ 
   \bN\big|_{\rm type\ II}  =\bN_P=N- \ha \nf =0  \ ,  \ \ \ \ \ \ \ 
  \bN\big|_{\rm type\ 0} =\bN_A=  N+ \fo \nf =12  \ .   
\la{4}   }
 Here $N=D-2$ is the number of transverse coordinates  ($N=24$ for the  bosonic and $N=8$   for the  type 0 and type II strings)
  and $\nf=2N=16$ is the number of  physical  components of the fermionic $\theta$ coordinates (after $\k$-gauge fixing). 
  $\bN_{P/A}$   correspond to  periodic/antiperiodic $S^1$ b.c. for the fermions. 
  $m_0$   is the ground state mass. 
  Expanding the  square root expression for $\E$  gives 
  \al{ \la{5} 
  \E= \sqrt{ (2\pi R\, T_1)^2 + m^2_0}\   =    \  2\pi R\, T_1 + { m^2_0 \ov 4 \pi R\, T_1} -  { m^4_0 \ov 64 \pi^3 R^3 T_1^3} + ... \ . 
  }
   Starting with the Nambu or GS  string in the static gauge
     the subleading  terms  in \rf{5} should  correspond  to the 1-loop, 2-loop, etc. 
   corrections to  the world-volume  free energy    computed in the  inverse string tension expansion. We shall   verify this explicitly  below.
   In contrast  to the 2-loop world-sheet S-matrix  case  in \ci{Conkey:2016qju,Beccaria:2025xry}, 
   here    the  direct agreement between the static gauge  and the   orthogonal 
    gauge  results   is a   consequence of the  vanishing of the  potential $\int K^4$, etc.,   counterterms  on the wrapped string background.


  \subsection{Summary of  results }

Starting  from the GS string analogs of \rf{11},\rf{12}  we will   find the following   finite 1-loop and 2-loop 
 corrections to the string energy $\E$ for the  periodic/antiperiodic
  b.c. for the fermions along   $X^{10}=\s^2\equiv \s^2 + 2\pi R$
\al{
\f^{}_{{\mathbb R\times S^1_{P}}}= &
2\pi T_1R- \frac{N - \ha \nf }{12R}
- \frac{(N-\ha \nf)^2}{576\pi  T_1R^{3}}+...\label{115}\ , \\
\f^{}_{{\mathbb R\times S^1_{A}}}= &
2\pi T_1R- \frac{N + \fo \nf}{12R}
- \frac{(N+ \fo \nf)^2}{576\pi  T_1R^{3}}+...\label{116}
\ . 
}
These  expressions   match the square root  expansion \rf{5} for the  type II  and type 0 strings,   respectively (cf. \rf{4}).\foot{As we shall discuss below,  it is  possible 
to show  that   starting with the non-linear  Nambu  string action in the static gauge 
 the exact square root  formula for  the energy 
\rf{5}    corresponds 
 to summing  up the contributions of all  higher loop bubble  diagrams.
This is  consistent   with   the  expected equivalence to  the  orthogonal gauge  approach 
and    does  not  explicitly  use hidden integrability of the theory  (i.e.  TBA  as in \ci{ Dubovsky:2012wk}).}
This  checks the   consistency of the $\zeta$-function regularization  with the  non-linear structure of the GS action beyond the 1-loop level. 
The  absence of corrections  in the  periodic BPS case \rf{115} ($N=\ha \nf=8$)
 may  be attributed to the residual world-sheet supersymmetry  
present in the GS action in the static  and adapted $\k$-symmetry gauge.\foot{The  same formally applies also to the GS 
 string defined in  target space dimensions $D=3,4,6$, where the condition $\nf=2N$ is also satisfied.}
 

The   energy  density  of an  M2   brane  wrapped on  the $X^{11}$   circle   
with the other two world-volume directions  being noncompact, 
i.e.  for the  membrane  on  the  cylinder  $\mathbb R^2 \times \tilde S^1_{P/A}$,   can be found  by taking the limit $R\to \infty$ in the expression for the  energy 
on the torus $\mathbb R \times S^1_P \times  \tilde S^1_{P/A}$  given below 
 but it is  useful to display it explicitly first.
The direct 2-loop computation  from \rf{11},\rf{12}   gives 
\al{
\f_{{\mathbb R^2\times \tilde S^1_{P}}}
&
=2\pi T_2\tilde R-\big(N-\ha \nf\big)\frac {\zeta(3)}{8\pi ^3 \tilde R^2}
-\big(N-\ha \nf\big)^2\frac{3\zeta(3)^{2}}{256\pi^{7} T_2\tilde R^{5}}+...\  =
 \  2\pi T_2\tilde R   \ ,\label{18}
 \\
\f_{{\mathbb R^2\times \tilde S^1_{A}}}&
=2\pi T_2\tilde R-\big(N+\tfrac{3}{8} \nf\big)\frac { \zeta(3)}{8\pi ^3 \tilde R^2}
-\big(N+\tfrac{3}{8} \nf\big)^2\frac{3\zeta(3)^{2}}{256\pi^{7} T_2\tilde R^{5}}+...\no \\
&= 2\pi \tR T_2\Big(1
-\frac {7 \zeta(3)}{2\pi ^2 \rm T_2}
-\frac{147\zeta(3)^{2}}{8\pi^{4} \rT_2^2}+...\Big) \ , \ \ \ \ \  \ \ \ \ \   \rT_2 = (2\pi)^2 \tR^3  T_2 \ , \label{114}
}
where  we used  that $N=\ha \nf=8$ and  the definition of dimensionless tension $\rT_2= (\td R/\lp)^3$  in \rf{7}.
As in the string case, all  quantum corrections  in 
 the periodic   supersymmetry-preserving   choice of b.c. in  \rf{18}   vanish  for the  physical 
  values of $N$ and $\nf$. This  may be attributed to the presence of an 
   effective 3d  world-volume supersymmetry 
   in the gauge-fixed BST action.\foot{The same applies also  to the  supermembrane  action  not only in $D=11$ but also in    $D=4,5,7 $ where
  the $\k$-symmetric BST action also exists  \ci{Achucarro:1987nc} (see  also \ci{Tseytlin:2025dae}):  in all of these cases  one 
   has $N=D-3= \ha \nf$.}
   The 2-loop   computation thus checks a  consistency of the $\z$-function regularization of   the M2  brane  theory with the underlying supersymmetry.

 Note that   if one includes only the  1-loop correction in \rf{114}  the   energy density $\E$  changes sign 
 when  $\rT_2^{(0)}\big|_{1}=\frac {7\zeta(3)}{2\pi^2} = 0.426...$.  Once the 2-loop   correction is included  
 this  value is modified to  $\rT_2^{(0)}\big|_{2}  ={7(1+\sqrt7)\zeta(3)\over4\pi^2} =0.777...$.  This  suggests   that   one cannot  
 draw   a  reliable conclusion  about  the value  of $\td R/\lp$   for which $\E$   vanishes  based  just on the knowledge of the  first few terms in the expansion
 \rf{114}.   We shall  elaborate on  this issue  below. 

   \subsubsection*{M2  brane on 2-torus}
  
 For the  M2  brane on a 2-torus  $S^1 \times\td  S^1$   with periodic  b.c.  
  we find that  as  at  the 1-loop    order \ci{Duff:1987cs},  the 2-loop   correction    vanishes  for $N= \ha \nf$.    $\f_{\mathbb R\times S^1_{P}\times \tilde S^1_P}= (2\pi )^2 R \td R T_2$
  should hold  to all orders. 

 The  2-loop expressions with antiperiodic b.c. in one circle, i.e.
   $\f_{\mathbb R\times S^1_{P}\times \tilde S^1_A}$ and $\f_{\mathbb R\times S^1_{A}\times \tilde S^1_P}$,  
 are  related   by interchanging $R\leftrightarrow \tR$. 
 \  $\f_{\mathbb R\times S^1_{A}\times \tilde S^1_P}$ 
   may be  expressed  in terms of infinite  sums of modified Bessel functions 
    $K_\nu (2\pi mn{ R\ov\tilde R}),\,m,n\in \mathbb Z$. The latter 
 admit asymptotic expansions for  ${R\ov \tilde R}\gg 1$ given by   exponentially suppressed terms 
  with the leading one being $\Oe{2}$. 
  While  ${R\ov \tilde R}\gg 1$ expansion of $\f_{\mathbb R\times S^1_{P}\times \tilde S^1_A}$  is the same as  the 
  ${\tR\ov  R}\gg 1$ expansion of  $\f_{\mathbb R\times S^1_{A}\times \tilde S^1_P}$, the 
  ${R\ov \tilde R}\gg 1$ expansions of the two are inequivalent  and given by 
  \al{
\f_{\mathbb R\times S^1_{A}\times \tilde S^1_P}\Big|_{_{R\gg  \tilde R}}&
=
\frac{R}{\tilde R^2}\Big\{\rT_2 -\tfrac{1}{12 } \big(N+\fo  \nf\big)\frac{\tilde R^2}{R^2}-\big(N-\ha \nf\big)\frac{ \zeta(3) }{4 \pi^2 }+\Oe{2}\nonumber
\\
 &\qquad
-\frac 1{\rT_2}\Big[\tfrac{1}{288 } \big(N+\fo \nf\big)^2\frac{\tilde R^4}{R^4}
+\big(N-\ha \nf\big)^2\frac{3\zeta(3)^{2}}{32\pi^{4} }+\Oe{2}\Big]+...\Big\} \label{121} 
\\
&  =
\frac{R}{\tilde R^2}\Big\{\rT_2 -\frac{\tilde R^2}{R^2} + \Oe{2}
-\frac{1}{\rT_2 }\Big[ \ha \frac{\tilde R^4}{R^4}
+\Oe{2}\Big] +...\Big\}\ , \no \\
\f_{\mathbb R\times S^1_P\times \tilde S^1_{A}}\Big|_{_{R\gg  \tilde R}}&=\frac R {\tilde R^2}\Big\{\rT_2-\tfrac{1}{12 }N \frac{\tilde R^2}{R^2}-\big(N+\tfrac{3}{8} \nf\big)\frac{ \zeta(3) }{4 \pi^2 }+\Oe{}\nonumber\\
&\qquad \qquad \ -\frac 1\rT_2 \Big[ \tfrac{1}{288  }N^2\frac{\tR^4}{ R^4}+\big(N+\tfrac{3}{8} \nf\big)^2\frac{3\zeta(3)^{2}}{32\pi^{4} }+\Oe{}\Big]+...\Big\} \  \label{122}
\\
&=\frac R {\tilde R^2}\Big\{\rT_2-\tfrac{2}{3 }\frac{\tilde R^2}{R^2}-\frac{ 7\zeta(3) }{2 \pi^2 } + \Oe{}
- {1\ov \rT_2 } \Big[\tfrac{2}{9} \frac{\tR^4}{ R^4}+\frac{147\zeta(3)^{2}}{8\pi^{4} }+\Oe{}\Big] +...\Big\} \ . \no
}
In   the   second equality in  \rf{121} and \rf{122} we specialized  to the physical values of $N=8,  \ \nf=16$. 
Here  $\rT_2 = ({\tR\ov \lp})^3= {  \gs^2}$  as in \rf{7},\rf{144}  and it is assumed that one   first expands for  ${\tR \gg   \lp} $ (i.e. $\rT_2 \gg 1$) and then  for  ${R\gg   \tR}$.

 The  inverse M2  brane tension expansion  thus  corresponds to the strong coupling or $1/\gs^2$  expansion from the 10d string theory point of view. It is important to note that the IIA  string  interpretation of the semiclassical membrane expansion 
depends on which geometric scale multiplies the inverse tension. For a 
membrane wrapped on the 11d circle in flat space the relevant scale is 
$R_{11}=\gs\ls$ itself, and since $T_2R_{11}^3=\gs^2/(2\pi)^2$ the loop 
expansion in $(T_2R_{11}^3)^{-1}$ is a strong-coupling expansion in 
$1/\gs^2$. In the AdS$_4\times S^7/\mathbb{Z}_k$ case the fluctuation 
scale is the curvature radius $L$, and 
$(T_2L^3)^{-1}=\pi/\sqrt{2kN}=4\pi^2 \gs \ls^3/L^3$. Then  each membrane 
loop  contribution  is proportional to  $\gs$ for fixed  effective string tension  ${\rm T_1}= {L^2\ov 2 \pi \ls^2}$,
explaining why in \ci{Giombi:2023vzu,Giombi:2024itd,Beccaria:2025npl} the membrane
semiclassical expansion reproduces the weakly coupled type IIA string
expansion.

Decompactifying the periodic circle ($R\to \infty$)  in \rf{122}  we  find   that  $(2\pi R)^{-1} \f_{\mathbb R\times S^1_P\times \tilde S^1_{A}}$   reduces to $\f_{\mathbb R^2\times \tilde S^1_{A}}$ in \rf{114}. 
Shrinking the periodic 11d  circle, i.e.  taking  $\tR$ to zero 
  for fixed $T_1= 2\pi \tR T_2 $   the expression in \rf{121}  reduces to the  type 0  string one  in \rf{116} or \rf{5}.  One can also  give  a formally equivalent  finite temperature interpretation: 
if   $X^{10}$ is interpreted as  a Euclidean time direction,  
 the antiperiodic   string  free energy  \rf{116}    determines  the  tachyon mass  value     which is 
  related to the  Hagedorn temperature $\T_H$  
   \ci{Atick:1988si}.
  If one   considers a  non-zero  $\tR$  so that   the value of type IIA   string coupling $\gs$ is increased, 
   one may conjecture that 
   $ \f_{\mathbb R\times S^1_{A}\times \tilde S^1_P}$  may be used to  determine   $\gs$ corrections to the 
  energy of the corresponding ``thermal''  string state    and thus to  the critical temperature $\T = ( 2 \pi R)^{-1}$  \ci{Russo:2001vh}. 

  While  \rf{121}    gives the expansion of free energy  for ${\tR\ov R} =  \gs {\ls \ov R}  \ll 1$  
   the expansion  for ${\tR\ov R}  \gg 1$    can be found from \rf{122}   by interchanging $R$ and $\tR$ in the 1-loop and 2-loop   terms
   there. Dropping   subleading $\frac{R}{\tR}$ and  $e^{-\pi \tR/R}$   terms we get 
   \al{
\f_{\mathbb R\times S^1_{A}\times \tilde S^1_P}\Big|_{_{R\ll  \tilde R}}
=\frac R {\tilde R^2}\Big[\rT_2   
-\frac{ 7\zeta(3) }{2 \pi^2 }  + ... 
+{1\ov \rT_2 } \Big(
-\frac{147\zeta(3)^{2}}{8\pi^{4} }+...\Big)  
+\OO\big( {1\ov \rT_2^{2}}\big)\Big] \ .  \label{102}
}
This  expression  generalizes the 1-loop  expression  in  \ci{Russo:2001vh}  to 2-loop order.\foot{The 1-loop  computation in   \ci{Russo:2001vh}    used l.c. gauge 
  formulation   dropping the quartic   interaction  term in the M2  brane Hamiltonian \ci{deWit:1988wri}. It would be interesting to rederive the 2-loop correction found here  in the static gauge   in the
   l.c. Hamiltonian   approach. 
}
It is  also formally equivalent   to \rf{114} as $\tilde S^1_P$ decompactifies in this limit.

As already   mentioned   after  \rf{114}   one cannot  draw  reliable  conclusions  about the  critical radius (or temperature) just on the 
 basis of the first  few orders  of  inverse  tension expansion. 
 The same  applies to the   conjectured type 0 interpretation   of M2 brane free energy  on  11d antiperiodic circle, i.e.  
   the  strong-coupling  value of 
 type 0 tachyon mass \ci{Bergman:1999km}  discussed at the 1-loop level  in  \ci{Russo:2001vh,Russo:2001tf} (cf.  \ci{Baykara:2026gem}).

   Let  us note  also that while the   double  dimensional reduction   of the M2  brane action on the    periodic 11d  circle  gives  the usual 10d   GS  action \ci{Duff:1987bx}, in the case when the 
   fermions  are  antiperiodic  in $\sigma^3=X^{11}$, their  zero mode $\theta(\sigma^1, \sigma^2)$    drops out  
    and thus   one ends  up with  just the  bosonic 
  (Nambu)   string action  in 10d.\foot{This need not, a priori,  be in conflict with the 
  conjecture \ci{Bergman:1999km} that  type 0A   string  may be related to 
  11d theory on antiperiodic circle: 
   in the non-supersymmetric case one may
   not be able to directly  connect via a naive reduction  the  weak  and strong coupling (M-theory)  descriptions as one may   need to account also  for the 
   ``twisted sector'' modes  (cf. \ci{Russo:2001tf}). 
   A similar remark applies to an apparent  mismatch between  massless  field  content 
   in type 0  and 11d   low-energy effective actions (cf. \ci{Baykara:2026gem}). 
    }
   This   is the reason    why  the  1-loop $N \frac{\tilde R^2}{R^2}$ and the 2-loop $ N^2\frac{\tR^4}{ R^4 }  $ terms 
  in \rf{122}   do not   depend on $\nf$  as they are  purely bosonic  string mode contributions  (the same as $N$ and $N^2$ terms in \rf{18} or \rf{114}).

   One can similarly  compute  the M2  brane energy    in the  case of antiperiodic b.c. in both circles. Taking the limit  ${R\ov \tR}\gg 1$ and dropping exponentially suppressed 
   terms  the resulting  expansion of $\f_{\mathbb R\times S^1_A\times \tilde S^1_{A}}$ is found to have the same form as  \rf{122}.

 \subsubsection*{3-loop  correction and resummation of higher loop bubble diagrams}
 
\begin{figure}[H]
    \centering
        \centering
        \includegraphics[width=0.79\textwidth]{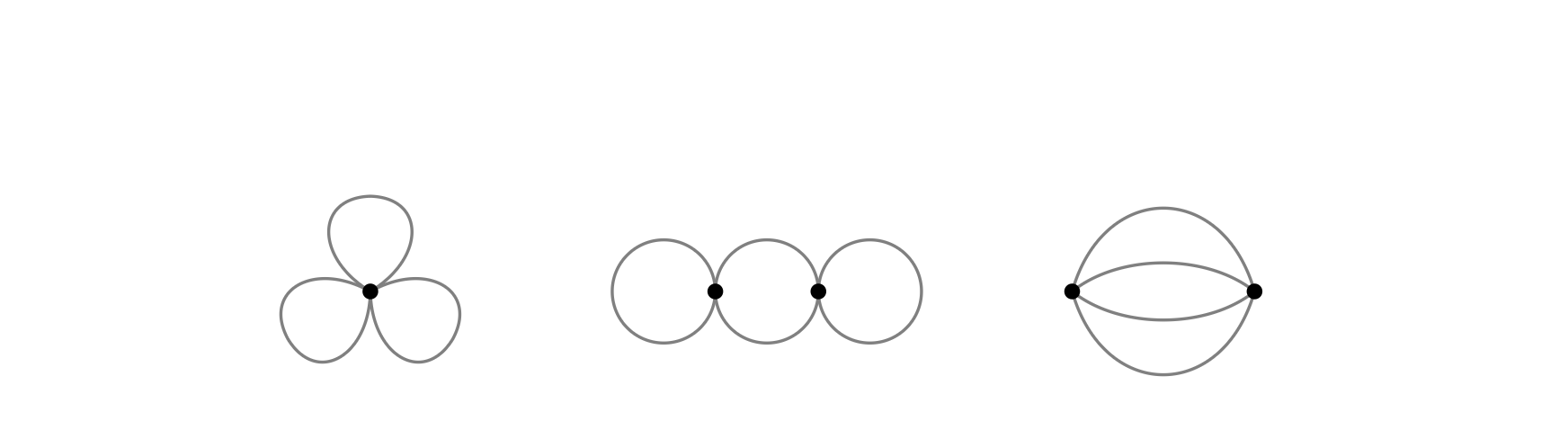}
    \caption{\small 
   3-loop  diagrams
    }
    \label{fig1}
\end{figure}

 Below  we will also  compute the next 3-loop terms  in the  large  effective tension expansion of the  energy of the 
   bosonic string  on $\mathbb R \times S^1$  and  the bosonic 
 membrane on $\mathbb R^2 \times \td S^1$. These   will  generalize  the    $N$-dependent  terms in \rf{115} or \rf{116}  and \rf{18} or \rf{114}.  The relevant diagrams are given    in Fig. \ref{fig1}.
 In the bosonic  string case   we find  that only the middle (``chain'')   diagram 
 is  nonvanishing   and finite 
 when computed in the standard $\z$-function regularization. It is proportional to $N^3$ 
 so that  to 3-loop order 
 \al{
\f^{B}_{{\mathbb R\times S^1}}= &
2\pi T_1R- \frac{N }{12R} - \frac{N^2}{576\pi  T_1R^{3}}
-      \frac{N^3}{13824\pi^2  T_1^2 R^{5}}   +    ...= 2\pi T_1R\  \bE (u) \  ,\label{h15} \\
   & \ \ \    \bE(u) = 1 - u  - \tfrac{1}{2}u^2  -\tfrac{1}{2} u^3 + ... \ , \qquad \ \ 
\ \  u\equiv   \frac{N }{24 \pi T_1 R^2 } \ .\la{h16}
 }
 This   matches the expansion of the square root  expression in \rf{5}  where in the bosonic  string case $m_0^2=- {1\ov 3} \pi T_1  N$  (cf. \rf{2}).

\begin{figure}[H]
    \centering
        \centering
        \includegraphics[width=0.6\textwidth]{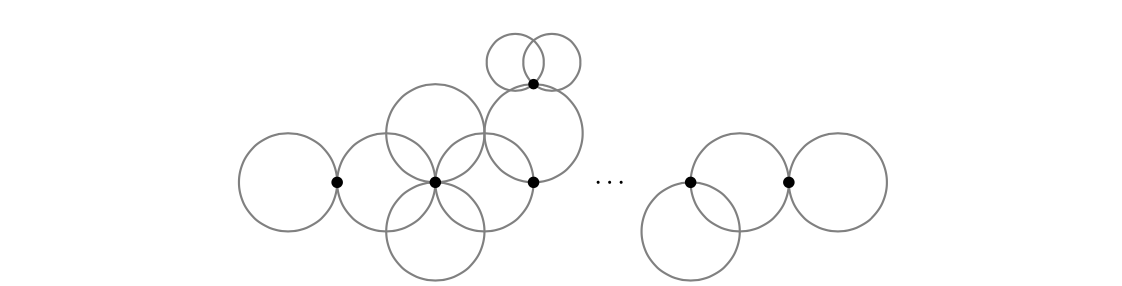}
    \caption{\small 
   Higher loop  bubble or   cactus  diagrams generalizing the  first two diagrams in Fig. \ref{fig1}. 
   A generic $L$-loop cactus diagram is a tree of one-loop bubbles glued at quartic, 6-vertex,  etc.,  vertices.
    }
    \label{fig2}
\end{figure}

  
 Making the assumption  that only the  ``bubble'' or ``cactus''   diagrams  in Fig. \ref{fig2} 
   give  non-vanishing   contributions to $\E$ 
    also at higher loop orders  one can actually prove  that the exact   expression for \rf{h15}  should be given by the   square root  expression  in \rf{5}. 
 \begin{figure}[H]
    \centering
        \centering
        \includegraphics[width=0.6\textwidth]{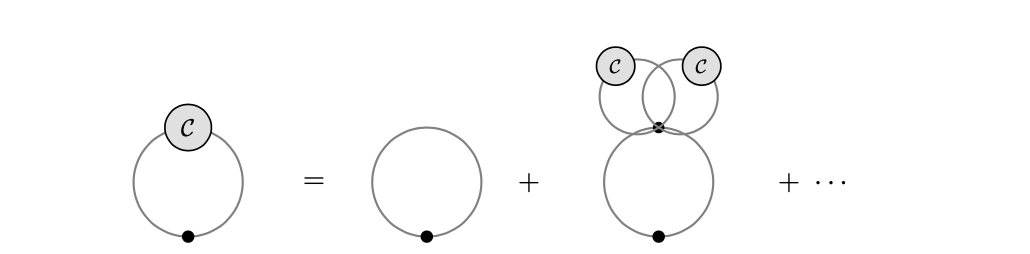}
    \caption{\small 
   Recursion relation:  attaching cactus $\cal C$  as a  vertex generates all cactus diagrams.  This is a diagrammatic
representation  of the large $N$ saddle    equation  \rf{a62}  
  whose solution  generates  the whole set of such graphs.
    }
    \label{fig22}
\end{figure}
 Indeed, one may  use the observation   that  the ``cactus'' 
    diagrams  give dominant contribution  at each loop order  in the large $N$ limit.
    Then we may  sum them up by 
     considering   the formal   large $N$ limit in the Nambu path integral
       (see  Fig. \ref{fig2} and Fig. \ref{fig22}).
       Such  large  $N=D-2$ limit was   first considered in \ci{Alvarez:1981kc}.
        Introducing a Lagrange  multiplier $\l_{ab}$ to  write the Nambu action 
 $T_1\int d^2\s\, \sqrt{\det( \delta_{ab} + \del_a X^i \del_b X^i)}$   
  in the 
 form quadratic  in $\del_a X^i$\ ($i=1, 2, ..., N$), integrating  over  $X^i$  and then   solving for $\l_{ab}$   in the large $N$  saddle point approximation  
  \ci{Alvarez:1981kc,Polyakov:1987ez}
  one finds  
  for the resulting free energy density or function $\bE$ in \rf{h16}  (see    \rf{a699})
 \be 
 \bE^2-1 =  -2u \ , \ \ \ \qquad \bE= \sqrt{ 1 - 2u} \ , \qquad \ \ \ \ \   u=- \frac{N\,\zeta(-1)  }{2 \pi T_1 R^2 } = \frac{N }{24 \pi T_1 R^2 }\ , \la{h7}
 \ee
 where $u$ is  proportional to the 1-loop   correction. 
 The resulting $\E= 2 \pi R T_1 \bE$   is then the   same as in \rf{5}.
 This  derivation has a straightforward generalization to the 
  GS   string case with  the periodic (type II) or antiperiodic (type 0) 
    b.c.  where $N\to  N-\hal \nf$ or $N\to N + {1\ov 4} \nf$ as  in \rf{2},\rf{4}, i.e.    with   $u=- \hal {m_0^2 \ov (2\pi R T_1)^2}$. 
 
 
 In the case of the bosonic membrane on the cylinder $\mathbb R^2 \times \td S^1$
 all the  three  diagrams   in Fig. \ref{fig1} give non-zero finite contributions
 (there are no log UV divergences). 
     As a result, 
 we will  find      that  
 including  the 3-loop correction \rf{5f}  the bosonic terms in \rf{18},\rf{114} generalize to
 \al{ \f^B_{{\mathbb R^2\times \tilde S^1}}&
=2\pi T_2\tilde R-   \frac {N \zeta(3)}{8\pi ^3 \tilde R^2}
-  \frac{3N^2\zeta(3)^{2}}{256\pi^{7} \tR^{5}T_2}
-\frac{16 N^3 \zeta(3)^{3} +  N^2 \big[385 \z(9) -96 \z(3)^3\big]}
{8192\,\pi^{11}\tR^{8}T_2^{2}}
 +... \ , \la{h9}
}
The first  diagram in Fig. \ref{fig1}  contributes a  term $\sim (N^3 +2N) \z(3)^3$, the second 
 diagram a term $\sim N^3 \z(3)^3$, while the third ``basketball'' diagram  produces terms 
 $\sim (12 N^2 + N  ) \z(3)^3$ and $\sim N^2 \z(9)$.\foot{Note that 
 among the a priori expected 
 weight-nine constants  that could appear at 3-loop order
 only $\zeta(9)$ and $\zeta(3)^3$  are  actually present:  the
coefficients of the $\zeta(2)\zeta(7)$, $\zeta(4)\zeta(5)$ and $\zeta(6)\zeta(3)$ terms  vanish.}

The  analog of \rf{h9}  in the M2  brane   with periodic b.c.  for the fermions 
 is expected to  have the same structure as in \rf{18}, i.e.  with the replacement 
 $N\to N -\ha \nf$ applied also to the 3-loop term in \rf{h9}. In the  antiperiodic case 
 the 3-loop generalization of \rf{114} may  have  coefficients of the $\z(3)^3$ and $\z(9)$ terms  being  different  combinations  of $N$  with  $\nf$.

As in the string case above, one  can  sum up all  bubble  diagrams  in Fig. \ref{fig2}
that  generalize  the  first two  graphs in Fig.\ref{fig1}
 by observing  that they contribute the highest   power 
$N^L$  terms  at each loop order $L$.
  Thus  their  contribution 
  can be
again   found  in closed form  using a  large $N$ saddle point argument
like in  \rf{h15},\rf{h7}.  Such large $D=N+3$   approximation 
 for the bosonic membrane  was   suggested  in \ci{Floratos:1988jh}. 
Here we apply it to  the case of a  membrane on ${{\mathbb R^2\times \tilde S^1}}$ 
and the resulting relations  are  similar but  not equivalent to the ones in 
\ci{Floratos:1988jh}. Also, our interpretation is different: 
we use  saddle point argument as a tool  to sum a subclass of all bubble graphs 
 in a finite $N$ theory.

We  find that   the bubble graph   contribution to the membrane energy $\bE$  in \rf{h15} 
  satisfies not a quadratic  as in \rf{h7}  but  a {cubic}  \ci{Floratos:1988jh}
   equation:  
 \al{ \f^B_{{\mathbb R^2\times \tilde S^1}}\Big|_{\rm bubble}&
=2\pi \td R\, T_2\ \bE(u) \ , \qquad \qquad \bE\, (\bE^2-1) = -2 u  \ , \qquad \qquad 
u=  \frac {N\, \zeta(3)}{16\pi ^4 T_2\tR^3}\ , 
\la{h11}
}
where  $u$ is proportional to  the 1-loop  contribution in \rf{h9}. 
The  relevant real   solution of  this cubic  equation  that reproduces the small $u$ expansion of \rf{h9} 
 may be written in a closed form as
 \al{
\bE= \te {2\ov \sqrt 3} \cos \Big [{1\ov 3} { \arccos} (-3\sqrt 3 \, u) \Big]
   = \te  1-u -\frac32u^{2}- 4 u^{3}     +...\  .\la{h19}
}
 The   $u^3$ term  here   is precisely   the $N^3 $  part of the 3-loop  contribution   in \rf{h9}.
 The same relations \rf{h11}, \rf{h19}   should   sum also   bubble graph contributions to $\bE$  in the M2 brane case, with $N$ in   the 1-loop  correction $u$  replaced  by $N-\ha \nf$ in  type II case and $N+ {3\ov 8} \nf$   (cf. \rf{18},\rf{114}).
  
  The  presence of an additional  order $N^2 $ part in the 3-loop term in \rf{h9}
  involving $\z(9)$   indicates that  in contrast to the  string case   a closed   all-order expression for  the  membrane energy 
  $\f^B_{{\mathbb R^2\times \tilde S^1}}$  should be non-trivial to find. 
   This is not  surprising given that the membrane theory 
   cannot be reduced to a free theory in any gauge. 
     
\begin{figure}[H]
    \centering
    \begin{subfigure}{0.4\textwidth}
        \centering
        \includegraphics[width=\textwidth]
        {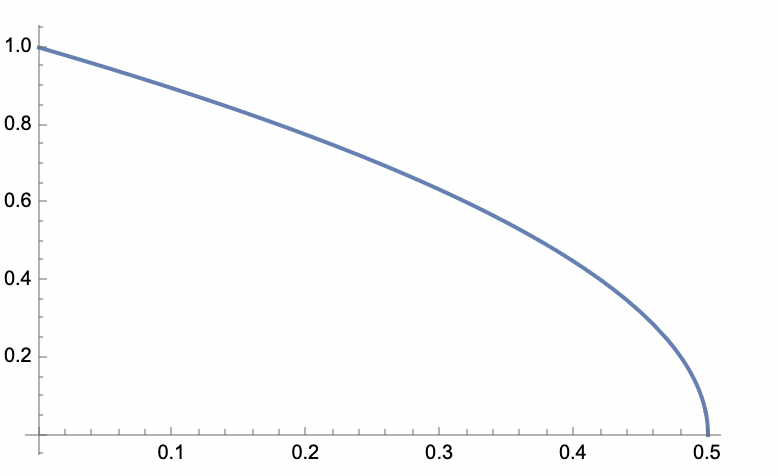}
        \caption{}
        \label{fig4}
    \end{subfigure}
     \hfill
    \begin{subfigure}{0.4\textwidth}
        \centering
        \includegraphics[width=0.9\textwidth]
        {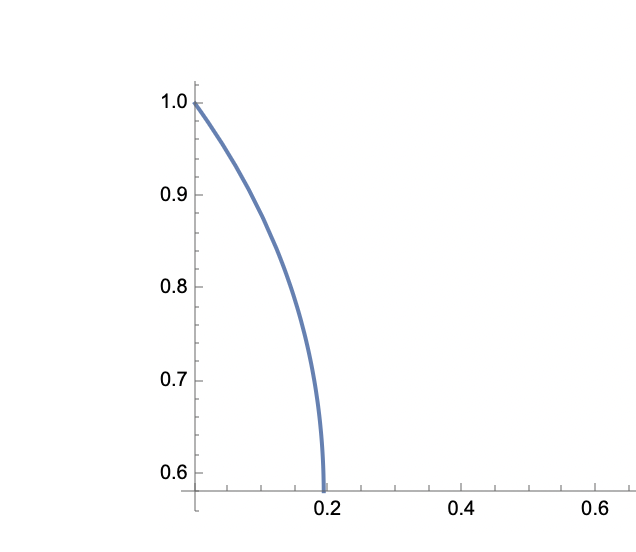}
        \caption{}
        \label{fig55}
    \end{subfigure}
    \caption{\small 
    Plots of  $\bE(u)$: (a) string case  in \rf{h7} and (b) membrane case in   \rf{h19}. 
    }
    \label{fig5}
\end{figure}

 It is interesting to note that the shape of the function $\bE(u)$  in \rf{h19}
 is similar to the square root one  in \rf{h7} (see Fig. \ref{fig55}).  
However, in contrast to $\bE$ in \rf{h7}   that vanishes  for  $u=\hal$ 
the only real zero of the  function $\bE$ that solves the cubic  equation in \rf{h11} 
is  at $u=0$.  The relevant  positive  $\bE$ branch  in \rf{h19}  is 
 for $0< u \leq {1\ov 3 \sqrt 3}$ and thus  
  $ {1\ov  \sqrt 3}\leq \bE \leq 1$. 
That means that unless the behaviour of $\bE$   is  qualitatively changed  by  contributions of  other  non-bubble diagrams  (like  the third one in Fig. \ref{fig1}
contributing to the  ``subleading''  
$N^2$ part of the 3-loop term in \rf{h9}) one cannot reach $\bE=0$  by  changing  the value of $\tR$ relative to $\lp$, i.e. of dimensionless tension $\rT_2 = \tR^3/\lp^3$. 
 As the same   should be  true  also for the  M2 brane on antiperiodic circle,
    that  appears not to support   the conjecture that the wrapped M2 brane state  may be representing 
 an  analog of type 0 string   ground state  which  is tachyonic at weak coupling but 
  becomes massless  at some critical    value of  $\rT_2 = \gs^2$ (cf. \rf{144}).


 \subsubsection*{Twisted    11d background}

Finally,  we will  also consider the M2  brane  free energy in  flat  11d  background  with   ``twisted'' 3-space part (denoted as $\mathbb R^2 \times \td S^1_q$)   which is 
related to 10d KK  Melvin solution \cite{Dowker:1993bt}. 
It was discussed  at the string theory level in \ci{Russo:1995ik,Tseytlin:1995zv}
and also in the  11d context in   \ci{Russo:1998xv,Costa:2000nw,Russo:2001tf}. 
This  background  effectively  
 ``interpolates''  between the  cases of  periodic and antiperiodic   b.c. for the fermions   around the  11d circle. The   corresponding 
  metric is  
\al{
&d s^2_{11}=d X^m dX^m+d r^2+r^2(d \varphi+ q\,   d\psi)^2+  \tR^2 d\psi^2 \ ,\ \ \ \ \ \ \ \   X^{11} = \tR\,  \psi\ , \qquad \quad m=1,2,...,8\ ,\label{d51}
\\& \qquad \qquad \qquad 
\varphi\equiv \varphi+2\pi \ ,\qquad \psi\equiv \psi+2\pi \ ,\qquad 0\leq q<2\ , \la{52}
}
where $q$ is a real dimensionless parameter. The bosonic part of the 
M2 brane  action is invariant under $q\to q+1$  while the fermionic part 
 -- under $q\to q+2$. The  supersymmetry is thus  broken  for any $q\not=2n$. 
  The corresponding  1-loop  correction to the M2  brane free energy  was  discussed (but not computed explicitly) in 
 \ci{Russo:2001tf}. Including also the 2-loop  correction     we find that 
\al{\la{22}
\f_{\mathbb R^2 \times \tilde S_{q}^1}&=2\pi \tR T_2 -\frac {\U(q)}{8\pi^3 \tR^2 }-\frac {3\,[\U(q)]^2}{256\pi^7 \tR^5 T_2}+\OO ( T_2^{-2})\ , 
}
where   both  the 1-loop  and 2-loop  corrections  are  expressed in terms of a single   function $\U(q)$
\al{\la{23}
&\U(q)=(N-2)\,\zeta(3)+2\UU(q)-\ha \nf\UU(\tfrac{1}2q)\ ,\qquad \qquad 
\UU(q)\equiv \re \li _3(e^{2\pi i q})= \sum_{n=1}^{\infty} \frac {\cos (2 \pi n q)}{n^3}\ .
}
Here  again  $N=8$ and $\nf=16$. 
For  the  zero twist $q=0$ (or any $q=2n$)  we have   $\UU(0)=\zeta(3)$   so that 
 the free energy in  \rf{22}   reduces  to  the periodic-fermion  expression in \rf{18}, i.e. all quantum corrections 
vanish. 
 For $q=1$  we  recover the  antiperiodic fermion  expression in \rf{114}. 
 One may also   generalize \rf{22} to the case of several  twisted planes  with parameters $q_r$.

\

The rest of this  paper is organised as follows. 
In section \ref{sect2}   we compute the 2-loop  free energy of  bosonic  Nambu 
   string  on circle and  membrane on a  torus. 
 In section \ref{sect3} we account for  the  2-loop  contributions of  the GS and M2 brane  fermions.
 Then  in section \ref{sect4}  we  combine these results 
   to get the total 2-loop  expression  for the  energy in various  cases. 
 In section \ref{s5}  we  describe the computation of  the 3-loop  correction to 
 the energy of  bosonic string  and membrane wrapped 
  on a circle.  Section \ref{s6}  presents the resummation of all  bubble   graph contributions to this energy. 
  In section \ref{s7} we  discuss the generalization  of the 2-loop M2 brane energy   expression to the  case of   flat  11d   background  with one or two  twist parameters. 
 Some  basic definitions and technical details  are delegated to Appendices  \ref{A1},\ref{A2},\ref{A3} and \ref{ap4}.


\section{Free energy of  bosonic membrane}
\label{sect2}

The starting point is the  Dirac  action for a  membrane in  flat   space 
($a,b=1,2,3; A,B=1,...,D$)
\al{
&S_B=T_2 \int d^{\,3} \sigma \; \sqrt{h}\ ,\qquad\qquad  h_{a b}=\partial_a X^A \partial_b X^B \delta_{AB},\label{232}
}
Considering a  membrane  stretched along $l=1$ or 2  non-compact and  wrapped 
  on other  $3-l$  circular directions 
and fixing the static gauge as in \rf{14} we   get the expansion of \rf{232} as in \rf{11}. 

In general, for a   scalar field   obeying 
 boundary conditions characterised by a shift $\eta$ (with $\eta=0$   corresponding to  periodic and $\eta=\ha$ -- to  antiperiodic  b.c.) 
  the 1-loop free energy is proportional to ($V$ is the  volume of non-compact directions)   
\be
 \Omega^{(\eta)}=\frac {1}{V} \sum_n \text{Tr}\, \ln (-\partial^2)=\sum_{n_1,...,n_{3-\c}} \int \frac{d^\c p}{(2\pi)^\c}\ \ln  P^2\ .\quad\label{233}
\ee
We shall also consider the  second derivative  of the free  propagator  at coinciding points defined as 
 \al{
 &
\Delta^{(\eta)}_{ab}
=\langle \partial_a X (\sigma)\partial'_b X(\sigma') \rangle\Big|_{_{\sigma=\sigma'} }=
\frac 1 {V_c}\sum_{n_1,...,n_{3-\c}}\int \frac{d^\c p}{(2\pi)^\c}\frac {P_a P_b}{P^2}\label{d26}\  ,\\
&P\equiv \big(p, k_m^{(\eta)}\big)\ , \qquad \ \ 
P^2=\big(k_m^{(\eta)}\big)^2+p^2\ ,
\qquad
k_m^{(\eta)}=\frac{m+\eta}{\tR}\ , \qquad m\in \mathbb Z \la{204}
\ . }
Here \rf{204}    corresponds to the case when $l=2$ (so that $p$ is 2-component  continuous momentum  and $\tR$  is the radius of a circular dimension); in the  case of 2 circular dimensions $k_n$  has 2 components. 
$V_c$ is the volume of the compact directions, i.e. 
 $ V_c=2\pi \tilde R $ for $ \mathbb R^2 \times \tilde S^1$, and $V_c=4\pi^2 R\tilde R $ for $\mathbb R\times S^1\times \tilde S^1$.
In this section   we consider only periodic scalars $X^i$    and define 
\be \la{205} \Omega\equiv \Omega^{(0)}\ , \qquad \qquad 
\Delta_{ab}\equiv \Delta_{ab}^{(0)} \ . \ee
The classical and  bosonic 1-loop   contributions to the free energy density in \rf{6} are 
\be
 \f_0 =V_c T_2\ ,\qquad \ \ \ \   \f_1^B=\ha   N  \Omega\ ,  \label{206}
 \ee
where $N$  (equal to $D-2$ in the string case  and $D-3$  in the membrane case)  is the number of transverse components $X^i$. The regularised values for $\Omega$ are given  in Appendices \ref{A2} and \ref{A3}.

The 2-loop contribution  comes from the quartic  term $\mathcal L_{4B}$  in \rf{11}, i.e. is   given by 
 the bubble diagram in figure \ref{fig0}:
\al{
\langle \mathcal L_{4B}\rangle=\frac1 {8T_2}\Big[N(N-2)(\Delta^{a}\,_a)^2-2N^2 \Delta^{}_{ab}\Delta^{ab}\Big]\ . \label{117} 
}
The $\z$-function   regularized value of $\Delta^{a}\,_a$   is zero  
\al{
&\Delta^{a}\,_a= \frac 1{V_c} \sum_{n_1,...,n_{3-l}}\int \frac{d^lp}{(2\pi)^l} =0\ , \qquad \ \ \ \ \ \ \   
\sum_{n=-\infty}^{\infty} 1 =1  + 2 \zeta(0) =0 \ ,  \la{207}
}
so that the  bosonic  contribution to the 2-loop free energy is  given by 
\al{
\f_2^B 
=-\frac{ 1} {4T_2}\, N^2\, V_c\,  \Delta_{ab}\Delta^{ab}\label{71}\ .
}
Here $\Delta_{ab}$  are defined   using  $\zeta$-function regularisation  and their finite values   are given 
in Appendices \ref{A2} and \ref{A3}. 
Note   that  the 2-loop  free energy  is free of log UV divergences.

\subsubsection*{String  on a circle}

The above discussion   can be repeated for the case of a string 
 with world   volume of    $\mathbb R \times S^1$. In this case   we find (see (\ref{c3}),(\ref{c7}),(\ref{c10}),(\ref{c132}))
\al{\la{89}
& V_c=2\pi R\ ,\qquad \Omega=-\frac 1{6R}\ ,\qquad \Delta_{11}^{}=\frac{1}{24\pi R^2}\ ,
\qquad
\Delta_{22}^{}=-\frac{1}{24\pi R^2}\ , 
\\ &
\f^B_{_{\mathbb R\times S^1}}
= 2\pi R\, T_1-\frac{N}{12R}-\frac{N^2}{576\pi T_1 R^3}+...= 2\pi  R\, T_1\,  \bE \ , \qquad 
\ \  \bE= 1 -\frac{N}{12\T_1}-\frac{N^2}{288 \T_1^2 }+...\ ,\label{41}\\ &\qquad \ \ \ 
\qquad \ \ \  \T_1\equiv 2\pi R^2 T_1 = {R^2\ov \ls^2} \ .\la{78} }
Eq. \rf{41}  agrees with the large $\T_1$  expansion of the  square-root  expression  for the string energy in the   l.c. gauge in \rf{5}, i.e. 
\be
\f^B_{_{\mathbb R\times S^1}}=\frac 1 R \sqrt{\T_1 ^2-\tfrac{1}{6}N \T_1 }\ .\label{b230}
\ee

\subsubsection*{Membrane   on  $\mathbb R^2 \times \td S^1$   cylinder}

For a bosonic  membrane on $\mathbb R^2\times \tilde S^1$  we have explicitly (see (\ref{c4}),(\ref{c11}),(\ref{c117}),(\ref{c8}))
\al{
&V_c=2\pi  \tR\ ,\qquad \Omega=-\frac { \zeta(3)}{4\pi^3 \tR^2}\ ,\qquad \Delta_{33}=-\frac{\zeta(3)}{8\pi^4 \tilde R^3},
\qquad
\Delta_{rs}=
\frac{\zeta(3)}{16\pi^4 \tilde R^3}\,\delta_{rs}\ , \qquad  r,s=1,2 \ , \la{412} \\
& \f^B_{_{\mathbb R^2\times \td S^1}}=2\pi\tR\,  T_2  -\frac {N \zeta(3)}{8\pi ^3 \tilde R^2}-\frac{3N^2\zeta(3)^2 }{256\pi^7 T_2  \tilde R^5}+...=\frac 1{2\pi\tilde R^2 }\Big( \rT_2-\frac {N \zeta(3)}{4\pi ^2}-\frac{3N^2\zeta(3)^2 }{32\pi^4 \rT_2 }+...\Big)\ , \la{512} 
}
where $\rT_2=(2\pi)^2 \tR^3 T_2$ is  dimensionless tension  as defined in  \rf{7}. 

\subsubsection*{ Membrane on  $S^1 \times \td S^1$  torus}

The 1-loop  term  in free energy  of membrane  with world-volume  $\mathbb R^{1}\times S^1\times \tilde S^1$ 
is given by 
\al{
\f^B_{_{1,\, \mathbb R^{1}\times S^1\times \tilde S^1}}=&\ha N\,  \Omega(R, \tR)\ ,\qquad 
\ \ \ 
\Omega= \sum_{n_1,n_2\in \mathbb Z}  \sqrt{\frac{n_1^2}{R^2}+\frac{n_2^2}{\tilde R^2}}=\Z_{0,0}(-\frac{1}{2})  \  , \la{216}\\
\Z_{0,0}(-\tfrac{1}{2})
=&-\frac{1}{6R}-\frac{\zeta(3)R}{2\pi^{2}\tilde R^{2}}-\frac{4}{\pi \tilde R}\, \B_{1}\big({R\ov \tR}\big)\ , \label{221}
}
where  $\Z_{0,0}(-\frac{1}{2})$ is the   $s=-\ha$ value  of the  Epstein zeta function. 
The function  $\B_{1}$ in \rf{221}   is expressed in terms of the sum of  modified Bessel functions $K_1(2\pi n_1n_2{R\ov \tR})$  (see Appendices \ref{A1} and \ref{A3}).

Using that the regularised values of the non-zero components of  $\Delta_{ab}$  in \rf{d26},\rf{205}    may be written as 
\al{\la{812} 
&\Delta_{11}
=-\frac{1}{8\pi^2 R\tilde R}\Z_{0,0}(-\ha)
,\qquad
\Delta_{22}
=-\frac{1}{8\pi^2  \tilde R}\partial_{R}\Z_{0,0}(-\ha)
,\qquad
\Delta_{33}
=-\frac{1}{8\pi^2 R}\partial_{\tilde R}\Z_{0,0}(-\ha)\ , 
}
the 2-loop contribution to the  free energy \rf{71}   may be written as 
\al{
&\f^B_{_{2,\mathbb R\times S^1\times \tilde S^1}}
=-\frac{N^2}{T_2}
\Big(
\frac{1}{1152\pi^2\tilde RR^3}
+
\frac{3\zeta(3)^2 R}{128\pi^6\tilde R^5}
+
\Ee_1\Big)\ ,\la{555}
\\
&\Ee_1=
\frac{\B_{1}}{24\pi^3 R^2\tilde R^2}
+
\frac{\C_{1}}{24\pi^2 R \tilde R^3}
-
\frac{3\zeta(3)R\,\C_{1}}{8\pi^4 \tilde R^5}
+
\frac{\B_{1}^2}{2\pi^4 R\, \tilde R^3}
+
\frac{\B_{1}\,\C_1}{\pi^3 \tilde R^4}
+
\frac{2R\,\C^2_1}{\pi^2  \tilde R^5}\ ,\label{E19}
}
where $\B_1$ and $\C_1$ are combinations of 
 modified Bessel functions $K_1$ and $K_0$ (see  (\ref{D7})-(\ref{D9})). 
 As a result,  the free energy for $\mathbb R^{1}\times S^1\times \tilde S^1$ membrane computed in 2-loop approximation 
   may be written as (cf. \rf{512}) 
\al{
\f^B_{_{\mathbb R^{1}\times S^1\times \tilde S^1}}
=&\frac R{\tilde R^2}\Big\{ \rT_2 
-N
\Big[
\frac{1}{12}\frac{\tR^2}{ R^2}
+
\frac{2\,}{\pi } \frac\tR R\,  \B_{1}
+
\frac{\zeta(3)}{4\pi^{2}}
\Big]
-\frac{1}{ \rT_2}N^2
\Big[
\frac{1}{288}\frac{\tR^4}{ R^4}
+
\frac{3\zeta(3)^2 }{32\pi^4}
+
\frac{4\pi^2\tilde R^5}{R}\Ee_1\Big]\Big\}
+...\ . \la{222}}
 The modified Bessel functions admit asymptotic expansions for  ${R\ov \tilde R}\gg 1$  given by  exponentially suppressed terms 
 with the leading   contribution being  proportional to  $\exp(-2\pi {R\ov \tilde R})$.
Dropping  such exponentially suppressed terms and expressing   the result
 in terms of the dimensionless string tension in \rf{78}, i.e.  
 \be \T_1 = 2\pi R^2 T_1 = (2\pi)^2 R^2 \tR T_2= {R^2\ov \tR^2} \rT_2 \ , \la{332} \ee  we find 
\be
\f^B_{_{\mathbb R^{1}\times S^1\times \tilde S^1}}\Big|_{R\gg \tR}
=\frac{1}{R}\Big[ \T_1  -\frac{N}{12 }-\frac{N \zeta(3)R^2 }{4 \pi^2 \tilde R^2} - {1\ov  \T_1  }\Big(
\frac{N^2}{288 }+\frac{3 N^2 \zeta(3)^2 R^4}{32 \pi^4  \tilde R^4}\Big)+...\Big] \ . \label{235}
\ee
Here the  terms without $\zeta(3)$   factors  are the same  as  in the string case  in 
\rf{41}, i.e. originate from the $n=0$  modes on $\td S^1$  while the $\z(3)$ terms  stand for the contributions  of the KK 
membrane ($n\not=0$)  modes. 


\section{Fermionic contribution  to  free  energy}
\label{sect3}

Let us    now  consider the 2-loop contribution to the free energy from the fermionic terms 
in the expansion \rf{12} of the M2 brane  action in the static  and $\k$-symmetry gauges \rf{14},\rf{15}. 
To account  for the possibility of either periodic or antiperiodic b.c.   for the fermions  we consider the fermionic  propagator $S^{(\eta)}_{\alpha \beta}(\sigma,\sigma') $    with  possible half-integer   shifts  $\eta$ in one or two    compact directions  
  and define like in \rf{d26}\al{
&\tD^{(\eta)}_{\alpha \beta}=\langle \theta_\alpha (\sigma) \theta_\beta(\sigma') 
\rangle\Big|_{_{\sigma=\sigma'}}\ ,\quad 
 \tD^{(\eta)}_{a,\alpha\beta}=\langle  \theta_\alpha (\sigma)\partial'_a \theta_\beta(\sigma') \rangle\Big|_{_{\sigma=\sigma'}}\ ,\quad \
\tD^{(\eta)}_{ab,\alpha \beta}=\langle \partial_a\theta_\alpha (\sigma)\partial'_b \theta_\beta(\sigma') \rangle\Big|_{_{\sigma=\sigma'}}\ ,\label{34}
 \\
& S^{(\eta)}_{\alpha\beta}(\sigma,\sigma')=\langle  \theta_\alpha (\sigma) \theta_\beta(\sigma') \rangle = \frac1{V_c}\sum_{n_1,...,n_{3-l}}\int \frac{d^l  p}{(2\pi)^l}\frac {i(\slashed P \mathcal PC^{-1})_{\alpha \beta}}{P^2}\ e^{iP\cdot(\sigma-\sigma')}\label{64}\ ,\\
& P=(k^{(\eta)}_n,p)\ ,\qquad P^2=\big( k^{(\eta)}_{n}\big)^2+p^2\ ,\qquad   k^{(\eta)}_{n}=\frac {n+\eta} {\tR} \ ,\qquad \eta=0,\ha \label{69}\ .
}
For notational   simplicity   in \rf{69} here we assumed  that there is   just one discrete momentum $k_n$  corresponding  to the  $\td S^1$ direction with radius $\td R$, i.e. that of  $l=2$; for a 2-torus   $k_n$   and $\eta$   will  be  2-vectors. 
 $\mathcal P$ is the spinor  projector  defined in (\ref{15})  and  $C$ is the charge conjugation matrix.\foot{We label Majorana  fermions
  as  $\theta_\alpha$  with 
  $\bar \theta^\alpha =C^{\alpha \beta}\theta_\beta$,   $\theta_\alpha=\bar \theta^\beta C_{\beta \alpha }$, where   $C^{\alpha \beta}C_{\gamma\beta}=\delta^\alpha_\gamma$ and $C_{\alpha\beta}=-C_{\beta\alpha}$.
One has   the identities $
(\Gamma^{a})^{T}=-C\Gamma^{a}C^{-1},\ \  (C\Gamma^{a})^{T}=C\Gamma^{a}$, 
 and also 
$
\text{Tr}(\Gamma^{a}\mathcal{P})=0,\quad \text{Tr}(\Gamma^{a}\Gamma^{b}\mathcal{P})={\nf}\delta^{ab},\ \ \text{Tr}(\Gamma^{c}\Gamma^{a}\Gamma^{d}\Gamma^{b}\mathcal{P})={\nf}(\delta^{ab}\delta^{cd}-\delta^{ac}\delta^{bd}+\delta^{ad}\delta^{bc}),$
where $\nf$ is the  number of independent  fermion components, i.e.  $\nf=16$ in 11d   case. }


\iffa 
For the fermions compactifying on $S^1$, there are two types of boundary conditions: periodic $(P)$ or antiperiodic  $(A)$. The periodic boundary condition is related to the usual Kaluza-Klein reduction. 
The antiperiodic boundary condition case is a special case of the Scherk-Schwarz compactification \cite{Scherk:1978ta}, extended to superstrings by \cite{Rohm:1983aq}. While the bosons remain periodically identified, the fermions are antiperiodic, with shifts in allowed eigenvalues of momenta. In dimensional reduction, keeping only zero KK modes, this gives a spontaneous breaking of the supersymmetry.
\fi 

The fermionic contribution to the  1-loop free energy is 
\al{\la{6778}
\f^F_1=&-\tfrac {1 }4\,\nf\, \Omega^{(\eta)}\ ,}
where  $\nf=16$ is the number of  independent components of the fermions   and $\Omega^{(\eta)}$ was defined in (\ref{233}). 
The 2-loop  correction  coming from the   quartic vertices in  (\ref{12}) is found to be 
\al{
\f_2^F=&{V_c\ov T_2} \Big[ \tfrac {1}{4} N  \Delta^{ab}\Sigma^{(\eta)}_{b,\alpha \beta}(C\Gamma_a\mathcal P )^{\alpha \beta }
  -\tfrac { 1} {16}(C\Gamma^a\mathcal P)^{\alpha \beta}(C\Gamma^b\mathcal P)^{\gamma\delta}\Big(
\tD^{(\eta)}_{b,\alpha\beta}\tD^{(\eta)}_{a,\gamma \delta}-\tD^{(\eta)}_{\alpha\gamma}\tD^{(\eta)}_{ba,\beta \delta}-\tD^{(\eta)}_{a,\alpha\delta}\tD^{(\eta)}_{b,\gamma\beta }
\Big) \Big] \ .
\label{82}
}
$\tD_{\alpha \beta}$ and $\tD_{ab,\alpha \beta}$ in \rf{34}   actually  vanish as momentum integrands 
in these cases   are  odd functions of momentum  so that  the 2-loop correction involves only $ \tD^{(\eta)}_{a,\alpha\beta}$.  
The term $\sim \epsilon^{abc}\partial_{a}X^{i}\partial_{b}X^{j}\bar{\theta}\Gamma_{ij}\partial_{c}\theta$ in \rf{12} does
  not contribute to the 2-loop   free energy due to $\delta^{ij} \G_{ij}=0$.
Furthermore, 
 $  \tD^{(\eta)}_{a,\alpha\beta} $   can be written in terms of the bosonic propagator constant 
 $\Delta^{(\eta)}_{ab}$ in \rf{d26} as  
\be\la{366}
\tD^{(\eta)}_{b,\alpha\beta}
=(\Gamma^{a}\mathcal{P}C^{-1})_{\alpha\beta}\ \Delta^{(\eta)}_{ab} \ .\quad
\ee
As a  result, \rf{82}   may be written 
as 
\al{
\f^F_2=&\frac{V_c}{T_2}\Big[\tfrac{1}{4}N \Delta^{ab}\Delta^{(\eta)}_{cb}\text{Tr}(\Gamma^{c}\Gamma_{a}\mathcal{P})
-\tfrac{1}{16}\Big(\Delta^{(\eta)}_{db}\Delta^{(\eta)}_{ca}\text{Tr}(\Gamma^{d}\Gamma^{a}\mathcal{P})\text{Tr}(\Gamma^{c}\Gamma^{b}\mathcal{P})+\Delta^{(\eta)}_{da}\Delta^{(\eta)}_{cb}\text{Tr}(\Gamma^{c}\Gamma^{a}\Gamma^{d}\Gamma^{b}\mathcal{P})\Big)\Big]\nonumber\\
=&\frac{V_c}{T_2}\Big(\tfrac{1}{4}N \nf \Delta_{ab}
-\tfrac{1}{16}\nf^2 \Delta^{(\eta)}_{ab}\Big)\Delta^{(\eta)ab}\ ,\la{887}
}
where $\Delta_{ab}=\Delta_{ab}^{(0)}$ is the periodic  bosonic  propagator  factor  as in  \rf{205}. 

An analogous  expression is  found  for the GS   string case on a circle  with periodic or antiperiodic b.c. for the fermions  leading to the following  fermionic counterparts of the bosonic string contribution in \rf{41}
\al{&
\f^F_{\mathbb R\times S^1_P}=\nf \frac {1}{24 R}+\Big(\tfrac{1}{4}N \nf -\tfrac{1}{16}\nf^2 \Big)\frac{1}{144\pi  T_1R^{3}}+... \ ,\la{380}
\\&
\f^F_{\mathbb R\times S^1_A}=-\nf \frac {1}{48 R}-\Big(\tfrac{1}{8}N\nf +\tfrac{1}{64}\nf^2 \Big)\frac{1}{144\pi  T_1 R^{3}}+...\ . \la{390} 
}
Similarly, for the fermionic contribution  to the M2  brane free energy on the circle with 
the bosonic  part given in \rf{512}   we get 
\al{&
\f^F_{{\mathbb R^2\times \td S^1_P}}=\nf \frac {\zeta(3)}{16\pi^3 \tilde R^2}+\Big(\tfrac{1}{4}N\nf-\tfrac{1}{16}\nf^2\Big)\frac{3\zeta(3)^{2}}{64\pi^{7} T_2\tilde R^{5}}+...\ ,\la{667} \\
&
\f^F_{{\mathbb R^2\times \td S^1_A}}=-\nf \frac{3\zeta(3)}{64\pi^{3}\tilde R^{2}}-\Big(\tfrac{1}{4}N\nf+\tfrac{3}{64}\nf^2 \Big)\frac{9\zeta(3)^{2}}{256\pi^{7} T_2\tilde R^{5}}+... \ .\la{377}
} 
For the M2  brane  on $S^1\times \tilde S^1$  torus 
(with the  bosonic part given in   \rf{555}--\rf{235}) there are  4 possible   choices 
of fermion b.c.  $PP, PA, AP$ and $AA$,   where $P$ stands for periodic ($\eta=0$) 
 and $A$ --  for antiperiodic ($\eta=\ha $)  choice in one of the two circles. 
 The  exact expressions in the $AP$ and $PA$ cases  are related by $R\leftrightarrow \td R$. 
If  expressed in terms of the functions $\B_\s$  and $\Ee_\s$ ($\s=\pm 1, \pm \hal$) of the  { same} 
  argument $R\ov \tR$  the fermionic 1-loop and 2-loop contributions are   given by 
\al{
\f^F_{\mathbb R\times S^1_P \times \td S^1_P}
   =&\nf \Big(\frac{1}{24R}+\frac{\zeta(3)R}{8\pi^{2}\tilde R^{2}}+\frac{\B_{1}}{\pi \tilde R}\Big) 
+{1\ov T_2} \Big(N\nf -\tfrac{ 1 }{4}\nf^2  \Big)\Big(\frac{1}{1152\pi^{2}R^{3}\tilde R}+\frac{3\zeta(3)^{2}R}{128\pi^{6}\tilde R^{5}}+{\Ee_1}\Big)+...\ ,\la{3121} \\
\f^F_{\mathbb R\times S^1_A \times \td S^1_P}   
=&
\nf\Big(-\frac{1}{48R}+\frac{\zeta(3)R}{8\pi^{2}\tR^{2}}+\frac{ \B_{-1}}{\pi \tR}\Big)
 \no \\ 
&\qquad + \frac 1{T_2}\Big[-\big(N \nf+\tfrac{1}{8}\nf^2\big)
\frac{1}{2304\pi^2 R^3 \tilde R}
+
\big(N \nf-\tfrac{1}{4}\nf^2\big)
\frac{3\zeta(3)^2R}{128\pi^6 \tilde R^5}+ {\Ee_{-1}}\Big]
+...\ , \la{3139}
\\
\f^F_{\mathbb R\times S^1_P \times \td S^1_A}   
=&\nf\Big(  - \frac{3\zeta(3)}{32\pi^{2}}\frac{R}{\tilde R^{2}} + \frac{\B_{\frac{1}{2}}}{\pi \tilde R}\Big) + {1\ov T_2} \Big[- \Big({N\nf}+\tfrac{3}{16}\nf^2\Big)\frac{9\zeta(3)^{2}R}{512\pi^{6} \tilde R^{5}} + \Ee_{\frac{1}{2}}\Big] +...\ , \la{3131}
\\
\f^F_{\mathbb R\times S^1_A\times \tilde S^1_{A}}=
&
\nf\Big(- \frac{3\zeta(3)}{32\pi^{2}}\frac{R}{\tilde R^{2}}+ \frac{\B_{-\frac{1}{2}}}{\pi \tilde R}\Big) 
+ {1\ov T_2} \Big[- \Big({N\nf}+\tfrac{3}{16}\nf^2\Big)\frac{9\zeta(3)^{2}R}{512\pi^{6} \tilde R^{5}} + \Ee_{-\frac{1}{2}}\Big] +... 
\ .  \label{d415}
}
The functions  $\B_1$ and   $\Ee_1$ are the same as in \rf{E19}   and $ \Ee_{-1}$, $\Ee_{\pm \hal} $ and $\B_{\pm \hal}$ are  
combinations  of modified Bessel functions $K_\nu(2\pi m n {R\ov \tR})$, 
which are defined in  Appendix \ref{A3} (see  \rf{e40}--\rf{e428}).


\section{Total 2-loop  M2 brane  free energy}
\label{sect4}

The total 2-loop  free energy is found  by   combining  the bosonic  contribution from section \ref{sect2}   with the fermionic one from section \ref{sect3}. 

 
 In the limiting string case on a circle   we find   the  expressions given in the Introduction  in \rf{115},\rf{116}
 which are consistent with the expansion of the  exact square root formula \rf{5}  for the energy. 
 In the case of  M2  brane on the cylinder $\mathbb  R^2 \times  \td S^1$    we get \rf{18} and \rf{114}. 

In the case  of the M2 brane on the torus   with 
 antiperiodic fermions in one  or two circles of   $S^1 \times  \td S^1$    the   
energy is a non-trivial function 
 of  dimensionless parameters $T_2\tR^3 $ and  $R/\tR$. The expansion for large $R/\tR$  is  given in \rf{122}. 
 Explicitly, we get from \rf{206},\rf{71} and \rf{6778},\rf{887}
 \al{
\f= V_cT_2
+
\tfrac{1}{2}\big(N\Omega-\ha \nf\Omega^{(\eta)}\big)
-\frac{V_c}{4T_2}\big(N\Delta_{ab}
-\ha \nf\Delta^{(\eta)}_{ab}\big)^2+...\ ,\qquad \qquad  V_c = (2\pi)^2 R \tR \ . \la{417}}
In the case of periodic  fermions $\eta=0$  so that  $\Omega^{(\eta)}=\Omega$    and $\Delta^{(\eta)}_{ab}= \Delta_{ab} $,
     so that  both  the 1-loop  and 2-loop 
 contributions are found to  be proportional to $N - \ha \nf$    and thus vanish, in agreement with the  BPS nature of  the corresponding configuration.

In the case of fermions   antiperiodic on one circle  we get from \rf{222}   and \rf{3139} 
\al{
\f_{\mathbb R\times S^1_{A}\times \tilde S^1_P}
&=
4 \pi^2 T_2 R\tilde R-\big(N+\tfrac{1}{4} \nf\big)\frac{1}{12 R}-\big(N-\tfrac{1}{2} \nf\big)\frac{ \zeta(3) R}{4 \pi^2 \tilde R^2}-\big(N\B_{1}-\tfrac{1}{2} \nf\B_{-1}\big)\frac{2}{\pi \tilde R}\no 
\\
 &
\ \ -\frac 1{T_2}\Big[\big(N+\tfrac{1}{4} \nf\big)^2\frac{1}{1152\pi^{2} \tilde RR^{3}}
+ \big(N-\tfrac{1}{2} \nf\big)^2\frac{3\zeta(3)^{2}R}{128\pi^{6} \tilde R^{5}} + N^2\Ee_1 
- \Ee_{-1}  \Big] +...
\ ,  \label{4233}
\\
N^2\Ee_1 - \Ee_{-1} &=
\big(
N+\tfrac{1}{4} \nf
\big)
\big(
N\B_1-\tfrac{1}{2} \nf\B_{-1}
\big)
\frac{1}{24\pi^3 R^2 \tilde R^2}
+\big(
N+\tfrac{1}{4} \nf
\big)
\big(
N\C_1-\tfrac{1}{2} \nf\C_{-1}
\big)
\frac{1}{24\pi^2 R \tilde R^3}\nonumber
\\
&\quad
- \big(
N-\tfrac{1}{2} \nf
\big)
\big(
N\C_1-\tfrac{1}{2} \nf\C_{-1}
\big)
\frac{3R\zeta(3)}{8\pi^4 \tilde R^5}
+
\big(N\B_1
-\tfrac{1}{2} \nf\B_{-1}
\big)^2
\frac{1}{2\pi^4 R \tilde R^3}\nonumber
\\
&\quad
+\big(N \B_1-\tfrac{1}{2} \nf\B_{-1}\big)
\big(N \C_1-\tfrac{1}{2} \nf\C_{-1}\big)
\frac{1}{\pi^3  \tilde R^4}
+\big(N \C_1-\tfrac{1}{2} \nf\C_{-1}\big)^2
\frac{2R}{\pi^2 \tilde R^5}\ ,\label{9944}
}
where $\B_1,\B_{-1},\C_1,\C_{-1}$ are combinations  of modified Bessel functions $K_\nu(2\pi m n {R\ov \tR})$
defined in   Appendix \ref{A3} in equations (see (\ref{D7})--(\ref{D11})). 
Expanding them in the limit ${R\ov \tR} \gg 1$  and keeping only the leading exponentially suppressed term one finds 
(see  (\ref{f7}))
\al{
&N\B_{1}-\tfrac{1}{2} \nf\B_{-1}=\big(N+\tfrac{1}{2} \nf\big)K_1\big(2\pi \frac{R}{\tilde R}\big)+...=\frac{1}{2}\big(N+\tfrac{1}{2} \nf\big) \big(\frac\tR R\big)^{1/2}e^{-2\pi R/\tilde R}+...\ ,\label{429}\\
& N\C_{1}-\tfrac{1}{2} \nf\C_{-1}=\big(N+\tfrac{1}{2} \nf\big)K_0\big(2\pi \frac{R}{\tilde R}\big)+...=\frac{1}{2}\big(N+\tfrac{1}{2} \nf\big) \big(\frac\tR R\big)^{1/2}e^{-2\pi R/\tilde R}+...\label{430}\ .
}
As a result, we get from \rf{4233}   the  asymptotic expression given  in \rf{121}.

  The expression  in the case  of fermions    antiperiodic in $\td R$  circle  and  periodic in $R$ one   is found  by  interchanging  $R\lra \td R$   but it is  useful to write it in terms of the same    functions of $R/\tR$ as in \rf{4233}  to allow  for a   straightforward expansion in ${R\ov \tR} \gg 1$.
   Then we find 
    \al{
\f_{\mathbb R\times S^1_{P}\times \tilde S^1_A}
&=
4 \pi^2 T_2 R\tilde R-  N\frac{1}{12 R}-\big(N+\tfrac{3}{8} \nf\big)\frac{ \zeta(3) R}{4 \pi^2 \tilde R^2}-\big(N\B_{1}-\tfrac{1}{2} \nf\B_{\frac{1}{2}}\big)\frac{2}{\pi \tilde R}\no 
\\
 &
\ \ -\frac 1{T_2}\Big[N^2\frac{1}{1152\pi^{2} \tilde RR^{3}}
+  \big(N+\tfrac{3}{8} \nf\big)^2\frac{3\zeta(3)^{2}R}{128\pi^{6} \tilde R^{5}} + N^2\Ee_1 
- \Ee_{\frac{1}{2}}  \Big] +...
\ ,  \label{933}
\\
N^2\Ee_1 - \Ee_{\frac{1}{2}} &=
N
\big(
N\B_1-\tfrac{1}{2} \nf\B_{\frac{1}{2}}
\big)
\frac{1}{24\pi^3 R^2 \tilde R^2}
+ N \big(
N\C_1-\tfrac{1}{2} \nf\C_{\frac{1}{2}}
\big)
\frac{1}{24\pi^2 R \tilde R^3}\nonumber
\\
&\quad
- \big(
N+\tfrac{3}{8} \nf
\big)
\big(
N\C_1-\tfrac{1}{2} \nf\C_{\frac{1}{2}}
\big)
\frac{3R\zeta(3)}{8\pi^4 \tilde R^5}
+
\big(N\B_1
-\tfrac{1}{2} \nf\B_{\frac{1}{2}}
\big)^2
\frac{1}{2\pi^4 R \tilde R^3}\nonumber
\\
&\quad
+\big(N \B_1-\tfrac{1}{2} \nf\B_{\frac{1}{2}}\big)
\big(N \C_1-\tfrac{1}{2} \nf\C_{\frac{1}{2}}\big)
\frac{1}{\pi^3  \tilde R^4}
+\big(N \C_1-\tfrac{1}{2} \nf\C_{\frac{1}{2}}\big)^2
\frac{2R}{\pi^2 \tilde R^5}\ .\label{944}
}
The  corresponding asymptotic expansion was given in \rf{122}.

  Finally, in the case when fermions are antiperiodic in both circles we get  formally 
  the same  expressions as in \rf{933},\rf{944}    but with $\B_{\frac{1}{2}} \to \B_{-\frac{1}{2}}$, $\C_{\frac{1}{2}} \to \C_{-\frac{1}{2}}$, i.e. 
     \al{
\f_{\mathbb R\times S^1_{A}\times \tilde S^1_A}
&=
4 \pi^2 T_2 R\tilde R-  N\frac{1}{12 R}-\big(N+\tfrac{3}{8} \nf\big)\frac{ \zeta(3) R}{4 \pi^2 \tilde R^2}-\big(N\B_{1}-\tfrac{1}{2} \nf\B_{-\frac{1}{2}}\big)\frac{2}{\pi \tilde R}\no 
\\
 &
\ \ -\frac 1{T_2}\Big[N^2\frac{1}{1152\pi^{2} \tilde RR^{3}}
+   \big(N+\tfrac{3}{8} \nf\big)^2\frac{3\zeta(3)^{2}R}{128\pi^{6} \tilde R^{5}} + N^2\Ee_1 
- \Ee_{-\frac{1}{2}}  \Big] +...
\ ,  \label{733}
\\
N^2\Ee_1 - \Ee_{-\frac{1}{2}} &=
N
\big(
N\B_1-\tfrac{1}{2} \nf\B_{-\frac{1}{2}}
\big)
\frac{1}{24\pi^3 R^2 \tilde R^2}
+ N \big(
N\C_1-\tfrac{1}{2} \nf\C_{-\frac{1}{2}}
\big)
\frac{1}{24\pi^2 R \tilde R^3}\nonumber
\\
&\quad
- \big(
N+\tfrac{3}{8} \nf
\big)
\big(
N\C_1-\tfrac{1}{2} \nf\C_{-\frac{1}{2}}
\big)
\frac{3R\zeta(3)}{8\pi^4 \tilde R^5}
+
\big(N\B_1
-\tfrac{1}{2} \nf\B_{-\frac{1}{2}}
\big)^2
\frac{1}{2\pi^4 R \tilde R^3}\nonumber
\\
&\quad
+\big(N \B_1-\tfrac{1}{2} \nf\B_{-\frac{1}{2}}\big)
\big(N \C_1-\tfrac{1}{2} \nf\C_{-\frac{1}{2}}\big)
\frac{1}{\pi^3  \tilde R^4}
+\big(N \C_1-\tfrac{1}{2} \nf\C_{-\frac{1}{2}}\big)^2
\frac{2R}{\pi^2 \tilde R^5}\ .\label{9440}
}
In the   limit $R\gg \tR$ the Bessel function dependent terms  become  exponentially suppressed so  that the 
 resulting asymptotic expansion of \rf{733}    is the  same as  of \rf{933} given in 
 \rf{122}.

\section{3-loop  contribution} 
\la{s5}

In this section we will discuss how to extend  the above 2-loop  computation to 3-loop level.
We will focus on the  simplest cases of the bosonic  string    on a cylinder $\mathbb R \times  S^1$  ($d=1$)  and  the membrane on $\mathbb R^2 \times \td S^1$  ($d=2$). 
The  corresponding diagrams  are  given in Fig. \ref{fig1}.  The 6-vertex is determined from the 
expansion of the  static-gauge   Lagrangian  that generalizes \rf{11}  ($i=1, ..., N$)
\begin{align}
{\cal L}_B=&T_d\sqrt{\det(\delta_{ab}+T_d^{-1}\J_{ab})}
=T_d+\tfrac12\tr \J+\frac1{T_d}\Big[\tfrac18(\tr \J)^2-\tfrac14\tr \J^2\Big]
\nonumber\\
&\qquad\qquad +\frac1{T_d^2}\Big[\tfrac1{48}(\tr \J)^3+\tfrac16\tr \J^3
-\tfrac18(\tr \J)(\tr \J^2)\Big]+O(T_d^{-3}) , \qquad\qquad  \J_{ab}\equiv \partial_aX^i\partial_bX^i \ . 
\label{6L}
\end{align}
The   contributions to  the energy density   in \rf{6} of the three diagrams (``6-vertex'', ``chain'' and ``basketball'') in Fig. \ref{fig1}
may be written as\foot{Here $V_c$ is the length of the $S^1$.  In the membrane 
 case  we denote  its radius as $\td R$   by analogy with the M2 brane  wrapped 
 on 11d circle discussed above. Also, $P=(p, k_n)$  where  $k_n= {n\ov R}$ in the  string case and $ {n\ov \tR}$ in the membrane case.}
\al{& \qquad \qquad \qquad \qquad \qquad 
\cE^B_3=\cE_{\rm 6v}+\cE_{\rm ch}+\cE_{\bb} 
\ , \la{522}\\
\cE_{\rm 6v}&=\frac{V_c}{6T^2_d}\,(N^3+2N)\,
\Delta_{ab}\Delta^{b}{}_{c}\Delta^{ca} ,
\qquad \qquad 
\cE_{\rm ch}=-\frac{1}{4T_d^2}N^3 \sum_{n}\int\frac{d^d p}{(2\pi)^d}\,
\frac{\big(P_aP_b\Delta^{ab}\big)^2}{(P^2)^2} ,
\label{cha}\\
\cE_{\bb}&=-\frac{V_c}{16T_d^2}\int d^{d+1}\s\Big[
(3N^2-2N)\big(\tr D^2\big)^2- (2N^2-4N)\,\tr D^4\Big] ,\la{bb} \\
&D_{ab}(\s)=\del_a \del'_b  G(\s, \s')\big|_{\s'=0} \ , \ \ \ \   \ 
G(\s,\s') = \langle X(\s)\, X(\s' )\rangle
\ , \qquad  \Delta_{ab}\equiv  D_{ab} (0) \ , \qquad  \Delta^a_a = 0\ , 
\label{b1}
}
where we  assume that $\Delta_{ab}$  is defined using $\z$-function regularization as in  \rf{d26},\rf{207}.\foot{Here we  assume  that bosons are periodic on  the circle 
so  we set  the twist $\eta=0$  but generalization to  $\eta\not=0$ is   straightforward.}

Note that   $D_{ab}(\s)$ with  non-zero $\s$, i.e.  at {separated} points,  is 
 symmetric and traceless. Then    in 2  and  3 dimensions  it satisfies  the identity
\begin{equation}
\tr D^4=\tfrac12(\tr D^2)^2  
\  .
\label{ne}
\end{equation}
 As we   will  discuss  below, the  contact ($\s=0$) term contribution
  can be  ignored in  the string  case but not in the membrane case.

\subsection{String on $\mathbb R\times S^1$}

In this case $\Delta_{ab}$  is  given by \rf{89}   and we find  that 
\begin{equation}
\Delta_{ab}\, \Delta^{b}{}_c\Delta^{ca}=\Delta_{11}^3+\Delta_{22}^3=0 ,
\qquad {\rm i.e.} \qquad \ \ \cE_{\rm 6v}=0 \ .  \la{577}
\end{equation}
In $\E_{\rm ch}$ in \rf{cha} we have $P_aP_b\Delta^{ab}=(p^2-k_n^2)/(24\pi R^2)$   
and $\frac{(p^2-k_n^2)^2}{(p^2+k_n^2)^2}=1-\frac{4k^2_n p^2}{(p^2+k_n^2)^2} $  and 
using $\sum_n 1=0$ (that  discards the power divergent part of the  $p$-integral) 
 we  are left with
$- 4 k_n^2 \int\frac{dp}{2\pi}\frac{p^2}{(p^2+k_n^2)^2}=-|k_n| $. 
Thus using that for $k_n = { n\ov R}$   one has 
$\sum_{n}|k_n|=2R^{-1}\z(-1) = - {1\ov 6} R^{-1}$   and thus 
\begin{equation}
\cE_{\rm ch}=\ -\frac{N^3}{13824\,\pi^2R^5T_1^2}\  .
\label{che}
\end{equation}
To compute  the basketball   diagram   contribution  \rf{bb} we may 
 map the cylinder $\s= \s_1 + i \s_2$ ($\s_2\equiv \s_2 + 2 \pi R$)  to the plane by $z=e^{\s/R}$.  From  $G(\s,0)=-\frac1{4\pi}\ln\big|2\sinh\frac{\s}{2R}\big|^2$ one has,
for $\s\neq0$,
$\tr D^2=\frac{1}{32\pi^2R^4}
\big|\sinh\frac{\s}{2R}\big|^{-4}=\frac{1}{2\pi^2R^4}\frac{|z|^2}{|1-z|^4} ,
$
so that, using \rf{ne},  and  modulo possible contact term   contribution, we get\foot{We shall confirm this assumption below, see  remark  below \rf{I2}.} 
\begin{equation}\la{59b}
\cE_{\bb}=-\frac{N^2}{16  \pi^3R^5 T_1^2} \int_{\mathbb C}d^2z\,  |z|^2\, |1-z|^{-8} .
\end{equation}
The integral here  can be defined using an analytic  continuation (to remove power divergences) as complex Beta-function $B(2, -3)$  which is 
proportional to $1/\Gamma(-1)=0$  and hence\foot{Let us note that  the chain diagram contribution \rf{che} may also be  computed 
in the coordinate space using the same complex-plane analytic continuation
as for the basketball diagram. Writing
$D=\partial_\sigma^2G$ and $\bar D=\partial_{\bar\sigma}^2G$, the relevant
contraction reduces to
$\tr(\Delta D\Delta D)=(12\pi R^2)^{-2}(D^2+\bar D^2)$.
Under the cylinder-to-plane map $z=e^{\sigma/R}$, 
$
\int d^2\sigma\,D^2=\frac{1}{16\pi^2R^2}
 \int_{\mathbb C}d^2z\, \frac{z}{\bar z}(1-z)^{-4}$
and similarly for $\bar D^2$. Analytic continuation of the generalized
complex Beta-function  integral gives
$\int_{\mathbb C}d^2z\,\frac{z}{\bar z}(1-z)^{-4}
=\int_{\mathbb C}d^2z\,\frac{\bar z}{z}(1-\bar z)^{-4}=\frac{\pi}{6}$
and 
$\int d^2\sigma\,(D^2+\bar D^2)=(48\pi R^2)^{-1}$  which leads again to
(\ref{che}).}
\begin{equation}\la{b51}
\cE_{\bb}=0  \ . 
\end{equation}
As a result, the chain  diagram contribution \rf{che} itself 
 gives the total   expression for $\E^B_3$ in the string case
 \be
 \cE^B_{3}\Big|_{\mathbb R\times S^1}  =\ -\frac{N^3}{13824\,\pi^2R^5T_1^2} \ .
 \la{b45}\ee
  This is  the 3-loop term  
 in \rf{h15}  which matches the expansion of the 
exact square root expression  in \rf{h7}.

Let us note also that  the  generalization to twisted b.c. is found by replacing (see \rf{c3},\rf{c13})
\be 
N\ \ \to\ \  6 N B_2 (\eta) =  N ( 6\eta^2 - 6 \eta + 	1)  \ .  \la{5122}\ee
The generalization to the GS  superstring case  is   straightforward  and we will  skip the details. 
One finds that the only   non-vanishing 3-loop 
contribution comes  again   from the chain diagram   and is found by replacing $N \to N -\ha \nf$ in the periodic case, and $N \to N +\four  \nf$ in the  antiperiodic case, in 
agreement with the expansion of \rf{5}.

Assuming  that  this pattern continues  to   higher loops,
 i.e. that   the only non-vanishing contributions   come  from bubble or cactus graphs 
 in Fig. \ref{fig2},  in section \ref{s6}
   we will present the  derivation of the exact relation in \rf{h7}.

\subsection{Membrane on $\mathbb R^2\times\tilde S^1$}

Here we have from \rf{412},\rf{cha}  ($V_c=2\pi \td R$)
\al{  & \Delta_{ab}={\rm diag}(\a,\a, -2\a)\ , \qquad \ \ \  
\a=\frac{\z(3)}{16\pi^4\Rt^3} \ , \qquad \ \  \  \Delta_{ab}\Delta^b{}_c\Delta^{ca}=-6\a^3\la{5123} \ ,\\
& \qquad 
\cE_{\rm 6v}=-(N^3+2N)\frac{\z(3)^3}{2048\,\pi^{11}\Rt^{8}T_2^2} \ , \la{513}
}
i.e.  the 6-vertex   contribution is non-vanishing here.

For the integral in the chain term in \rf{cha} we get  as in \rf{c11}
(using that  $P_aP_b\Delta^{ab}=\a\,(p^2-2k_n^2)$ and 
 removing power divergences, e.g.,   by  dimensional regularization $d=2 \to 2+2\epsilon$)
\begin{equation}
\int\frac{d^d p}{(2\pi)^d}\frac{(p^2-2k_n^2)^2}{(p^2+k_n^2)^2}
=\frac{3}{2\pi}k_n^2 \Big(\tfrac1\epsilon+\gamma_E+\tfrac32-\ln4\pi+\ln k_n^2\Big)+O(\epsilon)\ , \la{515b} \qquad k_n = {n\ov \td R}\ .
\end{equation}
The pole and  related constants multiply $\sum_n k_n^2= 2\tR^{-2} \zeta(-2)=0$, while
$\sum_n k_n^2\ln k_n^2= -4\tR^{-2} \z'(-2)=\pi^{-2} \tR^{-2} \z(3)$, so that we finish with 
\begin{equation}
\cE_{\rm ch}=-3N^3\frac{\z(3)^3}{2048\,\pi^{11}\Rt^{8}T_2^2} \ .
\label{6vc}
\end{equation}
Thus  the two bubble-diagram (6-vertex   and chain) 
  3-loop contributions  follow the pattern  of the 1-loop and 2-loop
ones in \rf{512}: they are powers  of  the same  $\z(3)$ constant. 
This   will no longer  be the case for the basketball   contribution  in \rf{bb}. 

To compute the integral in \rf{bb} 
one  may  (i) use an explicit UV 
 cutoff  (like $e^{-\ve\,  P^2}$)  in each  propagator $D_{ab}$ factor; 
 (ii)  subtract the identically regulated uncompactified zero-winding graph;
 (iii) discard the remaining local power divergences; 
  (iv) finally, 
  take $\ve \to 0$ and  do  $S^1$ mode  sums using  the $\z$-function regularization. 
One should  carefully  account   for   contact region whose width shrinks with $\ve\to 0$ as it may leave a finite remainder.  We shall   describe  instead an  approach  based on  coordinate space representation that leads to the same result.

Let us   introduce  the notation for the  two integrals    in  \rf{bb} 
\al{
\la{516}
I_1=\int_{\mathbb R^2\times S^1}\!\!d^3\s\,\big(\tr D^2\big)^2\ ,\qquad \qquad 
\qquad
I_2=\int_{\mathbb R^2\times S^1}\!\!d^3\s\;\tr D^4\  ,\\
D_{ab}(\s)=-\partial_a\partial_b  G(\s) \ , \qquad\ \ \  \ \ \ \    G=\sum_{m\in\mathbb Z}\,
\frac{1}{4\pi \sqrt{\s_1^2 +\s^2_2 + (\s_3 + 2 \pi \td R\,  m)^2}}\ . \la{5p} }
Here we used  the  representation of the scalar 
propagator  on $\mathbb R^2 \times \td S^1$ as a sum over images.
As we show in Appendix \ref{ap4}
$I_1$   may be computed  using the value of  $D_{ab}$ at separated points,  
while $I_2$    receives a contact contribution so that  (cf. \rf{ne})
\begin{equation}
I_2-\tfrac12 I_1=-8\a^3 =-\frac{\z(3)^3}{512\,\pi^{12}\Rt^{9}}\  .
\label{I2}
\end{equation}
 Note that in  the case of  two world-volume dimensions
 such  contact terms vanish because of  $\tr\Delta^{3}=0$ in \rf{577}. This is why the string theory 
result  \rf{b51}  found above  ignoring  contact terms is, in fact,    prescription-independent.


To evaluate $I_1$  one is   to  isolate  its  most singular part (see \rf{d13},\rf{d14}).
One finds that in \rf{d13} there is  no $\s^{-3}$ term  and thus 
 there is  no logarithmic  UV  divergence in $I_1$ and hence   also in  the  full 3-loop  correction.  Introducing  a short-distance  cutoff ($\s >\varepsilon$)  
we   may write $I_1$ as 
\begin{equation}
I_1 (\varepsilon)
=\frac{1}{16\pi^3\varepsilon^9}
+\frac{42\,\a^2}{5\,\pi \varepsilon^3}
+\frac{4\,  \beta}{\pi^{12}\Rt^9}+O(\varepsilon)\ ,\la{5231}
\end{equation}
where constant $\beta$  in the finite part is  to be determined.  

To compute the  finite  $I^{\rm ren}_1$ part   of $I_1$ 
we need first to subtract  the  $\s\to 0$ singular parts 
from the integrand in \rf{516}.  Splitting the radial 
 integral into the  $\s \leq L$  and  $\s > L$  region (where  we set $u=\s^{-1}$) 
 we get
\begin{align}
I_1^{\rm ren}=J_1+J_2\ ,\qquad 
J_1&=4\pi\!\int_0^{L}\, d\s\, \s^2\Big[\angl{ ({\tr}D^2)^2}-   \frac{9}{64\pi^4\s^{12}}
-\frac{63\,\a^2}{10\pi^2\s^6}  \Big]
-\frac{1}{16\pi^3L^9}-\frac{42\alpha^2}{5\pi L^3}\ ,\label{521}\\
J_2&=2\pi\!\int_0^{1/L}\!\frac{du}{u^3}\int_0^1\!dt\;     ({\tr}D^2)^2\Big|_{\s=u^{-1},\ \cos\theta={Lu}t}\ .
\end{align}
Evaluating $I_1^{\rm ren}$   numerically (for  fixed $L$  and upper limit in the sum over $m$) 
 we find  that 
\begin{equation}
\beta= 0.0258890533... \ . 
\qquad
\label{Cbb}
\end{equation}
Making  a natural assumption that  being a  3-loop coefficient 
 $\b$   should  be given 
 by a sum   of transcendental constants 
   in weight-nine  basis 
 $
\big\{\zeta(9),\ \z(2)\zeta(7),\ \z(4)\zeta(5),\ \z(6)\zeta(3),\ \zeta(3)^3\big\} 
$
 with rational coefficients 
   we find, to a high precision,  a unique solution\foot{We did not manage to derive this  expression 
    analytically so it  may be  treated as (a very plausible) conjecture.
     An analytic derivation 
       may be possible  using  methods  developed for  loop diagram  computations in 
 finite temperature  field theory \ci{Arnold:1994eb,Andersen:2000zn,Hofmann:2009ru,Schroder:2012hm}.}
\begin{equation}
{\;\beta =\frac{385\,\zeta(9)-100\,\zeta(3)^{3}}{8192}\;} \ . 
\label{clos}
\end{equation}
Assuming \rf{clos}
 the renormalized values of $I_1$  and $I_2$  in \rf{I2}   are  then given  by 
\begin{equation}
I_1^{\rm ren}=\frac{385\,\zeta(9)-100\,\zeta(3)^{3}}{2048\,\pi^{12}\Rt^{9}} \ ,\ \ \ 
\qquad\ \ \  \ \ \ I_2^{\rm ren}=\frac{385\,\zeta(9)-108\,\zeta(3)^{3}}{4096\,\pi^{12}\Rt^{9}} \ .
\la{ii}
\end{equation}
As a result,  the  basketball  graph expression  in \rf{bb}   is   given by 
\begin{equation}\la{526}
\cE_{\bb}=-\frac{385\,N^2\zeta(9)}{8192\,\pi^{11}T_2^2\Rt^{8}}
+\frac{N(12N+1)\,\zeta(3)^3}{1024\,\pi^{11}T_2^2\Rt^{8}} \ .
\end{equation}
Taking into account  the contributions \rf{513} and \rf{6vc} of two other  diagrams 
 the  3-loop  term in the bosonic  membrane   energy  density  may be written as 
\begin{equation}
{
\cE^B_{3}\Big|_{\mathbb R^2\times\tilde S^1}
=-\frac{N^{2}}{8192\,\pi^{11}T_2^{2}\Rt^{8}}
\Big[385\,\zeta(9)+16(N-6)\,\zeta(3)^{3}\Big]} \ . 
\label{5f}
\end{equation}

\iffa 
 the entire $\Phi^3$ structure, including the
\begin{equation}
\cE^B_{3}=-\frac{N^{2}}{\pi^{11}T_2^{2}\Rt^{8}}
\left[\beta_\bb(\eta)+\frac{4N+1}{2048}\,\Phi(\eta)^{3}\right] ,
\label{generaleta}
\end{equation}
with $\beta_\bb(0)$ as in \rf{closed}. For $\eta=\tfrac12$ the same
evaluation gives $\beta_\bb(\tfrac12)=0.0215538459826235348\ldots$, which does
\emph{not} lie in the ordinary weight-nine basis --- as expected, since
$D^{(1/2)}_{L}=2D^{(0)}_{2L}-D^{(0)}_{L}$ mixes two periods and generates
level-two Euler sums.
\fi
Let us  stress   again that  no  logarithmic UV   divergences appear
in any of the three 3-loop graphs. This  was 
checked directly using  the short-distance expansion above, and, also, 
independently, by continuing the noncompact $\mathbb R^2$ part 
 to $d=2+\epsilon$ and verifying
that the integral of the ``evanescent'' (cf. \rf{ne})
 combination $\tr D^4-\tfrac12(\tr D^2)^2$  does not   contain an 
$1\ov \epsilon $ pole term. This is  consistent with the observation  in the Introduction 
 that all  local covariant
counterterms  which  are built from $K=\partial\partial X$
vanish on the wrapped-brane background.


A generalization  of \rf{5f} to include the   fermionic contributions, i.e.
 to  the  M2 brane case  should have the same structure with only $\z(9)$ and $\z(3)^3$
  terms present in the antiperiodic fermion case. Confirming this  should be, in principle, 
    straightforward   but is 
    more involved  so we   will  not attempt  to do this    here.
 
Let us  comment on the structure of the transcendental constants appearing in
\rf{h9},\rf{5f}. For a brane  with $d=1,2,...$ spatial dimensions 
 wrapped on a circle with the remaining
world-volume directions non-compact the 1-loop correction is proportional to
$\zeta(d+1)$, cf. \rf{c4},  and each loop order is to add
 transcendental weight
$d+1$. Thus  the $L$-loop term is expected to have weight $L(d+1)$. In the string
case $d=1$ this gives weight $2L$, i.e. only even zeta values or, equivalently,
rational multiples of $\pi^{2L}$. This is why no new constants can appear in
\rf{h15}. 

For the membrane ($d=2$) one gets weight $3L$, i.e. weight 3, 6 and 9 at
one, two and three loops. Hence  a non-trivial weight-nine basis  should   be relevant  at the third order in the inverse tension expansion.\foot{Note that in a generic massless four-dimensional theory the weight usually grows by
 two units per loop, so that weight-nine constants first  appear 
only at five or six loops.}
As was mentioned above, the  weight-nine  basis  contains 
$
\big\{\zeta(9),\ \z(2)\zeta(7),\ \z(4)\zeta(5),\ \z(6)\zeta(3),\ \zeta(3)^3\big\} 
$
but only $\zeta(9)$
and $\zeta(3)^3$ appear in \rf{h9},\rf{5f}.
The same happened one order lower: weight six  basis is
$\{\zeta(6),\zeta(3)^2\}$  but  only $\zeta(3)^2$ occurs in the 2-loop term in
\rf{512}.

 Curiously, in  both cases the surviving constants are precisely the
\emph{single-valued} zeta values, i.e. the image of the single-valued map
$\mathrm{sv}$ of \ci{Brown:2013gia}, for which
$\mathrm{sv}\,\zeta(2n)=0 , \ \ \  \mathrm{sv}\,\zeta(2n+1)=2\,\zeta(2n+1), 
$
i.e.   the single-valued  basis  at weight-six and weight-nine  is  given by  $\zeta(3)^2$ and by
$\{\zeta(9),\,\zeta(3)^3\}$ respectively. This is the same pattern as in the tree-level four-point amplitude of the closed
superstring:  coefficients  in its low-energy expansion are single-valued multiple zeta values
\ci{Schlotterer:2012ny,Stieberger:2013wea}.\foot{Explicitly, one has: 
$\frac{\Gamma(-\frac{1}{4} \a' s)\Gamma(-\frac{1}{4} \a' t)\Gamma(-\frac{1}{4} \a' u)}{\Gamma(1+\frac{1}{4} \a' s)\Gamma(1+\frac{1}{4} \a' t)\Gamma(1+\frac{1}{4} \a'  u)}
=-\frac{1}{stu}\,\exp\Big[2\sum_{n=1}^{\infty}(\frac{\a'}{4}  )^{2n+1} \frac{\zeta(2n+1)}{2n+1}
\big(s^{2n+1}+t^{2n+1}+u^{2n+1}\big)\Big] $  where  $s+t+u=0$.  
Thus only odd zeta values appear, and  at total degree nine the only
partitions of $9$ into odd parts $\ge 3$ are $9$ and $3+3+3$.  The corresponding
low-energy  expansion  contains terms with coefficients $\a'^3  \z(3)$, $\a'^6 \z(3)^2$, 
$\a'^{9} \zeta(9) + \a'^{9} \zeta(3) ^3$, etc., 
 exactly as in \rf{h9},\rf{5f}. We thank  J. Russo   for this  remark
 and pointing out possible connection to \ci{Russo:1997mk}. 
}

For twisted boundary conditions  the   factors of 
 $\z(3)$  in  \rf{513} and \rf{6vc}   should be 
 replaced by $\Phi(\eta)$ defined in \rf{c5}, but we  did not manage to find the 
  corresponding guess  for $\b(\eta)$ in \rf{5231},\rf{clos}. The  above  analogy with 
  single-valued  zeta's  suggests  a natural ansatz.
Let  $\Phi_n={\rm Re}\,{\rm Li}_n(e^{2\pi i\eta})$ and
$\Psi_n={\rm Im}\,{\rm Li}_n(e^{2\pi i\eta})$, with the single-valued polylogarithms
\ci{Brown:2004ib} on $|z|=1$ being $\Phi_n$ for odd $n$ and $\Psi_n$ for even $n$. 
$\beta(\eta)$ should be a weight-nine polynomial in these, containing an even
number of $\Psi$ factors as it is even in $\eta$. Since $\Psi_{2n}(0)=0$ while
$\Phi_{2n+1}(0)=\zeta(2n+1)$, such an ansatz  reduces to a
combination of $\zeta(9)$ and $\zeta(3)^3$ at $\eta=0$.

\section{Resummation of  bubble graphs }
\label{s6}

Let us  discuss   how to  sum up  all  the bubble graph contributions 
to the energy.  This is equivalent  to summing up 
  the  leading in $N$  terms  at  each order in the  inverse  tension expansion 
in  the  energy  $\bE$  in \rf{h15} and \rf{h11}, i.e.  
$\bE= \sum_{L=0}^\infty T^{-L}_d  ( c_L N^L + ...)$,   where $d=1$ for the string and $d=2$ for the membrane case.
 These $N^L$ terms  come from  the    graphs in Fig. \ref{fig2}.
 
  To this end  we may formally  take   the limit when 
both $N$ and  the dimensionless tension $T_d  R^d$   are   large  with their  ratio  being fixed. 
The large $N$  saddle point   approach we will use 
  is  essentially  the same as in \ci{Alvarez:1981kc,Floratos:1988jh}.  
We shall  start  with  the static-gauge  bosonic  brane action 
\begin{equation}
S=T_d \int d^{d+1}\sigma\;\Big[\,\sqrt{h}
-\tfrac12\lambda^{ab}\big(h_{ab}-\delta_{ab}- \partial_aX^i\partial_bX^i \big)\Big] ,
\label{aux}
\end{equation}
where $\lambda_{ab}$ is a Lagrange multiplier field. 
Integrating over  $N$   fields $X^i$  gives an effective action  (cf. \rf{6})
\begin{equation}
S_{\rm eff}[h, \l] =T_d \int d^{d+1}\sigma\Big[\sqrt{ h}
-\tfrac12 \lambda^{ab}(h_{ab}-\delta_{ab})\Big]
+\tfrac{1}{2}N \,{\rm Tr}\, \ln\big(-\lambda^{ab}\partial_a\partial_b\big)\  . \la{au2}
\end{equation}
If we  formally assume  that $T_d \sim  N $  and take  $N$ large  with $N/T_d$ fixed 
then the remaining path integral  over $h_{ab}$ and $\l_{ab}$   will be  dominated by the saddle  point
in $\l_{ab}$ and $h_{ab}$.   The dominant  extremum   of \rf{au2}  can be found by assuming that $\l_{ab}$ and $h_{ab}$  are  {constant}  and thus satisfying\foot{Here  we   denote the   number of non-compact dimensions  as $d$ instead of $l$  used in \rf{64}.}
\al{
&\lambda^{ab}=\sqrt{ h}\;h^{ab} \ , \ \ \ \ \qquad \ \ \  
h_{ab}=\delta_{ab}+ \frac{N}{T_d V_c}\frac{\partial\Omega[\l]}{\partial\lambda^{ab}} \ , \la{a62} \\
& \Omega[\lambda]=\sum_n \int\frac{d^{d}p}{(2\pi)^{d}}\ln\big(\lambda^{ab}P_aP_b\big)\  .
\label{gap}
}
Here   the world-volume is $\mathbb R^{d} \times S^1$. 
$\Omega$ is proportional to the 1-loop correction to the  free energy as in \rf{233},\rf{206}.

That this resums precisely the    bubble graphs in  Fig. \ref{fig2} 
  can be  seen by iterating the second
equation in \rf{a62}:
 to lowest order $h_{ab}=\delta_{ab}+\frac{N}{T_d}\Delta_{ab}$  where $\Delta_{ab}$   was defined  in \rf{d26}.
  This corresponds to insertion of  an effective  vertex 
$- \frac{N}{2T_d}\Delta^{ab}\partial_aX^i\partial_bX^i$ into \rf{aux}. It  
 generates the double bubble  graph  in Fig. \ref{fig0}  at two loops   and the first  two 3-loop diagrams 
  in Fig. \ref{fig1}, etc.  (cf.  Fig. \ref{fig22}). 
  The resulting  expression  for the energy $\bE$    as a function of $N/T_d$ is found  by evaluating \rf{au2} 
  at the saddle-point   values of $h_{ab}$ and $\l_{ab}$. 
  
  Explicitly, in the string  ($d=1$)  case  \ci{Alvarez:1981kc}
   we may  use  rotational symmetry  to 
  choose the matrix  $h_{ab}$ and  thus $\l_{ab}$  to be diagonal so that in \rf{a62}\foot{Here we assume that   label 1   corresponds to a compact  direction.}
\begin{equation}
h_{ab}={\rm diag}(h_1,h_2)\ , \qquad \qquad \l^{ab}=\diag(\l, \l^{-1} )\ , \ \ \qquad \ \ \  \ \  \l= \sqrt{h_1^{-1} h_2} \ . \la{a65}
\end{equation}
For a single  scalar $\Omega\big|_{\l_{ab}=\delta_{ab}}\equiv \Omega(R)$ is given by  \rf{c3} or in the periodic ($\eta=0$) case in  \rf{89}
so that 
\begin{equation}
\Omega[\lambda]={\sqrt{\lambda}}\,\Omega \big(\tfrac{R}{\sqrt{\lambda}}\big)
=- \frac{\l}{6R}= \l \, \Omega(R) \ .\la{a66}
\end{equation}
Then the saddle-point  equations \rf{gap} give (here $\T_1=2\pi R^2T_1,\ \   V_c=2\pi R$)
\al{  &h_1=1-u \  ,
\qquad
h_2=1 +u\, h_1^{-1}{h_2}  \ , \qquad \qquad
 u \equiv-\frac{N }{4\pi R T_1}  \Omega(R) ={N\ov 12 \T_1} 
\ ,  \la{a67}
\\
& h_2={(1-u )( 1-2u)^{-1}}\ , \ \ \qquad    \ \ \     h_1^{-1}h_2=(1-2u)^{-1} \ , \la{a69}
}
As a result, the saddle point value of \rf{au2} is given by\foot{Here at the saddle point 
 $ \lambda^{ab}(h_{ab}-\delta_{ab})=0$.  $V$ is the volume of non-compact  direction.}
\begin{equation}
F\big|_{\rm saddle}= V \E \ ,\qquad \quad  \E= V_c T_1 \Big( \sqrt{h_1 h_2}  -  \sqrt{h_1^{-1} h_2 }\ u\, \Big) 
= 2\pi RT_1\bE \ , \qquad \bE=\sqrt{1-  2u }
\ . \la{a699}
\end{equation}
The  sum of all bubble diagrams thus reproduces the exact  square root expression for the energy in 
\rf{5},\rf{h7}, implying   that all other non-trivial diagrams like  the third one in Fig. \ref{fig1} 
should give    vanishing contributions under a proper regularization 
 that ensures the equivalence of the results  in the static and the 
 orthogonal gauge.\foot{Note that the large $N$ 
  saddle point argument  above   is  also similar in spirit to a  heuristic  justification    of  the equivalence of the Nambu 
   and Polyakov string path integrals in \ci{Polyakov:1987ez}.}

In the case of the membrane on $\mathbb R^2 \times \td S^1$  ($d=2$) 
 we may look, as in  \ci{Floratos:1988jh} 
   for a saddle point  solution   with\foot{Here we assume  that the  
  direction $1$  corresponds  to the compact direction.}
\begin{equation}h_{ab}={\rm diag}(h_1,h_2,h_2)\ , \qquad 
\l^{ab}={\rm diag}(\l_1,\l_2,\l_2)\ , \qquad 
\lambda_1=\frac{h_2}{\sqrt{h_1}}\ ,\qquad\qquad  \lambda_2={ \sqrt{h_1}} \ .\la{610}
\end{equation}
Here the 1-loop correction  for a single scalar  is  given in \rf{412}:
$\Omega(\tR)=-{\z(3)\ov  4\pi^{3}\Rt^{2}}$  so that $\Omega[\lambda]$ in \rf{gap} is found to be (see \rf{c4})
\begin{equation}
\Omega[\lambda]=\frac{1}{{\lambda_2}}\,
\Omega \big(\tfrac{\Rt}{\sqrt{\lambda_1}}\big)
=-\frac{\lambda_1}{\lambda_2}\;\frac{\z(3)}{4\pi^{3}\Rt^{2}} \ . \la{611}
\end{equation}
With (cf. \rf{h11}) 
\begin{equation}  V_c=2\pi\Rt\ , \qquad  \T_2=(2\pi)^2\Rt^3T_2\ , \qquad \qquad \la{612}
u \equiv\frac{N\, \z(3) }{16\pi^{4}T_2\Rt^{3}}
=\frac{N\, \z(3) }{4\pi^{2}\T_2}\ ,
\end{equation}
the  equations \rf{a62} become
\begin{equation}
h_1=1-{2u}h_1^{-1/2}\ ,\qquad \ \ \ 
h_2=1+u\,{h_1^{-3/2}}h_2 \ \ \to  \ \   h_2=h_1^{3/2} (h_1^{3/2}-u)^{-1} \ . 
\label{6133}
\end{equation}
In  \rf{au2}   one has  again  $\l^{ab} ( h_{ab}-\delta_{ab}) =0$  
and also 
$\ha N \Omega[\l]=- u V_cT_2\,h_2 h_1^{-1}$. As a result, the analog of $\cE$ in  \rf{a699} 
  is
\al{
&\cE=V_cT_2\big(h_1^{1/2} h_2 -h_1^{-1}h_2 \, u \big)
=V_cT_2\,{h_2}{h_1^{-1}}\big(h_1^{3/2}- u \big)= 2\pi\Rt\,T_2\, \bE \ ,\la{614} \\
&  \ \ \bE=h_1^{1/2}  \ , \ \qquad \ \ \ \  \bE^3 = \bE -2 u \ , \la{615} 
}
Here we used the   equations  in \rf{6133}. Thus $\bE$ is subject, as in  \ci{Floratos:1988jh}, 
 to a cubic equation already given in   \rf{h11}. 
The relevant positive branch is  for $u \le\frac{1}{3\sqrt3}$, i.e. $\T_2 \ge\ \frac{3\sqrt3}{4\pi^{2}}\,N\z(3)$  and thus  the resulting $\bE$ never reaches 0. 

To find similar expressions in the case of  membrane on 2-torus  one is just to replace 
 the 1-loop  coefficient $u$ in \rf{612},\rf{615} by its  corresponding value in \rf{216},\rf{221}.\foot{The same  applies also  to cases of other topologies of  membrane  discussed in 
 \ci{Floratos:1988jf,Floratos:1989vk,Odintsov:1990pia}.}

It is also   straightforward to generalize the above  argument to the presence of fermions, i.e. to the GS  string   and the  BST  membrane cases. 
If we  formally  consider  the number of  the 
fermionic components  $\nf$   to be of the same order  as $N$, then  
  $N$ in the   1-loop  correction in \rf{au2}   will be  replaced   by $N -\ha \nf$  in the case of the periodic fermions as in \rf{115},\rf{18}, and by $N +{1\ov 4} \nf$  and  by  $N +{3\ov 8} \nf$  
 in the case of the antiperiodic fermions as in \rf{116} and \rf{114}.\foot{The ``$\sqrt h$'' terms in the GS and BST actions
  can be written    in a similar  form as in \rf{aux},  while  the 
   WZ terms    will 
  contribute only to subleading at large $N\sim \nf$   terms in the free energy.}


\section{M2 brane in twisted 11d  background}
\la{s7}

One may also   compute 2-loop M2  brane  energy 
   in  flat  but topologically non-trivial 
11d backgrounds like   \rf{d51}  that   
allows
 to interpolate between the periodic and antiperiodic boundary  conditions for the fermions. 
 We shall assume that the M2  brane   is wrapped on the  11d circle $\td S^1$  with  its  two  other directions being non-compact  and  separate from  the  2-plane which is twisted with $\td S^1$.

\subsection{One-twist case } 

The metric \rf{d51} may be written as 
\be 
 d s^2_{11} =d X^m dX^m+2d Wd \overline W+ dX_{11}^2  \label{55}\ , \qquad    X_{11} = \tR \psi \ , \qquad 
 W=\tfrac {1}{\sqrt{2}}r\,  e^{i(\varphi+q\, \psi)}\ , \ \ \ \ \ \  0\le q< 2  \ . 
\ee
Here  the $W$-plane  is  chosen to be   $(X^{9},X^{10})$. 
Choosing  the static gauge like in \rf{14}, i.e. 
\be 
X^1=\s^1, \qquad X^2 =\s^2,\ \ \ \ \ \  \ X^{11}\equiv X^{11} + 2 \pi\tR=\s^3
\ , \la{778}\ee
the expansions of the bosonic   \rf{11}   and fermionic \rf{12}  parts of the M2 brane Lagrangian 
 read ($I=1, ..., 6$)\foot{In \rf{d519} we omitted  the term $\sim \epsilon^{abc}\partial_{a}X^{i}\partial_{b}X^{j}\bar{\theta}\Gamma_{ij}\partial_{c}\theta$  as it does
  not contribute to the 2-loop   free energy due to $\delta^{ij} \G_{ij}=0$. }  
\al{
\mathcal{L}_B=&\te T_{2}+\frac{1}{2}\partial_{a}X^{I}\partial^{a}X^{I}+\partial_{a}\overline{W}\partial^{a}W+{1\ov T_2} \Big( \frac{1}{8}\partial_{a}X^{I}\partial^{a}X^{I}\partial_{b}X^{J}\partial^{b}X^{J}+\frac{1}{2}\partial_{a}X^{I}\partial^{a}X^{I}\partial_{b}\overline{W}\partial^{b}W\no\\
&\te -\frac{1}{4}\partial_{a}X^{I}\partial_{b}X^{I}\partial^{a}X^{J}\partial^{b}X^{J}-\partial^{a}X^{I}\partial^{b}X^{I}\partial_{a}\overline{W}\partial_{b}W
-\frac{1}{2}\partial_{a}\overline{W}\partial_{b}W\partial^{a}\overline{W}\partial^{b}W
\Big)+...\la{556} \ , \\
\mathcal{L}_F=&\te -\frac{1}{2}\bar{\theta}\Gamma^{a}\partial_{a}\theta+\frac{1}{T_{2}}\Big( {1\ov 4} \partial_{a}X^{I}\partial_{b}X^{I}\bar{\theta}\Gamma^{a}\partial^{b}\theta+\frac{1}{2}\partial_{(a}\overline{W}\partial_{b)}W\bar{\theta}\Gamma^{a}\partial^{b}\theta-\frac{1}{16}\bar{\theta}\Gamma_{a}\partial_{b}\theta\bar{\theta}\Gamma^{b}\partial^{a}\theta\Big)+...\ .\label{d519}
}
The 8 fluctuations  $X^i=( X^I, W,\bar W)$  written in terms of  $\s^3=X^{11}$ mode expansion  are 
\al{
&X^I=\sum_n X_n(\sigma^r)\ e^{i\tfrac n \tR \sigma^3}\ , \qquad W=\sum_n W_n(\sigma^r )\ e^{i\tfrac{n+q}
\tR\sigma^3},\qquad  \overline W=\sum_n \overline W_n(\sigma^r)\ e^{i\tfrac{n-q}\tR\sigma^3}\ .\la{3355}
}
where $\s^r= (\s^1, \s^2)$ and the mode numbers 
  of $W,\overline W$   are  shifted by $\pm q$.  
Thus  the  $W$ contributions to the  1-loop $\Omega$ \rf{233}  and  propagator \rf{d26} 
  constants  will have  shift $\eta=q$.  
  
  Since $X^{11}\to X^{11} + 2 \pi \td R$  is  correlated with a rotation 
  in 2-plane   the 
    fermions 
pick up a phase  under $\s^3 \to \s^3 + 2\pi \tR$:  \ \  $\theta \rightarrow e^{\pi q \Gamma_9\Gamma_{10}}\theta$.\foot{An equivalent   way to see this is 
to use   polar coordinates  where  the fermion  kinetic term  gets a flat Lorentz connection term \ci{Russo:1995ik}.}  Splitting $\theta$ with respect to the projector 
$\ha (1 +i \Gamma_9\Gamma_{10})$  we have 
\al{\theta=\theta_+ + \theta_-\ , \qquad 
\theta_\pm  \rightarrow e^{\pm i \pi q }\theta_\pm \ ,\qquad \qquad 
\theta_\pm=\mp i \Gamma_9\Gamma_{10}\theta_\pm \ .  \la{255}}
The corresponding  non-vanishing  fermionic propagator constant \rf{34}   will be 
 $\tD^{(\eta)}_{a,\alpha\beta}$ with $\eta= \pm \ha q $.\foot{For a generic real twist $q$, demonstrating the vanishing of the  fermionic propagator constants 
  $\Sigma^{(\eta)}_{\alpha\beta}$ and  $\Sigma^{(\eta)}_{ab,\alpha\beta}$ 
requires some  care. Using  
dimensional and $\zeta$-function regularisation we  get  the following  sums of 
    $r=1$ and $3$  powers  of the discrete  momentum  $k_n = {n +q \ov \tR}$: \ \ 
$
\sum_{n\in\mathbb Z} (k_n)^r
=
\frac{1}{\tR^r}
\big[
\zeta(-r,q)+(-1)^r\zeta(-r,1-q)
\big]
=0$

\noindent
and $
\sum_{n\in\mathbb Z} (k_n)^r\log (k_n)^2
=
-\frac{2}{\tR^r}
\big[
\zeta'(-r,q)+(-1)^r\zeta'(-r,1-q)
\big].
$
For $r=1$  and $3$ the latter sum is odd under $q\leftrightarrow1-q$ and thus
cancels  in the  combination of  paired $+q$ and $-q$ fermionic sectors.}

The  bosonic   contribution to the  1-loop  energy density  may be written as (see \rf{233})
\al{
\f^B_{1}=&\ha (N-2)  \Omega+\Omega^{(q)}=-\frac 1{8\pi^3 \tR^2}\Big[ (N-2)\zeta(3)+2\UU(q)\Big] \ , \la{009}\\
\UU(q)=&\re \li _3(e^{2\pi i q})= \sum_{n=1}^{\infty} \frac{\cos (2 \pi n q)}{n^3} \ . \la{008}
}
The 2-loop  contribution coming from  the quartic vertex in  \rf{556}   is found to be (cf. \rf{117})
\al{
\langle\mathcal L_{4B}\rangle=&{1\ov T_2} \te \Big[ \frac 1{8 }\Big((N-2)(N-4)\ntD{a}\,_a\ntD{b}\,_b-2(N-2)^2 \ntD{}_{ab}\ntD{ab}\Big)\no \\
&\qquad \qquad \te +\frac1{2}\Big((N-2)\ntD{a}\,_a\tqD{b}\,_b-2(N-2)\ntD{}_{ab}\tqD{ab}\Big)-\tqD{}_{ab}\tqD{ab}\Big] \ \no \\
=&-\frac {1}{4 T_2}\Big[ (N-2)\ntD{}_{ab}+2\Delta^{(q)}_{ab}\Big]^2\ , \label{518}
}
where  in the second equality    we used that $\ntD{a}\,_a=0$ in the $\zeta$-function regularization. The  non-zero $W$-field propagator   2-derivative  constants  are  found to be 
\al{
&\Delta_{33}^{(q)}=-\frac{1}{8\pi^4 \tR^3}\UU(q)\ ,\qquad 
\qquad
\Delta_{11}^{(q)}=\Delta_{22}^{(q)}=
\frac{1}{16\pi^4 \tR^3}  \UU(q) \  ,  \la{707}
}
 and thus the  bosonic   2-loop  contribution  to  free energy   takes the form 
\al{
\f^B_{2}=&-\frac{3}{256\pi^7 \tR^5T_2}\Big[(N-2) \zeta(3)+2\UU(q)\Big]^2\ .\la{808}
}
The total  bosonic   contribution    has then    the same form  as in the untwisted case in \rf{512} 
with  the replacement $N\zeta(3)\rightarrow (N-2)\zeta(3)+2\UU(q)$.

\begin{figure}[H]
    \centering
    \begin{subfigure}{0.48\textwidth}
        \centering
        \includegraphics[width=\textwidth]{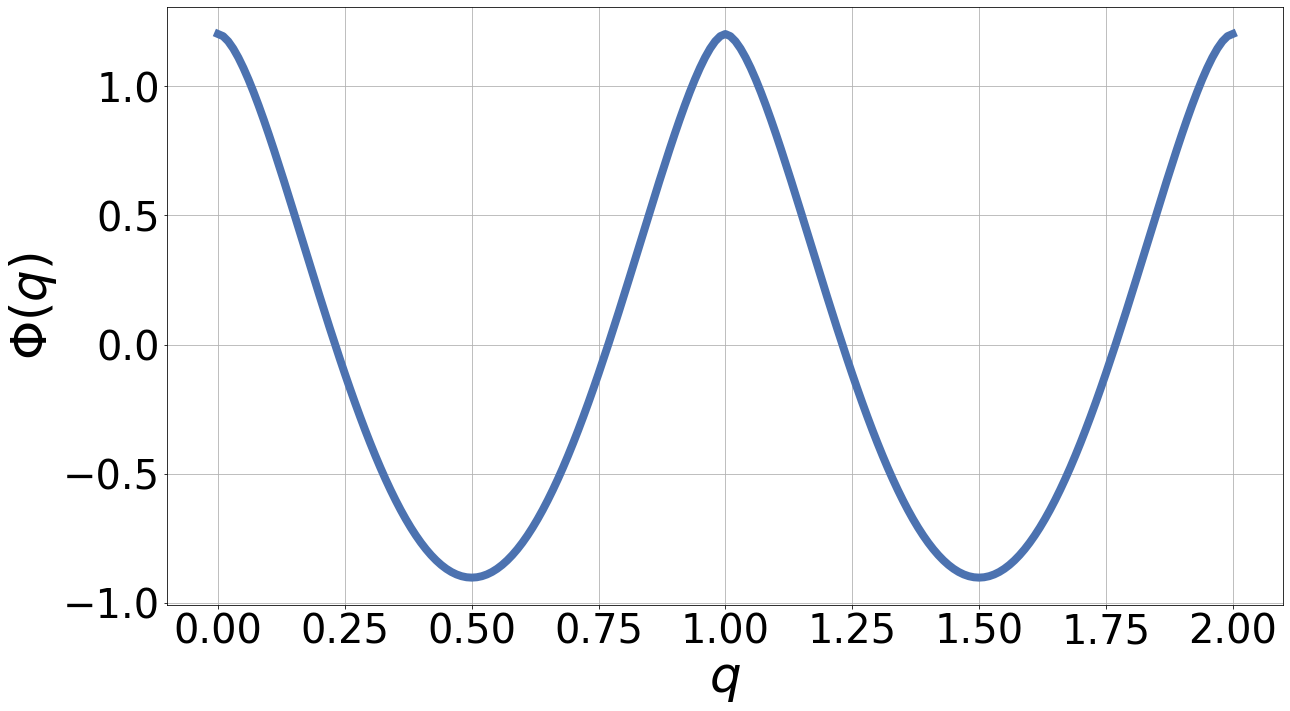}
        \caption{}
        \label{fig00-1}
    \end{subfigure}
    \hfill
    \begin{subfigure}{0.48\textwidth}
        \centering
        \includegraphics[width=\textwidth]{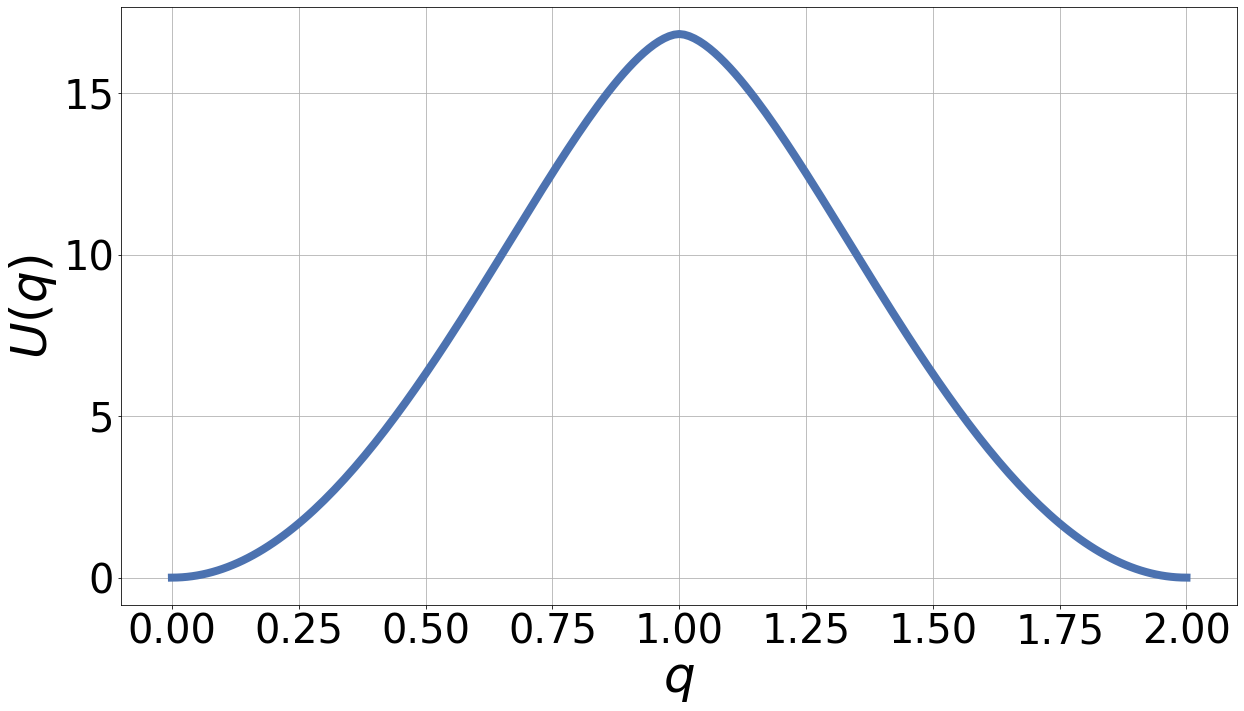}
        \caption{}
        \label{fig00-2}
    \end{subfigure}
    \caption{\small 
   Plots of functions  $\UU(q)$ and $\U(q)$   for  $ q\in (0,2)$   and $N=\ha \nf =8$.
    }
    \label{fig00}
\end{figure}

The fermionic  contribution to the 1-loop free energy is  given by (cf. \rf{6778})
\al{
\f^F_1 =\frac {1}{16\pi^3 \tR^2}\, \nf \UU(\te {q\ov 2})\ . \la{606}
}
From  the Wick contractions in the quartic term in  (\ref{d519})  we get for the  2-loop fermionic contribution (cf. \rf{887})
\al{
\f_2^F&=\frac{2\pi \tR}{T_2}\Big[\tfrac{1}{8}\nf\big[  (N-2) \Delta_{ab}+2\Delta^{(q)}_{ab}\big]\big(\Delta^{( q/2)ab}+\Delta^{(- q/2)ab}\big) -\tfrac{1}{64 }\nf^2 \big(\Delta^{( q/2)}_{ab}+\Delta^{(- q/2)}_{ab}\big)^2\Big]\no \\
&=\frac{3}{64\pi^7 \tR^5T_2}\Big(\tfrac{1}{4}\nf\Big[(N-2) \UU(0)+2\UU(q)\Big] \UU(\tfrac q2)-\tfrac{1}{16}\nf^2 [\UU(\tfrac q2)]^2\Big)\ . \la{2177}
}
In total, the  2-loop  M2  brane  free energy is  found to be given by \rf{22},\rf{23} (cf. \rf{114})
\al{
\f_{\mathbb R^2 \times \tilde S_q^1}&=2\pi \tR T_2 -\frac {\mathrm U(q)}{8\pi^3 \tR^2 }-\frac {3\, \big[\U(q)\big]^2}{256\pi^7 \tR^5 T_2}+...\ ,\label{626}\\
\mathrm U(q)&=(N-2) \zeta(3)+2\UU(q)-\tfrac {1}2\nf \UU(\tfrac{q}2)   \ \ 
\overset{N=8,\ \nf=16}{=}  \ \ 6\, \zeta(3)+2\UU(q)- 8  \UU(\tfrac{q}2) \ .\la{9966}
}
For
 generic $q$  the  fermions  obey twisted boundary conditions in $\s^3$  (cf. \rf{255}). 
Since  from \rf{008}  we have 
 \be\la{7788}
\Phi(0)=\zeta(3)\ ,\qquad\qquad  \Phi(\tfrac{1}{2})=-\tfrac {3}{4}\zeta(3)\ , 
\ee
  we conclude  that for  even integer $q$ when the  fermions  are periodic  
   we find that    \rf{626}  reduces to \rf{18}   and thus vanishes for $N=\ha \nf$. 
For odd integer  $q$  when the fermions are antiperiodic  
\rf{626}   reduces to \rf{114}.
Plots of the   functions $\UU(q)$   and $\mathrm U(q)$ are   given in  Fig. 
 \ref{fig00}.  

As discussed in \ci{Russo:1998xv}, the reduction of the 11d metric  \rf{d51} or \rf{55} to 10d 
gives a type IIA 
 non-supersymmetric 7-brane   background  supported by magnetic RR  vector
field (this background   is U-dual to the NS-NS   KK  Melvin solution).  The  GS  string action in this   background  can be found by the double-dimensional reduction \ci{Duff:1987bx} 
of the M2 brane action
provided one starts with the ``covariant''  form of the action written in the polar coordinates
 in which the corresponding  fermionic field $\vartheta$  appears  in the kinetic 
  term with a  covariant derivative   and   is periodic in $\s^3$  
(i.e. it is related  to  the  ``flat''  $\theta$ in \rf{255}   by the  rotation  $e^{
\frac  q 2\G_9 \G_{10}\psi }$). 
This   field  expanded in $\s^3$ 
then contains a zero-mode  part $\vartheta_0(\s^1,\s^2)$   which 
 is identified   with  the GS string  fermion.
 It corresponds to a lowest  ``twisted'' 
 mode  in the $\s^3$  expansion of $\theta$ in \rf{255}.
 This relation allows one  to identify the  quantum GS  string   contribution in the M2  brane partition function  by separating the   ``lowest-level''  mode parts (cf. \ci{Giombi:2023vzu}).

\subsection{Two-twist case}

 It is   straightforward   to generalize the above  computation 
  to other  similar  flat but  topologically non-trivial 11d backgrounds, 
  in particular, to the case   when there are 2  twist parameters  in the 
   two 2-planes (78) and (9\, 10)
  \cite{Russo:2001na} (cf. \rf{d51},\rf{55})
\al{
&d s^2=d X_h dX_h+\sum_{s}\Big[d r_s^2+r_s^2(d \varphi_s+q_s d \psi )^2\Big]+  \tR^2 d \psi^2 \  , \la{444}
\\&  h=1,2,...,6\ ,\qquad s=1,2\ ,\qquad \varphi_s \equiv \varphi_s+2\pi \ ,\qquad X_{11}=\tR \psi\equiv X_{11}+2\pi 
\tR \ . \la{777}
}
While the  single-twist   background  \rf{55} breaks  supersymmetry  for  generic $q$, 
here  half   of supersymmetry can be preserved  for a special relation between $q_1$ and $q_2$.  Indeed, the Killing spinor  equation   has the following solution 
\be\la{333}
\partial_{\psi}\epsilon=\tfrac 1{2}\big(   q_1 \Gamma_{78} +q_2 \Gamma_{9\, 10}      \big)\epsilon \ ,
\qquad \qquad 
\epsilon=e^{\frac 1{2}(  q_1 \Gamma_{78} +q_2 \Gamma_{9\, 10})\psi}\epsilon_0\ .
\ee
The boundary condition  $\epsilon(\psi+2\pi )\equiv \epsilon(\psi)$ then gives 
\be
e^{\pi(  q_1 \Gamma_{78} +q_2 \Gamma_{9\, 10})}\epsilon_0=\epsilon_0 \ .\label{75}
\ee
In addition to the trivial  solution   $q_1=q_2=0$
 there is a half-supersymmetric one   with $q_1=\pm q_2$ and $\epsilon_0 $ subject to   \cite{Russo:2001na}  
 \be  q_1= \ss\,    q_2 \ , \qquad \qquad \ss=\pm 1 \ , \qquad \qquad \la{79}
\mathrm P_ \ss\,   \epsilon_0=0\ ,\qquad  
\qquad \mathrm P_\ss \equiv \tfrac{1}{2}(1-\ss\,  \Gamma_{789\, 10}) \ . 
\ee
\iffa 
 For the 2-loop free energy, the bosonic part gives
\al{
\f_{2,\mathbb R^2 \times\tilde  S_{(q_1,q_2)}^1}^B&=2\pi \tR T_2-\frac 1{8\pi^3 \tR^2}\big((N-4)\zeta(3)+2\U(q_1)+2\U(q_2)\big)\nonumber\\
&\qquad\qquad\qquad -\frac{3}{256\pi^7 \tR^5T_2}\big((N-4) \zeta(3)+2\U(q_1)+2\U(q_2)\big)^2\ ,\quad 
}
 Combined with the fermionic part, we have 
 \fi
 For general $q_1, q_2$   the 2-loop free  energy  of the M2 brane  wrapped on $X_{11}$ 
   and  two other  non-compact directions $X^1$ and $X^2$ as in \rf{778}
   is found to be the following 
 generalization of the single-twist result  \rf{626}
\al{
\f_{\mathbb R^2 \times \tilde S_{(q_1,q_2)}^1}&=2\pi\tR T_2 -\frac {\mathrm U(q_1,q_2)}{8\pi^3 \tR^2 }-\frac {3\,\big[\mathrm U(q_1,q_2)\big]^2}{256\pi^7 \tR^5 T_2}+...\ , \la{a44}
\\
\mathrm U(q_1,q_2)&=(N-4)\zeta(3)+2\UU(q_1)+2\UU(q_2)-\tfrac{1}{8} {\nf}\sum_{\ss_1,\ss_2=\pm 1}\UU(\ss_1\tfrac{q_1}2+\ss_2\tfrac{q_2}2)\ .\la{a45}
}
As    in \rf{008}  one has $\UU(q)=\UU(-q)$
we  observe (using \rf{7788}) that 
 for   $q_1=\pm\, q_2$    the function $\mathrm U(q_1,q_2)$ in \rf{a45} 
 vanishes for the  physical values $N=\ha \nf=8$. Since the above wrapped M2 brane  configuration in  $\ha$-supersymmetric 11d background  preserves  (after $\kappa$-symmetry projection) 
 $1\ov 4$  of  the  maximal supersymmetry 
 this is  in agreement with the  expected absence of quantum  corrections 
  to the energy of  such supersymmetric brane state. 

\iffa
Also, the background itself preserves 16 supercharges for $q_1=\pm q_2$. The wrapped M2 imposes an additional commuting κ-projector, so it should preserve 8 supercharges: 1/4-BPS relative to the original 32, or “1/2-BPS relative to the supersymmetric fluxbrane background.” Thus the present phrase “a $1/2$-supersymmetric brane state” is ambiguous and, in the usual 11d counting, misleading.
\fi

\section*{Acknowledgements}
We   thank S. Kurlyand, R. Metsaev and J. Russo for  useful discussions and comments on the draft. 
Part of this work was done while AAT was attending 
the Workshop ``Integrable string sigma models''    (15-21.08.2026)  at 
Villa Garbald, Castasegna, Switzerland and he is grateful to the organizers -- S. Driezen and F. Seibold for the invitation  and  hospitality. 
ZW  thanks A. Tolley   for a  useful discussion.
 ZW is grateful to J. Minahan   for hospitality at Uppsala University while this work was in its final stages.
This work was supported by the STFC grant  
ST/X000575/1.



\

\appendix

\section{Epstein  $\zeta$-function and its special values } 
\label{A1}

Let us  consider the following example of Epstein $\zeta$-function\foot{We suppress its 
arguments $R$ and $\tR$.}
\al{
 \Z_{\lambda,\eta}(s)=\sum_{m,n\in \ZZ}\Big[\frac{(n+\lambda)^2}{R^2}+\frac{(m+\eta)^2}{\tilde R^2}\Big]^{-s} \ .}
 Its particular values    can be  computed as follows.  We first use  the   Mellin transform\footnote{$F^{-s}=\frac{1}{\Gamma(s)}\int_{0}^{\infty}dt\,t^{s-1}e^{-Ft}=\frac{1}{\Gamma(s)}\int_{0}^{\infty}dt\,t^{-s-1}e^{-F/t}$} 
\al{
 \Z_{\lambda,\eta}(s)=\frac{1}{\Gamma(s)}\int_{0}^{\infty}dt\,t^{s-1}\Big(\sum_{n}e^{-\left(n+\lambda\right)^{2}t/R_{}^{2}}\Big)\Big(\sum_{m}e^{-(m+\eta)^{2}t/\tilde R_{}^{2}}\Big)\ , \label{f21}
}
and then apply the Poisson summation\footnote{$\ensuremath{\sum_{n=-\infty}^{\infty}f(n+a)=\sum_{k=-\infty}^{\infty}\hat{f}(k)e^{2\pi ika}},\ \  \hat{f}(k)=\int_{-\infty}^{\infty}du\,f(u)e^{-i2\pi ku}$} to the sum over $n$ 
\al{
\sum_{n}e^{-(n+\lambda)^{2}t/ R_{}^{2}}=\sqrt{\tfrac{\pi}{t}}R_{}\sum_{k}\ e^{2\pi i k \lambda}\ e^{-k^{2}\pi^{2} R_{}^{2}/t}=\sqrt{\tfrac{\pi}{t}}R_{} \Big[1+2\sum_{k=1}^{\infty}\cos\left(2\pi\lambda k\right)e^{-k^{2}\pi^{2}R_{}^{2}/t}\Big]\ .
}
As a result,  (\ref{f21})  takes the form 
\al{
 \Z_{\lambda,\eta}(s)=\frac{\sqrt{\pi}R_{}}{\Gamma(s)}\sum_{m}\int_{0}^{\infty}dt\,t^{s-\frac 32}\Big[e^{-\left(m+\eta\right)^{2}t/\tR_{}^{2}}+2\sum_{k=1}^{\infty}\cos\left(2\pi\lambda k\right)e^{-\left(m+\eta\right)^{2}t/\tR_{}^{2}-k^{2}\pi^{2}R_{}^{2}/t}\Big]\ .\label{f24}
}
 Applying the Mellin transform 
 to the first term in the brackets in  (\ref{f24}) and using  that the second term   can be  expressed in terms of the modified Bessel function   
$K_{\nu}(z)=\frac{1}{2}(\frac{1}{2}z)^{\nu}\int_{0}^{\infty}\frac{\mathrm{d}t}{t^{\nu+1}} {\exp}(-t-\frac{z^{2}}{4t})$
  we find 
\al{
 \Z_{\lambda,\eta}(s)=&\frac{\sqrt{\pi} R_{}\Gamma(s-\hal)}{\tR_{}^{1-2s}\Gamma(s)}\Big[ \zeta(2s-1,1-\eta)+\zeta(2s-1,\eta)\Big]\nonumber \\
 &\qquad +\frac{4\sqrt{\pi} R_{}}{\Gamma(s)}\sum_{m}\sum_{k=1}^{\infty}\cos\big(2\pi\lambda k\big)\Big(\frac{|m+\eta|}{\pi k\tR R_{}}\Big)^{\hal-s}K_{\hal-s}\Big(2\pi k|m+\eta|\frac{R_{}}{\tR_{}}\Big)\ . \la{a55}
}
The  particular value  that is relevant  here is  $s=-\hal$ when 
 \al{ 
 & \Z_{\lambda,\eta}(-\ha )
 =-\frac{R}{2\pi^2 \tR^2}\Phi(\eta) -\frac 2{\pi \tR}\mathcal G_{\lambda,\eta}\ ,\label{f17}
 \\
& \UU(\eta)\equiv  \re\mathrm{Li}_3\big(e^{2 \pi i \eta}\big)
= \sum_{n=1}^{\infty} \frac{\cos (2 \pi n \eta)}{n^3}\ , \label{c5}\\
&\Bsumt_{\lambda,\eta}\equiv \Bsumt_{\lambda,\eta}\big( {R\ov \tR}\big)=\sum_{m}\sum_{k=1}^{\infty}\cos\left(2\pi\lambda k\right)\frac{|m+\eta|}{k}K_{1}\Big(2\pi k|m+\eta|\frac R\tR\Big)\label{E2}\ , 
}
where $K_1$ is  the modified Bessel function. 
Note that 
  $K_{1}\big(2\pi k|m+\eta|\frac R\tR\big)$ terms are  exponentially suppressed in  the limit  $R\gg\tR$  since  $K_\nu$ 
  has  the following asymptotic expansion  
\al{
&K_{\nu}\left(2\pi mnz\right)=\left({4mnz}\right)^{-\frac{1}{2}}\mathrm{e}^{-2\pi mnz}\sum_{k=0}^{\infty}a_{k}(\nu)\left({2\pi mnz}\right)^{-k},\label{f7}\qquad \qquad 
a_k(\nu)
=
\frac{\Gamma(k+\nu+\frac12)}
{2^k k!\,\Gamma(\nu+\frac12-k)}.
}
 Let us also define 
\al{
&\Bsumt_{\lambda}\equiv \Bsumt_{\lambda}\big( {R\ov \tR}\big)=\sum_{m,k=1}^{\infty}\cos\left(2\pi\lambda k\right)\frac{m}{k}K_{1}\Big(2\pi km\frac R\tR\Big)\label{f3}\ .
}
  The  derivatives   with respect to $\rho= R/\tR$  may be written as 
\al{
&\del_\rho \Bsumt _{\lambda,\eta}=-\rho^{-1} \, \Bsumt_{\lambda,\eta}-2\pi \Csumt_{\lambda,\eta},\qquad \del_\rho \Bsumt_{\lambda}=-\rho^{-1} \, \Bsumt_{\lambda}-2\pi \Csumt_{\lambda}\ ,\qquad \qquad \rho \equiv  {R \ov \tR} \ , \label{E4}\\
&\Csumt_{\lambda,\eta}(\r)=\sum_m\sum_{k=1}^\infty \cos(2\pi \lambda k)(m+\eta)^2
K_0\big(2\pi k|m+\eta|\r\big)\ ,\label{E6}\\
&\Csumt_{\lambda}(\r)=\sum_{m,k=1}^\infty \cos(2\pi \l k)m^2
K_0\big(2\pi km\r\big)\label{E777}\ .
}
Let us  also   list some special values $\B_r$ 
of the Epstein  $\zeta$-function 
(\ref{f17})  used  in the main text:
\al{&
\B_1 = \Bsumt_{0} \ , \qquad \ \B_{\frac 12}= \ha \Bsumt_{0,\frac 12} \ , \qquad \qquad 
\Bsumt_{0,\eta}=\sum_{m}\sum_{k=1}^{\infty}\frac{|m+\eta|}{k}K_{1}\big(2\pi k|m+\eta|\r \big)\ , \label{E7}
\\ 
 &\B_{-\frac 12} = \ha \Bsumt_{\frac 12,\frac 12} \ , \qquad \ \ \ \ \ \ 
\Bsumt_{\frac 12,\eta}=\sum_{m}\sum_{k=1}^{\infty}(-1)^k\frac{|m+\eta|}{k}K_{1}\big(2\pi k|m+\eta|\r\big) \ , \label{E8}\\
&\B_{-1} = \Bsumt_{\frac 12 } \ , \qquad\qquad \qquad  \Bsumt_\lambda = \ha 
 \Bsumt_{\lambda,0}- \frac \pi{4\rho}B_2(\lambda)\ ,\qquad \qquad 
 B_2(\l )=\l^2-\l +\tfrac{1}{6} \ . 
 \label{f4}
}
In \rf{f4} $\Bsumt_{\lambda,0}$  is split into the $m=0$  mode  contribution\foot{Note that 
$-\frac 2{\pi \tR}\lim _{\epsilon\rightarrow0}\sum_{k=1}^{\infty}\cos\left(2\pi\eta k\right)\frac{\epsilon}{k}K_{1}\left(2\pi k\epsilon\rho\right)=-\frac 1{\pi^2 R}\sum_{k=1}^{\infty}\frac{\cos(2\pi \eta k)}{k^2}
=-\frac {B_2(\eta)}{ R}$.}
$\frac \pi{2\rho}B_2(\lambda)$  proportional to the second Bernoulli polynomial  $B_2$   and 
$\Bsumt_\lambda$ which is  exponentially suppressed  for $\rho \gg 1$.


\section{Loop corrections to energy of  M2  brane on 
$\mathbb R^2 \times S^1$ }
\label{A2}

The 1-loop   contribution in \rf{233}  for a boson with twist $\eta$   can be defined using 
Hurwitz $\z$-function.\foot{In this Appendix we will use the same  notation $R$ for the radius  of $S^1$ in the string and the membrane case.} 
We have 
\al{
\sum_{n}\text{Tr}\ \ln(-\partial^{2})&= V \Omega^{(\eta)} 
= -V\sum_{n}\frac{\partial}{\partial s}\int\frac{d^{l}p}{(2\pi)^{l}}\Big[p^{2}
+ (k^{(\eta)}_n)^2\Big]^{-s}\Big|_{s=0}\ ,\label{c2} \qquad \ \ 
k^{(\eta)}_n = \frac{n+\eta}{R}\ , 
}
where $l=1$ for  string on $\mathbb R \times S^1$ and  $l=2$ for the membrane on $\mathbb R^2 \times S^1$. 
For $\mathbb R \times S^1$ case 
\al{
 \Omega^{(\eta)}
&
=
\frac{1}{
 R}\sum_{n}\big|n+\eta\big|=
\frac{1}{R}\Big[
\zeta(-1,1-\eta)+\zeta(-1,\eta)
\Big]=-\frac{1}{R}B_2(\eta)= -\frac{1}{R}(\eta^2 -\eta + \tfrac{1}{6})   \ ,\label{c3}
}
where $B_2$ is the second Bernoulli polynomial  as in \rf{f4}.
For $\mathbb R^2 \times S^1$ we get 
\al{
 \Omega^{(\eta)}
&
=-\frac{1}{4\pi R^{2}}      \sum_{n}   \big(n+\eta\big)^{2}   \ln\,   \frac{(n+\eta)^2}{R^2}  
+
\frac{1}{4\pi R^{2}}\sum_{n}\big(n+\eta\big)^{2}\nonumber\\
&=
\frac{1}{2\pi R^2}\Big[
\zeta'(-2,1-\eta)+\zeta'(-2,\eta)
\Big]
=-\frac{ 1}{4\pi^3 R^2}\UU(\eta) \ ,\label{c4}
}
where  $\UU(\eta)$    was defined in  \rf{c5}.

The 2-loop  correction involves the  products  of the  propagator constants  $\Delta_{ab}^{(\eta)}$ defined in \rf{d26},\rf{204}. 
For the component  corresponding to derivatives in compact  direction 
 $\Delta_{cc}^{(\eta)}$   we have
\al{
\Delta^{(\eta)}_{cc}=
\frac{1}{2\pi R}\sum_{n}\int \frac{d^l p}{(2\pi)^l}\frac {\big(k_n^{(\eta)}\big)^2}{p^2+\big(k_n^{(\eta)}\big)^2}=\frac{1}{2\pi R}\frac{\Gamma(1-\frac{l}{2})}{(4\pi)^{l/2}}
\sum_n |k^{(\eta)}_n|^l\ .\label{c9}
}
For $\mathbb R\times S^1$ ($c=2,\ l=1$) case   this gives    
\be
\Delta^{(\eta)}_{22}=\frac 1{4\pi R^2} \sum_n |n+\eta|=\frac 1{4\pi R^2} \Big[\zeta(-1,\eta)+\zeta(-1,1-\eta)\Big]=-\frac{1}{4\pi R^2}B_2(\eta)\ .
\label{c10}
\ee
For $\mathbb R^2\times S^1$ membrane  we get  ($c=3,\ l=2+2\varepsilon, \ \varepsilon \to 0$)
\al{
\Delta_{33}^{(\eta)}&=\lim_{\varepsilon\rightarrow0} \frac{\Gamma(-\varepsilon)}{(4\pi)^\varepsilon\, 8\pi^2 R}\sum_n|k_n^{(\eta)}|^{2+2\varepsilon}
=\lim_{\varepsilon\rightarrow0} \frac{1}{8\pi^2 R}\big(-\frac 1\varepsilon-\gamma_E\big)\sum_n|k_n^{(\eta)}|^{2}\big(1+2\varepsilon\ln|k_n^{(\eta)}| \big)\nonumber\\
&=-\frac{1}{4\pi^2 R}\sum_n|k_n^{(\eta)}|^{2}\ln|k_n^{(\eta)}|=-\frac{1}{8\pi^4 R^3}\UU(\eta)\ .\label{c11}
}
Here we used dimensional regularization to remove spurious power  divergences,  but  the result 
is well defined  if we use  just the $\z$-function regularization of the sum over $n$. Indeed, introducing a  UV cutoff $\Lambda$ in the  momentum $p$ integral in \rf{c9} we get 
 \al{
\Delta^{(\eta)}_{33} 
&=\frac 1{8\pi^2 R}\sum_n\big(k_n^{(\eta)}\big)^2\Big[\ln \Lambda^2+\ln\big(1+\tfrac {(k_n^{(\eta)})^2}{\Lambda^2}\big)-\ln\big( (k_n^{(\eta)})^2\big) \Big]\ .\la{4a4}
}
The dependence on $\Lambda$ here   goes away  
after   the sum over $n$ is  $\z$-function regularized since    
\al{
&  \sum_n\big(k_n^{(\eta)}\big)^2=\zeta(-2,\eta)+\zeta(-2,1-\eta)=0\ , \label{c13}\\
& \sum_n\big(k_n^{(\eta)}\big)^2\ \ln\big(1+\tfrac {(k_n^{(\eta)})^2}{\Lambda^2}\big)=\sum_{m=1}^\infty \sum_n \frac {(-1)^{m+1}}{m \Lambda^{2m}}\big(k_n^{(\eta)}\big)^{2m+2}\nonumber\\
&\qquad \qquad \qquad \qquad =\sum_{m=1}^\infty  \frac {(-1)^{m+1}}{m \Lambda^{2m}}\Big[\zeta(-2m-2,\eta)+\zeta(-2m-2,1-\eta)\Big]=0\ .\label{c14}
}
Here we used  that $\zeta(-2m,\eta)=- {1\ov 2m+1} B_{2m+1}(\eta)$, \ \  $B_{2m+1}(\eta)=-B_{2m+1}(1-\eta)$, where $B_{2m+1}$ are the  Bernoulli polynomials. 
As a result, we  get the same expression as in \rf{c11}. 

The $\Delta^{(\eta)}_{rs} \, (r,s=1,...,l)$  components are  given by 
\be
\Delta^{(\eta)}_{rs}=
\frac{1}{2\pi R}\sum_{n}\int \frac{d^l p}{(2\pi)^l}\frac {p_r p_s}{p^2+(k^{(\eta)}_n)^2}
=\frac{\delta_{rs }}{2\pi R}\frac{\Gamma(-\frac{l}{2})}{2(4\pi)^{l/2}}\sum_n |k^{(\eta)}_n|^l.\label{c12}
\ee
For $\mathbb R\times S^1$ ($l=1$)   this  gives (cf. \rf{c10}) 
\be
\Delta_{11}^{(\eta)}=-\frac{1}{4\pi R^2}\sum_n |n+\eta|=\frac{1}{4\pi R^2}B_2(\eta)=-\Delta_{22}^{(\eta)}.\label{c132}
\ee
For the  membrane on $\mathbb R^2 \times S^1$  ($l=2$)    we may use   dimensional 
regularization as in \rf{c11} or introduce a cutoff in 
 \rf{c12} as in \rf{4a4}. In the  latter case   we get 
\al{
\Delta^{(\eta)}_{rs}&=
\frac{1}{2\pi R}\sum_{n}\int \frac{d^l p}{(2\pi)^l}\frac {p_r p_s}{p^2+(k^{(\eta)}_n)^2}=-\frac{ \delta_{rs}}{16\pi^2 R}\sum_n (k_n^{(\eta)})^2\Big[\ln \Lambda^2+\ln\big(1+\tfrac {(k_n^{(\eta)})^2}{\Lambda^2}\big)-\ln (k_n^{(\eta)})^2\Big]\ , \la{cc33}
}
The first two terms  in the square   bracket  do not contribute due to  (\ref{c13}), (\ref{c14}) and  thus 
\al{
\Delta^{(\eta)}_{rs}&=\delta^{rs}\frac{1}{8\pi^2 R}\sum_n|k^{(\eta)}_n|^{2}\ln|k^{(\eta)}_n|=\frac{1}{16\pi^4 R^3}\UU(\eta)\, \delta^{rs} \ .\label{c117}
}
 We  may then  compute  the products of two $\Delta$'s  that enter the 2-loop correction
in \rf{71},\rf{887}. 
 For the  case of the  GS  string on $\mathbb R \times S^1$  we find 
\al{
\Delta_{ab}\Delta^{ab}=\frac 1{288 \pi^2 R^4},\qquad \Delta_{ab}\Delta^{(\hal)ab}=-\frac 1{576 \pi^2 R^4},\qquad \Delta_{ab}^{(\hal)}\Delta^{(\hal)ab}=\frac 1{1152 \pi^2 R^4} \ , \label{c7}
}
while   for the M2 brane on $\mathbb R^2 \times S^1$ 
\al{
\Delta^{}_{ab}\Delta^{ab}=\frac {3\zeta(3)^2}{128 \pi^8 R^6},\qquad \Delta_{ab}^{}\Delta^{(\hal)ab}=-\frac {9\zeta(3)^2}{512 \pi^8 R^6},\qquad \Delta_{ab}^{(\hal)}\Delta^{(\hal)ab}=\frac {27\zeta(3)^2}{2048 \pi^8 R^6}.\label{c8}
}

\section{Loop corrections to energy of  M2 brane  on  $\mathbb R \times S^1 \times \tilde  S^1$}
\label{A3}
Here  the analog of the 1-loop   contribution  \rf{c2} is 
\al{
&\Omega^{(\lambda,\eta)}=-\frac{\partial}{\partial s}{Z}(s)\Big|_{s=0}
=\sum_{m,n}\int\frac{dp}{2\pi}\ln\Big[p^{2}+\big(k_{n}^{(\lambda)}\big)^{2}+\big(\tilde k_{m}^{(\eta)}\big)^{2}\Big]\ ,\qquad \ \   k_{n}^{(\lambda)}= {n+\l\ov R},\ \  \ \td k_{m}^{(\eta)}= {m+\eta\ov \tR}\ , \no 
\\
& Z(s)
\equiv 
\sum_{m,n}
\int\frac{dp}{2\pi} 
\Big[p^2+\big(k_{n}^{(\lambda)}\big)^{2}+\big(\tilde k_{m}^{(\eta)}\big)^{2}\Big]^{-s}
=\sum_{m,n}\frac{1}{(4\pi)^{1/2}}\frac{\Gamma(s-\hal)}{\Gamma(s)}\Big[\big(k_{n}^{(\lambda)}\big)^{2}+\big(\tilde k_{m}^{(\eta)}\big)^{2}\Big]^{\hal-s}\label{214}\ .}
Thus 
\be\la{c22}
\Omega^{(\lambda,\eta)}=  
\sum_{m,n}\sqrt{\big(k_{n}^{(\lambda)}\big)^{2}+\big(\tilde k_{m}^{(\eta)}\big)^{2}}=\Z_{\lambda,\eta}(-\tfrac{1}{2})\ ,
\ee
 where $\Z_{\lambda,\eta}(-\frac{1}{2})$ is the 
  Epstein zeta function  given  in \rf{f17}.
  
\begin{figure}[H]
    \centering
    \begin{subfigure}{0.45\textwidth}
        \centering
        \includegraphics[width=\textwidth]{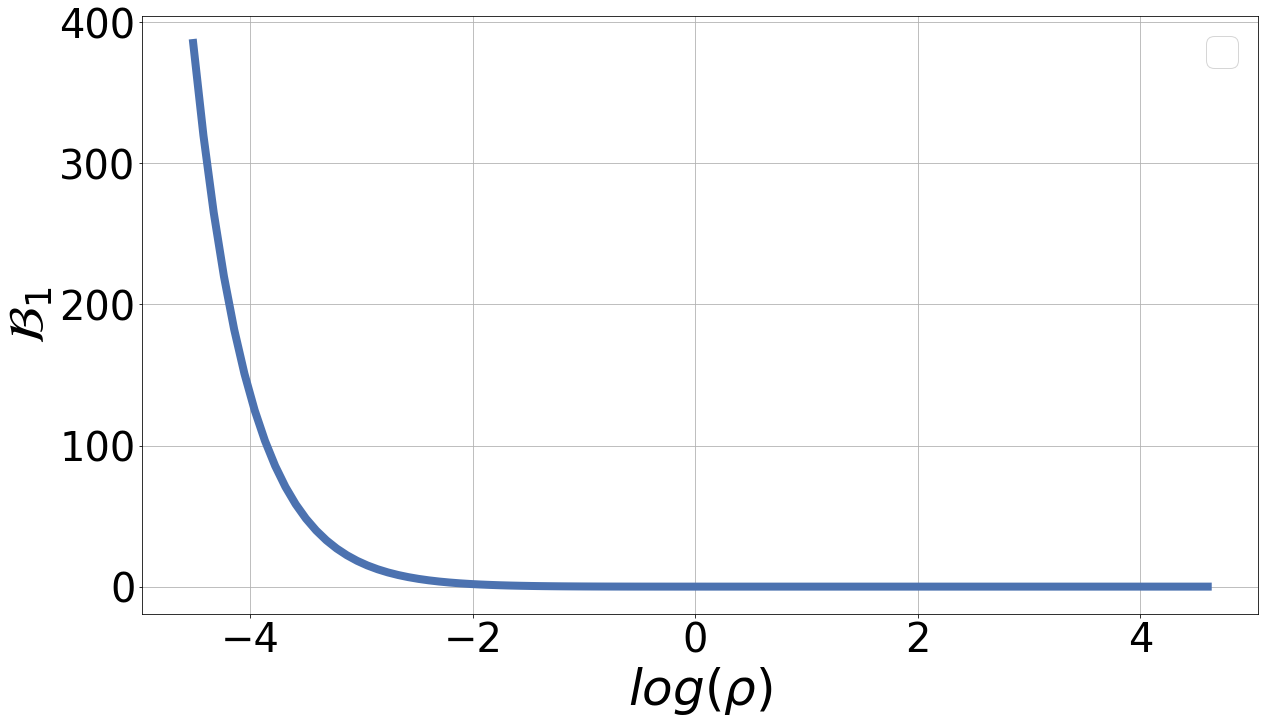}
        \caption{}
        \label{fig2-1}
    \end{subfigure}
     \hfill
    \begin{subfigure}{0.48\textwidth}
        \centering
        \includegraphics[width=\textwidth]{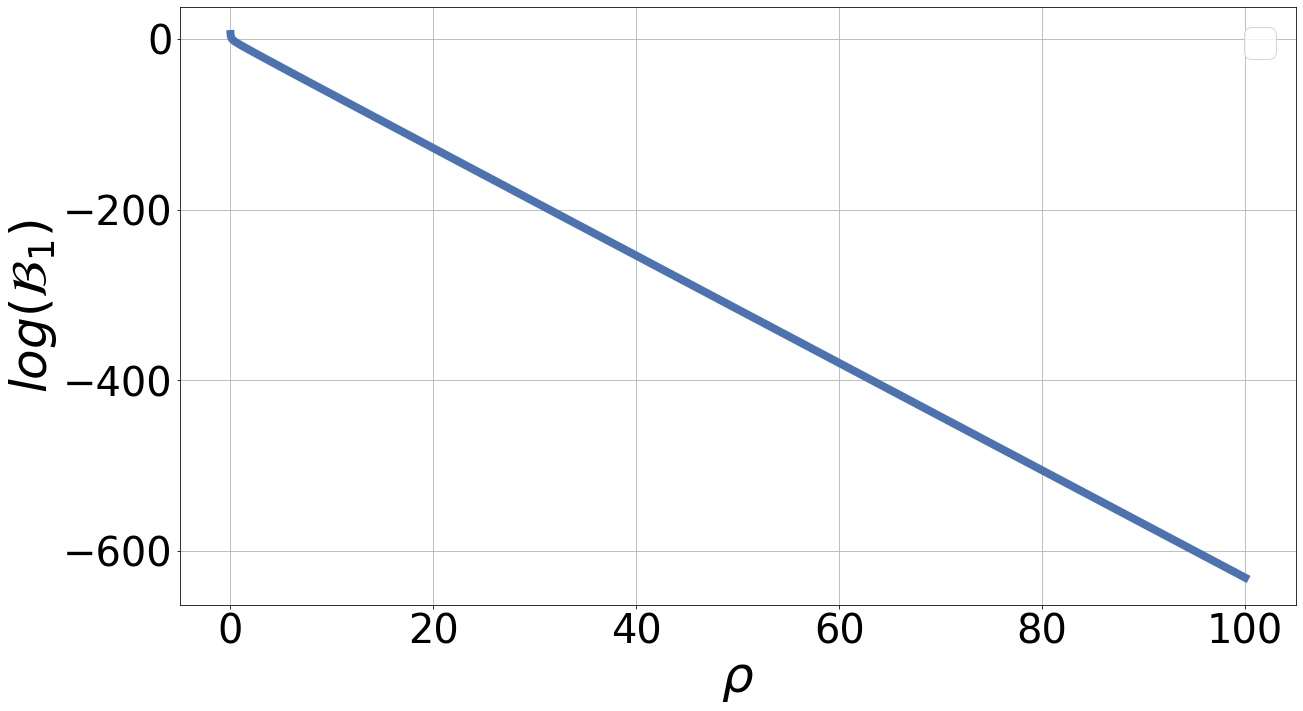}
        \caption{}
        \label{fig2-3}
    \end{subfigure}
    \caption{\small 
    Plots of  \(\mathcal B_1\) and \(\log\mathcal B_1\). 
    }
    \label{fig23}
\end{figure}

  Explicitly, for the $\mathbb R \times S^1_P \times \tilde  S^1_P$, 
  $\mathbb R \times S^1_{A} \times \tilde  S^1_{P}$,  $\mathbb R \times S^1_{P} \times \tilde  S^1_{A}$  and $\mathbb R \times S^1_{A} \times \tilde  S^1_{A}$
  cases expressed  in terms of the $\B_\s$ ($\s=\pm 1, \pm \hal$)  functions \rf{E7},\rf{E8},\rf{f4} 
  of 
  the same   argument $\r={R\over \tR}$ we  have 
  \al{
&\Z_{0,0}(-\tfrac{1}{2})
=-\frac{1}{6R}-\frac{\zeta(3)R}{2\pi^{2}\tilde R^{2}}-\frac{4\,\B_{1}}{\pi \tilde R}\ ,
\qquad\qquad  {\Z}_{\frac 12,0}(-\tfrac{1}{2})
=\frac{1}{12R}-\frac{\zeta(3)R}{2\pi^{2}\tilde R^{2}}-\frac{4\B_{-1}}{\pi \tilde R} \ , \label{C3}
\\
&{\Z}_{0,\frac 12}(-\tfrac{1}{2})=\frac{3\zeta(3)}{8\pi^{2}}\frac{R}{\tilde R^{2}}-\frac{4\,\B_{\frac 12}}{\pi \tilde R}
 \ , \qquad \qquad \qquad 
{\Z}_{\frac 12,\frac 12}(-\tfrac{1}{2})=
\frac{3\zeta(3)}{8\pi^{2}}\frac{R}{\tilde R^{2}}-\frac{4\,\B_{-\frac 12}}{\pi \tilde R}\label{C4}
 \ , 
\\
&\B_{1}=\sum_{n,m=1}^{\infty}\frac{m}{n}K_{1}\Big(2\pi mn\frac{R}{\tilde R}\Big)\label{D7}\ ,\qquad \qquad \qquad \ \ \ 
\B_{-1}=\sum_{n,m=1}^{\infty}(-1)^n\frac{m}{n}K_{1}\Big(2\pi mn\frac{R}{\tilde R}\Big)\ ,\\
&\B_{\tfrac 12}=\sum_{m=0}^{\infty}\sum_{n=1}^{\infty}\frac{m+\hal}{n}K_{1}\Big(2\pi n(m+\ha)\frac{R}{\tilde R}\Big)\ ,\qquad 
\B_{-\frac 12}=\sum_{m=0}^{\infty}\sum_{n=1}^{\infty}(-1)^n\frac{m+\hal}{n}K_{1}\Big(2\pi n(m+\ha)\frac{R}{\tilde R}\Big)\ .
}
The derivatives of the $\B_\s$ functions are given by\footnote{Here we used that  for  the  modified Bessel functions  one has 
$
K_1'(z)=-K_0(z)-\frac{1}{z}K_1(z)\ .
$
}
\al{
&\partial_R \B_\s=-\frac 1 R \B_\s-\frac{2\pi}{\tilde R}\C_\s,\qquad\qquad 
 \partial_{\tilde R} \B_\s=\frac 1{\tilde R}\B_\s+\frac{2\pi R}{\tilde R^2} \C_\s,\qquad \ \ \ \ \s=\pm 1, \pm \ha \ , \label{D11}
\\
&\C_{1}=\sum_{n,m=1}^{\infty}m^2K_{0}\Big(2\pi mn\frac{R}{\tilde R}\Big)\ ,\qquad 
\qquad \C_{-1}=\sum_{n,m=1}^{\infty}(-1)^nm^2K_{0}\Big(2\pi mn\frac{R}{\tilde R}\Big)\ ,\label{D9}\\
&\C_{\frac 12}=\sum_{m=0}^{\infty}\sum_{n=1}^{\infty}{(m+\ha)^2}K_{0}\Big(2\pi n(m+\ha)\frac{R}{\tilde R}\Big)\ ,\qquad 
\C_{-\frac 12}=\sum_{m=0}^{\infty}\sum_{n=1}^{\infty}(-1)^n{(m+\ha)^2}K_{0}\Big(2\pi n(m+\ha)\frac{R}{\tilde R}\Big)\ .\no 
}
Since the modified Bessel functions admit the asymptotic expansion (\ref{f7})
($K_\nu (z)\sim \tfrac 1{\sqrt z}e^{-z}+...,\,z\gg 1$)
 the  leading terms in the expansion of the $\B_\s$ and $\C_s$  functions 
 at ${R\ov \tilde R}\gg 1$  may be written as 
\al{
&\te \B_{1}=K_{1}\big(2\pi \tfrac{R}{\tilde R}\big)+\tfrac 52K_{1}\big(4\pi\tfrac{R}{\tilde R}\big)+...\ ,\qquad 
\B_{-1}=-K_{1}\big(2\pi \tfrac{R}{\tilde R}\big)-\tfrac 32K_{1}\big(4\pi\tfrac{R}{\tilde R}\big)+...\ ,\\
&\te \B_{1/2}=\frac 12 K_{1}\big(\pi \tfrac{R}{\tilde R}\big)+\frac 14K_{1}\big(2\pi\tfrac{R}{\tilde R}\big)+\frac 53K_{1}\big(3\pi\tfrac{R}{\tilde R}\big)+\frac 18K_{1}\big(4\pi\tfrac{R}{\tilde R}\big)+...\ ,\\
&\te \B_{-\hal}=-\frac 12 K_{1}\big(\pi \tfrac{R}{\tilde R}\big)+\frac 14K_{1}\big(2\pi\tfrac{R}{\tilde R}\big)-\frac 53K_{1}\big(3\pi\tfrac{R}{\tilde R}\big)+\frac 18K_{1}\big(4\pi\tfrac{R}{\tilde R}\big)+...\ ,\\
&\te \C_{1}=K_{0}\big(2\pi \tfrac{R}{\tilde R}\big)+ 5K_{0}\big(4\pi\tfrac{R}{\tilde R}\big)+...\ ,\qquad 
\C_{-1}=-K_{0}\big(2\pi \tfrac{R}{\tilde R}\big)- 3K_{0}\big(4\pi\tfrac{R}{\tilde R}\big)+...\\
&\te \C_{\hal}=\frac 14 K_{0}\big(\pi \tfrac{R}{\tilde R}\big)+\frac 14K_{0}\big(2\pi\tfrac{R}{\tilde R}\big)+\frac 52K_{0}\big(3\pi\tfrac{R}{\tilde R}\big)+\frac 14K_{0}\big(4\pi\tfrac{R}{\tilde R}\big)+...\ ,\\
&\te \C_{-\hal}=-\frac 14 K_{0}\big(\pi \tfrac{R}{\tilde R}\big)+\frac 14K_{0}\big(2\pi\tfrac{R}{\tilde R}\big)-\frac 52K_{0}\big(3\pi\tfrac{R}{\tilde R}\big)+\frac 14K_{0}\big(4\pi\tfrac{R}{\tilde R}\big)+...\ .
}
One can check numerically  that the  Bessel-function sums that define the functions 
 $\B_\s$ and $\C_\s$ in equations (\ref{D7})--(\ref{D9}) converge very fast
 and at large $\r={R\ov \tR}$  are exponentially suppressed.
  For example, one   may plot $\B_1$ as a function of $\r$  in the interval 
  $(0.01,100)$ (see Figure \ref{fig23}).
The building blocks of the 2-loop  correction are (cf. \rf{c9},\rf{c12})
\al{
&\Delta_{11}^{(\lambda,\eta)}
=-\frac{1}{8\pi^2 R \tilde R}\sum_{n,m}\Big[{\big(k_{n}^{(\lambda)}\big)^{2}+\big(\tilde k_{m}^{(\eta)}\big)^{2}}\Big]^{\frac 12}=-\frac{1}{8\pi^2 R\tilde  R}\Z_{\lambda,\eta}(-\frac{1}{2})\ ,\\
&\Delta_{22}^{(\lambda,\eta)}
=\frac{1}{8\pi^2 R \tilde R}\sum_{n,m}{\big(k_{n}^{(\lambda)}\big)^{2}}{\Big[{\big(k_{n}^{(\lambda)}\big)^{2}+\big(\tilde k_{m}^{(\eta)}\big)^{2}}\Big]^{-\frac 12}}=-\frac{1}{8\pi^2  \tilde R}\partial_{R}\Z_{\lambda,\eta}(-\frac{1}{2})\ ,\\
&\Delta_{33}^{(\lambda,\eta)}
=\frac{1}{8\pi^2 R \tilde  R}\sum_{n,m}\big(\tilde k_{m}^{(\eta)}\big)^{2}{\Big[{\big(k_{n}^{(\lambda)}\big)^{2}+\big(\tilde k_{m}^{(\eta)}\big)^{2}}\Big]^{-\frac 12}}
=-\frac{1}{8\pi^2 R}\partial_{\tilde R}\Z_{\lambda,\eta}(-\frac{1}{2})\ .
}
In the  periodic $\mathbb R \times S^1_P \times \tilde  S^1_P$
 case, the non-zero components of $\Delta_{ab}=\Delta_{ab}^{(0,0)}$  and their products are 
\begin{align}
&\Delta_{11}=\frac{1}{48 \pi^2 R^2 \tilde R}+\frac{\zeta(3)}{16 \pi^4 \tilde R^3}+\frac{\B_{1}}{2\pi^3 R\tilde R^2}\ ,\\
&
\Delta_{22}=-\frac{1}{48 \pi^2 R^2 \tilde R}+\frac{\zeta(3)}{16 \pi^4 \tilde R^3} -\frac{\B_{1}}{2\pi^3 R\tilde R ^2}-\frac{\C_{1}}{\pi^2 \tilde R^3}\ ,\qquad \label{56}
\quad 
\Delta_{33}=-\frac{\zeta(3)}{8 \pi^4 \tilde R^3}+\frac{\C_{1}}{\pi^2 \tilde R^3}\ ,\\
&\Delta_{ab}\Delta^{ab}
=
\frac{1}{1152\pi^4\tilde R^2R^4}
+
\frac{3\zeta(3)^2}{128\pi^8\tilde R^6}
+
\frac{\B_{1}}{24\pi^5 R^3\tilde R^3}
+
\frac{\C_{1}}{24\pi^4 R^2 \tilde R^4}
-
\frac{3\zeta(3)\C_{1}}{8\pi^6 \tilde R^6}
\nonumber\\
&\qquad \qquad \qquad 
+
\frac{\B_{1}^2}{2\pi^6 R^2 \tilde R^4}
+
\frac{\B_{1}\C_1}{\pi^5 R \tilde R^5}
+
\frac{2\C^2_1}{\pi^4  \tilde R^6}\ . 
\end{align}
In the  $\mathbb R \times S^1_{A} \times \tilde  S^1_{P}$  case we get 
\al{
& \Delta^{(\frac12,0)}_{11}=-\frac{1}{96 \pi^2 R^2\tilde R}+\frac{\zeta(3)}{16 \pi^4 \tilde R^3}+\frac{\B_{-1}}{2\pi^3 R\tilde R^2}\ ,\qquad \label{569}\\
&
\Delta^{(\frac12,0)}_{22}=\frac{1}{96 \pi^2 R^2\tilde R}+\frac{\zeta(3)}{16 \pi^4 \tilde R^3}-\frac{\B_{-1}}{2\pi^3 R \tilde R^2}-\frac{\C_{-1}}{\pi^2 \tilde R^3}\ , 
\qquad
\Delta^{(\frac12,0)}_{33}=-\frac{\zeta(3)}{8 \pi^4 \tilde R^3} +\frac{\C_{-1}}{\pi^2 \tilde R^3}\ , \\
&\Delta^{(\frac12,0)}_{ab}\Delta^{{(\frac12,0)}ab}
=
\frac{1}{4608\pi^4R^4\tilde R^2}
+
\frac{3\zeta(3)^2}{128\pi^8\tilde R^6}
-
\frac{\B_{-1}}{48\pi^5 R^3\tilde R^3}
-
\frac{\C_{-1}}{48\pi^4 R^2 \tilde R^4}
-
\frac{3\zeta(3)\C_{-1}}{8\pi^6 \tilde R^6}
\nonumber\\
&\qquad \qquad \qquad  \qquad 
+
\frac{\B_{-1}^2}{2\pi^6 R^2 \tilde R^4}
+
\frac{\B_{-1}\C_{-1}}{\pi^5 R \tilde R^5}
+
\frac{2\C^2_{-1}}{\pi^4  \tilde R^6}\\
&\Delta_{ab}\Delta^{{(\frac12,0)}ab}=-\frac{1}{2304\pi^{4}R^{4}\tilde R^{2}}+\frac{3\zeta(3)^{2}}{128\pi^{8}\tilde R^{6}}+
\frac{2\B_{-1}-\B_1}{96\pi^5 R^3\tilde R^3}
+
\frac{2\C_{-1}-\C_1}{96\pi^4 R^2\tilde R^4}
-
\frac{3\zeta(3)(\C_{-1}+\C_1)}{16\pi^6\tilde R^6}
\nonumber
\\
&\qquad \qquad \qquad  \qquad 
+
\frac{\B_{-1}\B_1}{2\pi^6 R^2\tilde R^4}
+
\frac{\B_{-1}\C_{1}+\B_1\C_{-1}}{2\pi^5 R\tilde R^5}
+
\frac{2\C_{-1}\C_1}{\pi^4 \tilde R^6}
\ . }
Similarly, in  the cases of 
 $\mathbb R \times S^1_P \times \tilde  S^1_{A}$   and $\mathbb R \times S^1_A \times \tilde  S^1_{A} $  expressed in terms of the same  functions $\B_\s$,$\C_\s$ of $\r={R\ov \tR}$   we have, respectively, 
\al{
&\Delta^{(0,\frac12)}_{11}
=-\frac{3\zeta(3)}{64\pi^4 \tilde R^3}
+\frac{\B_{\frac 12}}{2\pi^3 R\tilde R^2}\ , \\
&
\Delta^{(0,\frac12)}_{22}=
-\frac{3\zeta(3)}{64 \pi^4 \tilde R^3} 
-\frac{\B_{\frac12}}{2\pi^3R \tilde R^2 }
-\frac{\C_{\frac12}}{\pi^2 \tilde R^3}
\ , \qquad \quad 
\Delta^{(0,\frac12)}_{33}=
\frac{3\zeta(3)}{32 \pi^4 \tilde R^3}
+\frac{\C_{\frac12}}{\pi^2 \tilde R^3}\ ,
\\
&\Delta^{(0,\frac12)}_{ab}\Delta^{{(0,\frac12)}ab}
=
\frac{27\zeta(3)^2}{2048\pi^8\tilde R^6}
+
\frac{9\zeta(3)\C_{\frac12}}{32\pi^6 \tilde R^6}
+
\frac{\B^2_{\frac12}}{2\pi^6 R^2 \tilde R^4}
+
\frac{\B_{\frac12}\C_{\frac12}}{\pi^5 R \tilde R^5}
+
\frac{2\C^2_{\frac12}}{\pi^4 \tilde R^6}\ ,
\\
&\Delta_{ab}\Delta^{{(0,\frac12)}ab}=-\frac{9\zeta(3)^2}{512\pi^{8}\tilde R^{6}}
+
\frac{\B_{\frac12}}{48\pi^5 R^3\tilde R^3}
+
\frac{\C_{\frac12}}{48\pi^4 R^2\tilde R^4}
+
\frac{3\zeta(3)(3\C_{1}-4\C_{\frac12})}{64\pi^6\tilde R^6}\nonumber\\
&\qquad \qquad \qquad  \qquad 
+
\frac{\B_{1}\B_{\frac12}}{2\pi^6 R^2\tilde R^4}
+
\frac{\B_{\frac12}\C_{1}+\B_{1}\C_{\frac12}}{2\pi^5 R\tilde R^5}
+
\frac{2\C_{1}\C_{\frac12}}{\pi^4 \tilde R^6}
\ , \\ 
&\Delta^{(\frac{1}{2},\frac{1}{2})}_{11}
=-\frac{3\zeta(3)}{64\pi^4 \tilde R^3}
+\frac{\B_{-\frac12}}{2\pi^3 R\tilde R^2}
\ ,\\
&\Delta^{(\frac{1}{2},\frac{1}{2})}_{22}=
-\frac{3\zeta(3)}{64 \pi^4 \tilde R^3} 
-\frac{\B_{-\frac12}}{2\pi^3R \tilde R^2 }
-\frac{\C_{-\frac12}}{\pi^2 \tilde R^3}
,\qquad \label{564}
\qquad \  
\Delta^{(\frac{1}{2},\frac{1}{2})}_{33}=
\frac{3\zeta(3)}{32 \pi^4 \tilde R^3}
+\frac{\C_{-\frac12}}{\pi^2 \tilde R^3}
\ ,\\
&\Delta^{(\frac{1}{2},\frac{1}{2})}_{ab}\Delta^{{(\frac{1}{2},\frac{1}{2})}ab}
=
\frac{27\zeta(3)^2}{2048\pi^8\tilde R^6}
+
\frac{9\zeta(3)\C_{-\frac12}}{32\pi^6 \tilde R^6}
+
\frac{\B^2_{-1/2}}{2\pi^6 R^2 \tilde R^4}
+
\frac{\B_{-\frac 12}\C_{-\frac12}}{\pi^5 R \tilde R^5}
+
\frac{2\C^2_{-\frac12}}{\pi^4 \tilde R^6}\ ,
\\
&\Delta_{ab}\Delta^{{(\frac{1}{2},\frac{1}{2})}ab}=-\frac{9\zeta(3)^2}{512\pi^{8}\tilde R^{6}}
+
\frac{\B_{-\frac12}}{48\pi^5 R^3\tilde R^3}
+
\frac{\C_{-\frac12}}{48\pi^4 R^2\tilde R^4}
+
\frac{3\zeta(3)(3\C_{1}-4\C_{-\frac12})}{64\pi^6\tilde R^6}\nonumber\\
&\qquad \qquad \qquad  \qquad 
+
\frac{\B_{1}\B_{-\frac12}}{2\pi^6 R^2\tilde R^4}
+
\frac{\B_{-\frac12}\C_{1}+\B_{1}\C_{-\frac12}}{2\pi^5 R\tilde R^5}
+
\frac{2\C_{1}\C_{-\frac12}}{\pi^4 \tilde R^6}\ . 
}
Finally, let us   
give the explicit expressions for the functions $\Ee_\s$ in \rf{3121}--\rf{d415}:
\al{
&\Ee_{-1}=\Big[
\tfrac{ 1}{4}\nf N\big(2\B_{-1}-\B_1\big)
+\tfrac{1}{8}\nf^2\B_{-1}
\Big]
\frac{1}{24\pi^3 R^2 \tilde R^2}
+\Big[
\tfrac{ 1}{4}\nf N\big(2\C_{-1}-\C_1\big)
+\tfrac{1}{8}\nf^2\C_{-1}
\Big]
\frac{1}{24\pi^2 R \tilde R^3}\nonumber
\\
&\qquad \quad 
-\Big[
\tfrac{ 1}{2}\nf N\big(\C_{-1}+ \C_1\big)
-\tfrac{1}{4}\nf^2\C_{-1}
\Big]
\frac{3R\zeta(3)}{8\pi^4 \tilde R^5}
+\Big[
N \nf \B_{-1}  \B_1
-\tfrac{1}{4}\nf^2\B_{-1}^2
\Big]
\frac{1}{2\pi^4 R \tilde R^3}\nonumber
\\
&\qquad \quad 
+\Big[
\tfrac{ 1}{2}\nf N
\big(\B_{-1} \C_1+\C_{-1} \B_1\big)
-\tfrac{1}{4}\nf^2\B_{-1}\C_{-1}
\Big]
\frac{1}{\pi^3  \tilde R^4}
+\Big[
N \nf \C_{-1} \C_1
-\tfrac{1}{4}\nf^2\C_{-1}^2
\Big]
\frac{2R}{\pi^2 \tilde R^5}\label{e40}\ , \\
&\Ee_{\hal}=
\tfrac{1 }{2}\nf N\B_{\frac 12}
\frac{1}{24\pi^3 R^2 \tilde R^2}
+
\tfrac{1 }{2}\nf N\C_{\frac 12}
\frac{1}{24\pi^2 R \tilde R^3}\nonumber
\\
&\qquad \quad 
+\Big[
\tfrac{ 1}{2}\nf N\big(\tfrac 34\C_{1}- \C_{\frac 12}\big)
-\tfrac{3}{16}\nf^2\,\C_{\frac 12}
\Big]
\frac{3R\zeta(3)}{8\pi^4 \tilde R^5}
+
\Big[
N \nf \B_{1}  \B_{\frac 12}
-\tfrac{1}{4}\nf^2\B_{\frac 12}^2
\Big]
\frac{1}{2\pi^4 R \tilde R^3}\nonumber
\\
&\qquad \quad 
+
\Big[
\tfrac{ 1}{2}\nf N
\big(\B_{\frac 12} \C_1+ \B_1\C_{\frac12}\big)
-\tfrac{1}{4}\nf^2\B_{\frac 12}\C_{\frac 12}
\Big]
\frac{1}{\pi^3  \tilde R^4}
+
\Big[
N \nf \C_{1} \C_{\frac 12}
-\tfrac{1}{4}\nf^2\,\C_{\frac 12}^2
\Big]
\frac{2R}{\pi^2 \tilde R^5}\label{e42}
\ , \\
&\Ee_{-\frac 12}=
\tfrac{1 }{2}\nf N\B_{-\frac12}
\frac{1}{24\pi^3 R^2 \tilde R^2}
+
\tfrac{1 }{2}\nf N\C_{-\frac 12}
\frac{1}{24\pi^2 R \tilde R^3}\nonumber
\\
&\qquad \quad
+\Big[
\tfrac{ 1}{2}\nf N\big(\tfrac 34\C_{1}- \C_{-\frac12}\big)
-\tfrac{3}{16}\nf^2\,\C_{-\frac12}
\Big]
\frac{3R\zeta(3)}{8\pi^4 \tilde R^5}
+
\Big[
\nf N  \B_{1}  \B_{-\frac 12}
-\tfrac{1}{4}\nf^2\B_{-\frac 12}^2
\Big]
\frac{1}{2\pi^4 R \tilde R^3}\label{e428}
\\
&\qquad \quad
+
\Big[
\tfrac{ 1}{2}\nf N
\big(\B_{-\frac 12} \C_1+ \B_1\C_{-\frac12}\big)
-\tfrac{1}{4}\nf^2\,\B_{-
\frac12}\C_{-\frac 12}
\Big]
\frac{1}{\pi^3  \tilde R^4}
+
\Big[
\nf N  \,\C_{1} \C_{-\frac 12}
-\tfrac{1}{4}\nf^2\,\C_{-\frac 12}^2
\Big]
\frac{2R}{\pi^2 \tilde R^5}\nonumber\ . 
}

\def \tr {{\rm tr}}

\section{3-loop ``basketball'' diagram integrals }
\label{ap4}

The  integrals \rf{516} that   determine the  3-loop  basketball  diagram contribution (\ref{bb})  to the 
 membrane  energy  on $\mathbb R^2 \times S^1$  are 
\al{\la{d1}
I_1=\int d^3 \sigma\,\big(\tr D^2\big)^2, \qquad \qquad I_2=\int d^3 \sigma \ \tr D^4 \ ,\qquad \qquad  D_{a b}(\sigma)=-\partial_a \partial_b G(\sigma), 
}
where $D_{ab}$ is the second  derivative
of the scalar  propagator $G$   ($k_n = {n\ov \tR}$)
\iffa 
\footnote{The propagator $G(\sigma)$ is  found   by noticing the Poisson resummation properties  
$
\frac{1}{R}
\sum_{n\in\mathbb Z}
f\left(\frac{n+\eta}{R}\right)
=\sum_{m\in\mathbb Z}
e^{-2\pi i m\eta}
\int_{-\infty}^{\infty}dk\,
e^{2\pi i mRk}f(k).
$
that rewrites the compact $\mathbb R^2 \times \tilde S^1$ propagator in terms of the non-compact $\mathbb R^3$ propagators which is then evaluated by 
$
\int\frac{d^3P}{(2\pi)^3}
\frac{e^{iP\cdot x}}{P^2}
=\frac{1}{4\pi|x|}.$
}
\fi 
\al{
\quad G(\sigma)=\frac{1}{2\pi R}
\sum_{n\in\mathbb Z}
\int\frac{d^2p}{(2\pi)^2}
\frac{
e^{i\vec p\cdot\vec \sigma}
e^{ik_n\sigma_3}
}{
\vec p^{\,2}+k_n^2}= \sum_{m \in \mathbb{Z}} \frac{1}{4\pi\sqrt{\sigma_1^2+\sigma_2^2+\big(\sigma_3+2 \pi \tilde{R} m\big)^2}} \ .
}
Here   we wrote the propagator on $\mathbb R^2 \times S^1$  in terms of coordinates  on $\mathbb R^3$ 
using  sum over  images.
To isolate the singular part of $D_{ab}$ that  comes from the $m=0$ term in the sum    we note that in 3d 
\iffa 
The second derivative propagator contains extra distributional piece that we need to determine. Since $1/\sigma$ is singular and hence not ordinarily differentiable at $r=0$, while our integrals include the coincident point $\sigma=0$, its derivatives is then  defined distributionally, which accounts for the resulting contact terms. Since the Coulomb Green function satisfies the
distributional identity 
\fi
$\del^2\frac{1}{\sigma}=-4\pi\delta^{(3)}(\sigma)$  and 
\begin{equation}
\partial_a\partial_b\frac{1}{\s}
=\operatorname{P.V.}
\frac{3\sigma_a\sigma_b-\delta_{ab}\s^2}{\s^5}
-\tfrac {4\pi}3 \,\delta_{ab}\delta^{(3)}(\sigma)\ .
\end{equation}
Therefore we can represent $D_{ab}$ as a sum  of a contact term $S_{ab}$  and 
a ``separated-points''  term $D_{ab}^{\mathrm{sep}}$
\al{
D_{ab}(\sigma)
=&
S_{ab}
+
D_{ab}^{\mathrm{sep}}(\sigma),\label{bb4}\qquad\qquad \qquad S_{ab}=\tfrac{1}{3}
\delta_{ab}\delta^{(3)}(\sigma),\qquad \ \ \  
D_{ab}^{\mathrm{sep}}(\sigma) = D^{(0)}_{ab}+D^{\rm{reg}}_{ab}\ , 
 \\
D^{(0)}_{ab}=&
-\operatorname{P.V.}
\frac{
3\sigma_a\sigma_b-\delta_{ab}\sigma^2
}{
4\pi|\sigma|^5
}
\ , \qquad 
D^{\rm{reg}}_{ab}=-\sum_{m\neq0}
\partial_a\partial_b \frac{1}{4\pi\sqrt{\sigma_1^2+\sigma_2^2+\big(\sigma_3+2 \pi \tilde{R} m\big)^2}} \ .\la{d5}
}
For the integral  of a  function $F$ of $D$ we then have 
  \al{
\!\int\! d^3\s\,F[D]=\!\int\! d^3\s\,F[C+D^{\rm sep}]=\!\int\! d^3\s\,\Big(F[D^{\rm{sep}}]+S_{ab}\frac{\partial F}{\partial D_{ab}}+...\Big).
  }
  Since  $S_{ab}={1\ov 3} \delta_{ab}\delta^{(3)}(\sigma)$, the  terms in $F[D]$ 
   containing higher than first  power  in expansion in 
    $S_{ab}$ would not contribute if we set $\ \delta^{(3)}(0)=0$
    (using, e.g., dimensional regularization), i.e.
     \al{
\!\int\! d^3\s\,F[D]=\!\int\! d^3\s\,F[S+D^{\rm sep}]=\!\int\! d^3\s\,F[D^{\rm{sep}}]+\tfrac 13\delta_{ab}\frac{\partial F}{\partial D_{ab}}\Big|_{D\to D^{\rm sep} (0)} \ .  }
Using \rf{b1}   we then have  for the regularized   value of the integral 
\iffa
  \al{
  \frac{\partial F}{\partial D_{ab}}\Big|_{\sigma\to 0}=  \frac{\partial F}{\partial D_{ab}}\Big|_{D\to \Delta}\ ,
  }
  replacing the $D_{ab}^{\rm sep}(0)$ with coincident derivative propagator $\Delta_{ab}$,
  we therefore have the regulaised expression
  \fi
\begin{equation} \Big[\!\int\! d^3\s\,F[D]\Big]_{\rm reg} = 
\!\int\! d^3\s\,F[D^{\rm{sep}}]
+ \tfrac13\,\delta_{ab}\frac{\partial F}{\partial D_{ab}}\Big|_{D\to\Delta}\  , 
\label{bb19}
\end{equation}
where $\Delta_{ab}$ is given in \rf{512}. 
 For $F=(\tr D^2)^2$ one has
$\delta_{ab}\partial F/\partial D_{ab}\Big|_{D\to\Delta}=4\,\tr \Delta\,\tr \Delta^2=0$ (using  \rf{b1}), so that $I_1$  does not 
receive a contact contribution and thus may be computed  
using directly the value of $D$ at separated points. 
Then  using (\ref{bb19}), \rf{512}  and \rf{ne}
 we obtain for  $I_1$ and $I_2$ in \rf{d1}
 \al{&
  I_1=\!\int\! d^3\s\big[  \tr  (D^{\mathrm{sep}})^2  \big]^2,\qquad \qquad \qquad 
  I_2=\tfrac 12\!\int\! d^3\s\big[\tr (D^{\mathrm{sep}})^2\big]^2+\tfrac 43 {\tr}(\Delta^3)\ \la{d9},\\
  &\qquad I_2= \tfrac 12 I_1-\frac{\zeta(3)^3}{512 \pi^{12} \tilde R^9}\ .\la{d10}
}
It thus  remains only to evaluate $I_1$  depending on  $\tr (D^{\rm{sep}})^2$
that we may write as 
\al{
 &\tr( D^{\rm{sep}})^2  =A+B+C,\qquad \ \ \  \ \  
A=\tr[(D^{(0)})^2]=\frac 3{8\pi^2\sigma^6},\la{d11}\\
&B=2\tr(D^{(0)}D^{\rm reg})=-\frac{3\alpha}{2\pi \sigma^3}(1-3\cos^2\theta)+O(\sigma^{-1}),\qquad\ \ \ \ 
C=\tr[(D^{\rm reg})^2]=\tr(\Delta^2)+O(\sigma^2)\ , \la{d12}
}
where $\cos\theta = {\s_3\ov \s}$. 
The regular  part $D^{\rm reg}_{ab}(\sigma)$ in  \rf{d5} admits the expansion
\begin{equation}
D^{\rm reg}_{ab}(\sigma)
=
\Delta_{ab}+D^{[2]}_{ab}(\sigma)+D^{[4]}_{ab}(\sigma)+\cdots,
\qquad
D^{[2]}=\O(\s^2),\qquad D^{[4]}=\O(\s^4),\ \ \ldots .
\end{equation}
An explicit expansion of the resulting integrand $I_1$ shows that, after
angular averaging, all terms between $\s^{-12}$ and $\s^{-3}$ vanish except for
the $\s^{-6}$ term.  Thus the replacement
$D^{\rm reg}(\sigma)\to\Delta$ in \rf{d12} is sufficient for determining the UV
power divergences and  concluding about the absence of a logarithmic divergence.

Averaging over the angular dependence   we get 
\be
\left\langle[\tr( D^{\rm{sep}})^2 \right\rangle_{\rm angl} \equiv
\tfrac12\int_{-1}^{1}d\cos\theta\, \tr( D^{\rm{sep}})^2
=\frac{9}{64\pi^4\sigma^{12}}
+\frac{63\alpha^2}{10\pi^2\sigma^6}
+O(\sigma^{-2}). \la{d13}
\ee
Then integrating  \rf{d13} 
over the radial direction  near  $\s>\epsilon$  we get 
\be
I_1(\epsilon)=4\pi\int_\epsilon d\s\,\s^2
\Big[\frac{9}{64\pi^4\s^{12}}
+\frac{63\alpha^2}{10\pi^2\s^6}+O(\s^{-2})\Big] + ...=
\frac{1}{16\pi^3\epsilon^9}
+\frac{42\alpha^2}{5\pi\epsilon^3}
+O(\epsilon) + ... \ .\la{d14}
\ee
We evaluated the  integral $I_1$ for $D$ at separated points in \rf{d9}   numerically.  In the neighbourhood
of $\s=0$  we used  the 
 expansion\footnote{One uses  that the propagator can be written in terms of a harmonic function sum as 
 
 $\ \ \ \ G^{\rm reg}(\sigma,\theta)=
\sum_{k=1}^{\infty}
\frac{\zeta(2k+1)}
{2\pi(2\pi\tilde R)^{2k+1}}
\sigma^{2k}P_{2k}(\cos\theta)$ with $\cos\theta= \sigma_3/\sigma$. }
 of
$D^{\rm reg}$ and subtracted  the singular terms as in \rf{d14},\rf{521}
before numerical integration.   The remaining
radial and angular integrations were evaluated using the Gauss--Legendre
quadrature.  

\


\bibliography{bibt}
\bibliographystyle{JHEP}

\end{document}